\documentclass[usenatbib,onecolumn]{rasti}
\usepackage{newtxtext,newtxmath}
\usepackage[T1]{fontenc}

\DeclareRobustCommand{\VAN}[3]{#2}
\let\VANthebibliography\thebibliography
\def\thebibliography{\DeclareRobustCommand{\VAN}[3]{##3}\VANthebibliography}

\usepackage{graphicx}	% Including figure files
\usepackage{amsmath}	% Advanced maths commands
\usepackage{threeparttable}
\usepackage{booktabs}
\usepackage{enumerate}
\usepackage[shortlabels]{enumitem}
\usepackage{color}
\usepackage{multirow} 
\usepackage[table]{xcolor}  
\usepackage{makecell}
\DeclareSymbolFont{CMletters}{OT1}{cmtt}{m}{ui}
\DeclareMathSymbol{g}{\mathord}{CMletters}{`g}
\usepackage{caption}

\makeatletter
\newcommand\subsubsubsection{\@startsection{paragraph}{4}{\z@}{-2.5ex\@plus -1ex \@minus -.25ex}{1.25ex \@plus .25ex}{\normalfont\normalsize\itshape}}
\newcommand\subsubsubsubsection{\@startsection{subparagraph}{5}{\z@}{-2.5ex\@plus -1ex \@minus -.25ex}{1.25ex \@plus .25ex}{\normalfont\normalsize\itshape}}
\makeatother

\usepackage{listings}
\usepackage{xcolor}
\DeclareCaptionFormat{listing}{\textbf{#1}#2#3}
\definecolor{codegreen}{rgb}{0,0.6,0}
\definecolor{codegray}{rgb}{0.5,0.5,0.5}
\definecolor{codered}{rgb}{0.7,0.1,0.1}
\definecolor{codeblue}{rgb}{0,0,1}
\definecolor{codegreenblue}{rgb}{0.2,0.5,0.5}

\newcommand\pythonstyle{\lstset{
    language=Python,
    backgroundcolor=\color{white},   
    commentstyle=\color{codegreenblue}\itshape, 
    keywordstyle=\color{codegreen}\bfseries,    % Bold for primary   
    numberstyle=\tiny\color{codegray},
    stringstyle=\color{codered},
    basicstyle=\ttfamily,
    breakatwhitespace=false,         
    breaklines=false,                 
    captionpos=b,                    
    keepspaces=false,                 
    numbers=none,                    
    numbersep=5pt,                  
    showspaces=false,                
    showstringspaces=false,
    escapeinside={(*@}{@*)},
    showtabs=false,                  
    tabsize=2,
    emph={extract_broad, scipy_Voigt_profile, scipy_wofz_Voigt_profile},
    emphstyle={\color{codeblue}},
    emph={[2]max, round, values, astype},
    emphstyle={[2]\color{codegreen}},
    emph={[3]numpy, np, ac},
    emphstyle={[3]\color{codeblue}\bfseries},
    emph={[4]as},
    emphstyle={[4]\color{codegreen}\bfseries},
    literate=
    *{0}{{{\color{codegray}0}}}{1}
     {1}{{{\color{codegray}1}}}{1}
     {2}{{{\color{codegray}2}}}{1}
     {3}{{{\color{codegray}3}}}{1}
     {4}{{{\color{codegray}4}}}{1}
     {5}{{{\color{codegray}5}}}{1}
     {6}{{{\color{codegray}6}}}{1}
     {7}{{{\color{codegray}7}}}{1}
     {8}{{{\color{codegray}8}}}{1}
     {9}{{{\color{codegray}9}}}{1}
     {scipy.special}{{{\color{codeblue}\bfseries scipy.special}}}{11}
     {scipy.signal}{{{\color{codeblue}\bfseries scipy.signal}}}{11}
     {astropy.constants}{{{\color{codeblue}\bfseries astropy.constants}}}{15}
     {np.}{{{\color{black} np.}}}{3}
     {1j}{{{\color{codegray} 1j}}}{2}
     {0.1}{{{\color{codegray} 0.1}}}{3}
     {1.0}{{{\color{codegray} 1.0}}}{3}
     {296.0}{{{\color{codegray} 296.0}}}{5}
     {c2}{{{\color{black} c2}}}{2}
}
}

\lstnewenvironment{python}[1][]{
\pythonstyle \lstset{#1} }{}

\usepackage{color}

\newcommand{\blue}[1]{{\color{blue} #1}}

\makeatletter
\renewcommand{\maketag@@@}[1]{\hbox{\m@th\normalsize\normalfont#1}}%
\makeatother

\title[PyExoCross]{\textsc{PyExoCross}: a Python program for generating spectra and cross-sections from molecular line lists}

\author[Jingxin Zhang et al.]{
Jingxin Zhang\thanks{E-mail: jingxin.zhang.19@ucl.ac.uk (JZ); j.tennyson@ucl.ac.uk (JT)},
Jonathan Tennyson\blue{\footnotemark[1]} and Sergei N. Yurchenko
\\
Department of Physics and Astronomy, University College London, Gower Street, WC1E 6BT London, UK}

\date{Accepted 2024 April 22. Received 2024 April 2; in original form 2023 October 10}

\pubyear{2024}

\begin{document}
\label{firstpage}
\pagerange{\pageref{firstpage}--\pageref{lastpage}}
\maketitle

\begin{abstract}
\textsc{PyExoCross} is a Python adaptation of the \textsc{ExoCross} Fortran application, \textsc{PyExoCross} is designed for post-processing the huge molecular line lists generated by the ExoMol project and other similar initiatives such as the HITRAN and HITEMP databases. \textsc{PyExoCross} generates absorption and emission stick spectra, cross-sections, and other properties (partition functions, specific heats, cooling functions, lifetimes, and oscillator strengths) based on molecular line lists. \textsc{PyExoCross} calculates cross-sections with four line profiles: Doppler, Gaussian, Lorentzian, and Voigt profiles in both sampling and binned methods; a number of options are available for computing Voigt profiles which we test for speed and accuracy. \textsc{PyExoCross} supports importing and exporting line lists in the ExoMol and HITRAN/HITEMP formats. \textsc{PyExoCross} also provides conversion between the ExoMol and HITRAN data formats. In addition, \textsc{PyExoCross} has extra code for users to automate the batch download of line list files from the ExoMol database.  
\end{abstract}

% Include between one and six keywords.

%%%%%%%%%%%%%%%%%%%%%%%%%%%%%%%%%%%%%%%%%%%%%%%%%%

%%%%%%%%%%%%%%%%% BODY OF PAPER %%%%%%%%%%%%%%%%%%

\section{Introduction}

A major motivation of modern astronomy is the desire to characterize and model the atmospheres of the many extrasolar planets discovered over the previous three decades \citep{13TiEnCo.exo,19Madhus,jt853}. Doing this requires significant quantities of spectroscopic data with hot objects requiring billions of 
lines to accurately reproduce spectroscopic features of their atmospheres \citep{jt572}. The ExoMol project \citep{jt528} is systematically constructing comprehensive molecule line lists to aid the study of the atmospheres of exoplanets and other hot objects which contain molecules such as brown dwarfs, cool stars, and photon-dominated regions of the interstellar medium. ExoMol systematically
produces its data in a series of papers which are summarized in data release papers \citep{jt631,jt810}. 
Similar endeavours include several well-known  molecular spectroscopy databases used by astronomers such as
  HITRAN \citep{jt857}, HITEMP \citep{HITEMP2010,jt763},  GEISA \citep{jt504},  CDMS  \citep{muller2013cdms,CDMS}, the JPL spectroscopic catalogue \citep{JPL},
TheoReTS  \citep{rey2016theorets}, and NASA Ames \citep{21HuScLe}. 
A particular characteristic of ExoMol, which is shared by HITEMP, TheoReTS and NASA Ames, is that their line lists are
designed for studies involving hot molecules. The line lists  are often huge; the ExoMol database contains in excess of $10^{12}$ transitions. This means for many purposes it is desirable to have a means to rapidly and flexibly
post-process these transition data to provide them in forms, such as cross-sections, suitable for use in astronomical models and spectroscopic studies.

The ExoMol database is very widely used by scientists studying exoplanets. It has been used to create opacities
for a variety of exoplanet codes (see e.g. \citet{jt782,jt801,jt819}).  ExoMol data have been instrumental in many molecular detections in exoplanet atmospheres
such as the recent detection of CO$_2$ and SO$_2$ in the atmosphere of  WASP-39b by \citet{23RySiMuMa} using the JWST.
ExoMol is increasingly being used for studies of cool stars and brown dwarfs, as well as in a variety of non-astronomical applications.
\citet{jt708} provided a post-processing code \textsc{ExoCross} capable of treating data in both ExoMol and HITRAN/HITEMP formats. \textsc{ExoCross} is written in Fortran which makes it efficient for processing large datasets but somewhat inflexible. For example, it is hard to take part of the current \textsc{ExoCross} code and include as a utility in some other application; \textsc{PyExoCross} was developed in response to this issue. 

The main practical challenge that confronts us is processing huge amounts of data because of the extremely large datasets for polyatomic molecular hot line lists. In particular, many molecules have more than one Transition file in the ExoMol database, such as the parent isotopologue ${}^{23}\textrm {Na}{}^{16}\textrm O{}^1\textrm H$ of the molecule $\textrm {NaOH}$, its OYT5 dataset due to \citet{jt820} containing an excess of 400 billion transitions stored in 90 Transitions files, giving a total file size of 463 Gb. To overcome this problem, automation and batch data processing are important for improving efficiency and reducing errors caused by operational faults. In the \textsc{PyExoCross} program, the approach to working with tens of billion lines and optimizing the program for high throughput involves the use of efficient parallelization and vectorization.

\textsc{PyExoCross} is a general Python adaptation of the \textsc{ExoCross} Fortran application \citep{jt708} which aims to provide the same functionalities. Like \textsc{ExoCross}, \textsc{PyExoCross} is also designed to manage huge molecular line lists generated by the ExoMol project.  \textsc{PyExoCross} calculates spectra and spectral properties on the basis of the underlying spectroscopic data including temperature and pressure-dependent absorption and emission cross-sections, temperature-dependent cooling functions, state-dependent radiative lifetimes, oscillator strengths, and thermodynamic properties such as partition functions and specific heats. \textsc{PyExoCross} is designed to post-process data from the ExoMol database with the ExoMol, HITRAN, and HITEMP databases all serving as input databases for \textsc{PyExoCross}. \textsc{PyExoCross} exports the results in the ExoMol format which can also be read by effective
Hamiltonian spectral simulation program \texttt{PGOPHER} \citep{PGOPHER}. The ExoMol and HITRAN/HITEMP formats are the line lists formats that are initially offered; these databases are characterized by very extensive wavelength coverage and well-developed data structures.

There are several similar programs available for post-processing line lists including the HITRAN interface HAPI \citep{kochanov2016hitran,kochanov2018updates}, SPECTRAPLOT.COM \citep{goldenstein2017spectraplot}, and SPECTRA \citep{jt79,jt128}.  However these programs would struggle to process the huge line lists required for simulations of atmospheres at elevated temperatures. We 
note that 
RADIS \citep{21vaPaxx}, a fast line-by-line code for infrared emission and absorption spectra at equilibrium and non-local thermodynamic equilibrium (non-LTE) which works with HITRAN/HITEMP and ExoMol data, is designed for this purpose, as is the graphical processing unut (GPU) based opacity code of \citet{jt819}.
\textsc{PyExoCross} imports the input line lists in both ExoMol \citep{jt528,jt631} and HITRAN \citep{rothman2005hitran} formats and returns data in the ExoMol format. Future development will support more database formats in \textsc{PyExoCross}.
%%\textsc{ExoCross} and its Python version \textsc{PyExoCross} work more flexibly on this problem. 

\textsc{PyExoCross} performs a variety of functions. It can
 produce molecular absorption and emission cross-sections on a grid at a given set of temperatures and pressures  (see Section~\ref{sec:crosssection}); \textsc{PyExoCross} can apply a range of line profiles namely Gaussian, Lorentzian, and Voigt profiles (see Section~\ref{sec:profile}). \textsc{PyExoCross} also implements the computations of other useful functions including partition functions (Section~\ref{sec:partition}), specific heats (Section~\ref{sec:specificheat}), cooling functions (Section~\ref{sec:cooling}), radiative lifetimes (Section~\ref{sec:lifetime}), oscillator strengths (Section~\ref{sec:oscillatorstrength}), and line intensities in the form of spectra (Section~\ref{sec:stick spectra}). \textsc{PyExoCross} supports converting data formats between the ExoMol and HITRAN databases (Section~\ref{sec:conversion}). \textsc{PyExoCross} provides some additional functionalities according to the ExoMol database including automating the batch download of the ExoMol line list files, finding the uncertainty available datasets and comparing updates of the ExoMol database (Section~\ref{sec:additionlfun}). 
The paper is organized into six sections as follows. The main functionality of \textsc{PyExoCross} is described in Section~\ref{sec:functionality}. Section~\ref{sec:profile} discusses the line profile for calculating the cross-sections implemented in \textsc{PyExoCross}. Section~\ref{sec:protocol} gives the details of the calculation steps. The units used in \textsc{PyExoCross} are specified in Section~\ref{sec:units}. Finally, Section~\ref{sec:conclusions} presents some conclusions. The \textsc{PyExoCross} manual (\href{https://pyexocross.readthedocs.io/}{https://pyexocross.readthedocs.io/}), which is given in the supplementary data and in the \textsc{PyExoCross} GitHub repository (\href{https://github.com/ExoMol/PyExoCross.git}{https://github.com/ExoMol/PyExoCross.git}), provides comprehensive working instructions and use case examples.

\section{Main functionality}
\label{sec:functionality}

This section introduces the main functionalities of the \textsc{PyExoCross} program, including partition functions, specific heats, cooling functions, radiative lifetimes, oscillator strengths, intensities, stick spectra, and cross-sections. Some extra functionalities include filters, data format conversion, the batch download of line list files, data manipulation based on their uncertainty, and checking updates of the ExoMol database. The line profiles for calculating the cross-sections will be illustrated in Section~\ref{sec:profile}.

%\red{I think it would be logical to start here with the introduction of of the .trans and .states files. Otherwise they we introduce them  as part of discussion of other properties but they are  important by themselves.}

\subsection{ExoMol format}

ExoMol format \citep{jt548} assumes that a line list consists of a States file (\texttt{.states}) and Transitions files (\texttt{.trans}). A States file contains the state IDs, energy term values, total degeneracies,  total angular momenta, uncertainties, lifetimes, Lande-\'{g} factors, and any other quantum state descriptions. As an example, an excerpt from the recently updated States file for $^{27}$Al$^{1}$H \citep{jt732} is shown in Table~\ref{tab:AlHstates}. Transitions files consist of the upper IDs, lower IDs, the Einstein $A$-coefficient and (optionally) reference transition wavenumbers. An extract of a Transitions file for $_{}^{27}$Al$_{}^{1}$H is shown in Table~\ref{tab:transfile}. 

\begin{table}
\centering
\begin{threeparttable}  
\caption{Extract from the \texttt{.states} states file of the $^{27}$Al$^{1}$H line list \citep{jt732}.} 
\label{tab:AlHstates}
\begin{tabular}{crcccrcccccccrc}
\toprule
$i$ & \thead{$\tilde{E}$ (cm$^{-1}$)} & $g$ & $J$ & Unc (cm$^{-1}$) & \thead{$\tau$ (s)} & $+/-$ & e$/$f & State & $\varv$ & $|\upLambda|$ & $|\upSigma|$ & $|\upOmega|$ & \thead{Ecalc (cm$^{-1}$)} & Source label \\
\midrule
1  & 0.000000     & 12 & 0 & 0.000001 & inf        & + & e & X(1Sigma+) & 0 & 0 & 0 & 0 & 0.000000     & Ma \\
2  & 1625.061501  & 12 & 0 & 0.000597 & 4.9288E-03 & + & e & X(1Sigma+) & 1 & 0 & 0 & 0 & 1625.069321  & Ma \\
3  & 3194.213550  & 12 & 0 & 0.000365 & 2.6227E-03 & + & e & X(1Sigma+) & 2 & 0 & 0 & 0 & 3194.213685  & Ma \\
4  & 4708.808866  & 12 & 0 & 0.000196 & 1.8633E-03 & + & e & X(1Sigma+) & 3 & 0 & 0 & 0 & 4708.817022  & Ma \\
5  & 6170.189966  & 12 & 0 & 0.029900 & 1.4910E-03 & + & e & X(1Sigma+) & 4 & 0 & 0 & 0 & 6170.193041  & Ma \\
6  & 7579.564189  & 12 & 0 & 0.250000 & 1.2740E-03 & + & e & X(1Sigma+) & 5 & 0 & 0 & 0 & 7579.564189  & Ca \\
7  & 8938.046805  & 12 & 0 & 0.300000 & 1.1350E-03 & + & e & X(1Sigma+) & 6 & 0 & 0 & 0 & 8938.046805  & Ca \\
8  & 10246.644027 & 12 & 0 & 0.350000 & 1.0410E-03 & + & e & X(1Sigma+) & 7 & 0 & 0 & 0 & 10246.644027 & Ca \\
9  & 11506.245734 & 12 & 0 & 0.400000 & 9.7520E-04 & + & e & X(1Sigma+) & 8 & 0 & 0 & 0 & 11506.245734 & Ca \\
10 & 12717.633797 & 12 & 0 & 0.450000 & 9.2834E-04 & + & e & X(1Sigma+) & 9 & 0 & 0 & 0 & 12717.633797 & Ca \\
\bottomrule
\end{tabular}
$i$: State counting number; \\
$\tilde{E}$: Term value (in cm$^{-1}$); \\
$g_{\rm tot}$: Total state degeneracy; \\
$J$: Total angular momentum quantum number; \\
Unc: Estimated uncertainty of energy level (in cm$^{-1}$); \\
$\tau$: Lifetime (in s$^{-1}$); \\
$+/-$: Total parity; \\
e$/$f: Rotationless parity; \\
State: Electronic term value; \\
$\varv$: Vibrational quantum number; \\
$|\upLambda|$: Absolute value of the projection of electronic angular momentum; \\
$|\upSigma|$: Absolute value of the projection of the electronic spin; \\
$|\upOmega|$: Absolute value of the projection of the total angular momentum; \\
Ecalc: Calculated state energy (in cm$^{-1}$); \\
Source label: Method used to generate term value.           
\end{threeparttable}
\end{table}

\begin{table}
\centering
\setlength{\tabcolsep}{9.5mm}{
\begin{threeparttable}  
\caption{Extract from the \texttt{.trans} Transitions file of the $_{}^{27}$Al$_{}^{1}$H line lists \citep{jt732}.}
\label{tab:transfile}
\begin{tabular}{rrcc}
\toprule
$\thead f$ & $\thead i$ & $A_{fi} (\textrm{s}^{-1})$ & $\tilde{\nu}_{fi} (\textrm{cm}^{-1})$ \\
\midrule
 594  &        598 & 9.1644e-14 &       3.731850 \\
  60  &         28 & 1.7183e-05 &       4.209495 \\ 
  59  &         27 & 2.1667e-05 &       4.325081 \\
  58  &         26 & 3.2749e-05 &       4.605786 \\
  57  &         25 & 4.9130e-05 &       4.917133 \\
  56  &         24 & 7.0449e-05 &       5.228273 \\
  54  &         23 & 9.6527e-05 &       5.537220 \\
  53  &         22 & 1.2674e-04 &       5.844347 \\
  51  &         21 & 1.5996e-04 &       6.150075 \\
\bottomrule
\end{tabular}
$f$: Upper state counting number; \\
$i$: Lower state counting number; \\
$A_{fi}$: Einstein $A$-coefficient in s$^{-1}$; \\
$\tilde{\nu}_{fi}$: Transition wavenumber in cm$^{-1}$. (This column is optional).           
\end{threeparttable}   
}
\end{table}

\subsection{Partition functions}
\label{sec:partition} 

Although crucial for hot molecular models, partition functions are not always straightforward to determine \citep{jt169,jt571}. \citet{81Irwin.partfunc}, \citet{84SaTaxx.partfunc}, \citet{16BaCoxx.partfunc}, and \citet{jt692} present compilations of partition functions for key astrophysical species. The partition function $\textsl{Q}(T)$ can be defined as the sum over states: 
\begin{equation} 
    \textsl{Q}(T)=\sum_n g_n^{\mbox{\scriptsize{tot}}}e^{-c_2\tilde{E}_n/T},
    \label{eq:partition}
\end{equation}
where $g_n^{\mbox{\scriptsize{tot}}}=g_n^{\mbox{\scriptsize{ns}}}(2J_n+1)$ is the total degeneracy. $g_n^{\mbox{\scriptsize{ns}}}$ is the nuclear-spin statistical weight factor; note that ExoMol and HITRAN follow the \lq\lq physicist's\rq\rq{} convention and use the full nuclear spin degeneracy for $g_n^{\mbox{\scriptsize{ns}}}$, see \citet{jt777} for a discussion of this. $J_n$ is the corresponding total angular momentum. $c_2=hc/k_B$ is the second radiation constant (cm K) where $h$ is the Planck constant (erg s), $c$ is the speed of light (cm s$^{-1}$) and $k_B$ is the Boltzmann constant (erg K$^{-1}$). $\tilde{E}_n=E_n/hc$ is the energy term value (cm$^{-1}$). $T$ is the temperature (K). \textsc{PyExoCross} provides the functionality to compute molecular a partition function  with Eq.~\eqref{eq:partition} using the energy term values stored in the corresponding States file of a line list in question on a grid of temperatures. 
        
In spectroscopic applications, where a partition function $\textsl{Q}(T)$ is required, it can be read directly from an ExoMol partition function (.pf) file; this is recommended as in some cases these partition functions have improved accuracy compared to the simple sum of over the states in the States file. If the .pf file is not available, $\textsl{Q}(T)$ for a given temperature can be computed using Eq.~(\ref{eq:partition}) as part of the spectral simulations.  

\begin{figure}
    \centering
    \includegraphics[width=0.4\textwidth]{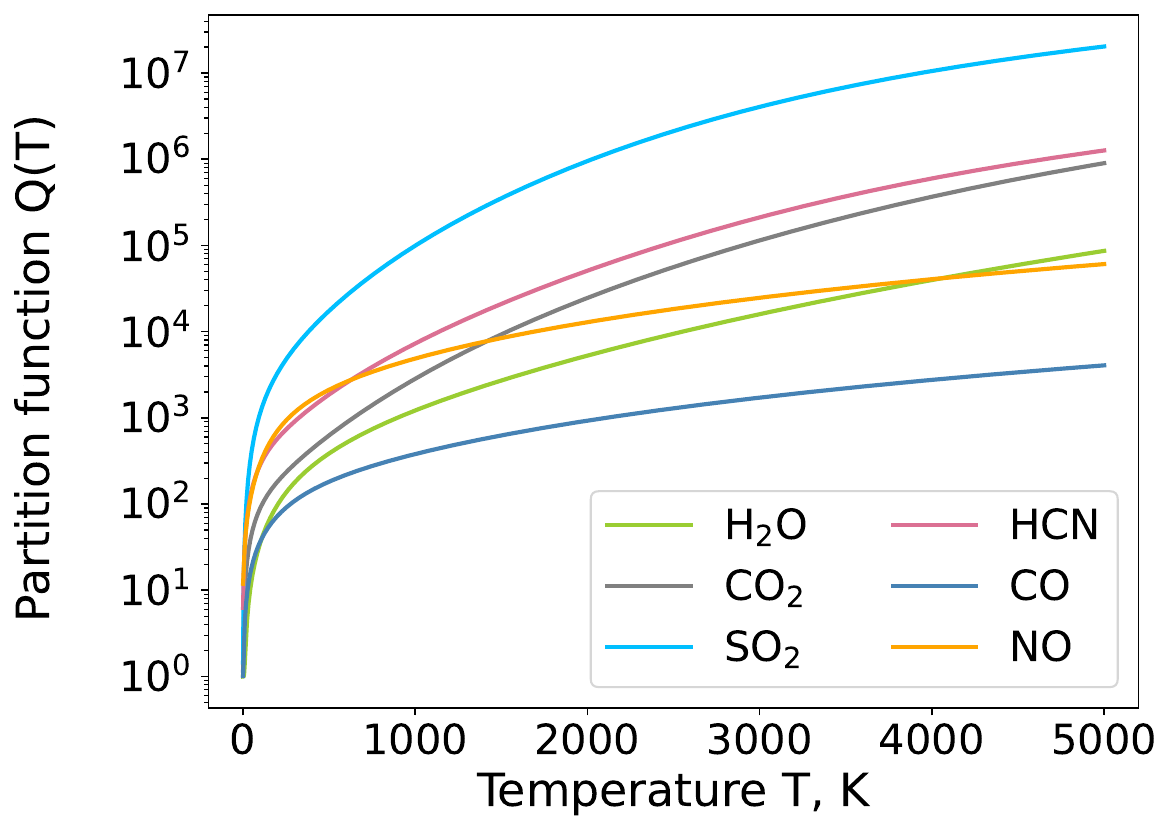}
    \caption{Partition functions of $^1$H$_2^{16}$O, $^{12}$C$^{16}$O$_2$, $^{32}$S$^{16}$O$_2$, $^1$H$^{12}$C$^{14}$N, $^{12}$C$^{16}$O, and $^{14}$N$^{16}$O with temperatures from 0 to 5000 K from the ExoMol database.}
    \label{fig:partition}
\end{figure}

Table~\ref{tab:pf} specifies the partition function file format and Table~\ref{tab:partitionfile} gives an example of the partition function file. Figure~\ref{fig:partition} displays examples of partition functions computed using the ExoMol line lists with temperatures from 0 to 5000 K.
Here and elsewhere examples are based on use of the following line lists: H$_2$O POKAZATEL \citep{jt734}, $^{12}$C$^{16}$O$_2$ UCL-4000 \citep{jt804}, $^{32}$S$^{16}$O$_2$ ExoAmes \citep{jt635,jt704}, $^1$H$^{12}$C$^{14}$N \citep{jt570}, $^{12}$C$^{16}$O Li2015 \citep{15LiGoRo.CO,21SoYuYa}, and NO XABC \citep{jt831}.

\textsc{PyExoCross} provides various ways to obtain the partition function: reading the local  \texttt{.pf} file or the ExoMol website directly,
computing the partition function from the ExoMol \texttt{.states} or by reading it from the HITRAN website.
We note that recent analysis by \citet{jt899} suggests that the maximum temperature quoted by HITRAN for a converged partition function is,
in many cases, too high.

\begin{table}
\centering
\setlength{\tabcolsep}{2.6mm}{
\begin{threeparttable}
\caption{Specification of the \texttt{.pf} partition function file format.}
\label{tab:pf}
\begin{tabular}{llll}
\toprule
Field & Fortran format & C format & Description\\
\midrule
$\textsl{T}$ & \texttt{F8.1} & \texttt{\%8.1f} & Temperature in K\\
$\textsl{Q}(T)$ & \texttt{F15.4} & \texttt{\%15.4f} & Partition function (dimensionless)\\
\bottomrule
\end{tabular}
Fortran format: \texttt{(F8.1,1x,F15.4)}. 
\end{threeparttable}
}
\end{table}
\begin{table}
\centering
\setlength{\tabcolsep}{20mm}{
\begin{threeparttable}
\caption{Extract from the \texttt{.pf} partition function file generated from the $_{}^{27}$Al$_{}^{1}$H line lists \citep{jt732}.}
\label{tab:partitionfile}
\begin{tabular}{cc} 
\toprule
$T$ & $\textsl{Q}(T)$ \\
\midrule
1.0 & 12.0000 \\
2.0 & 12.0042 \\
3.0 & 12.0858 \\
4.0 & 12.3885 \\
5.0 & 12.9621 \\
6.0 & 13.7647 \\
7.0 & 14.7312 \\
8.0 & 15.8065 \\ 
\bottomrule
\end{tabular}
$T$: Temperature in K; \\
$\textsl{Q}(T)$: Partition function (dimensionless).
\end{threeparttable}
}
\end{table}

\subsection{Specific heats}
\label{sec:specificheat}

\begin{figure}
    \centering
    \includegraphics[width=0.4\textwidth]{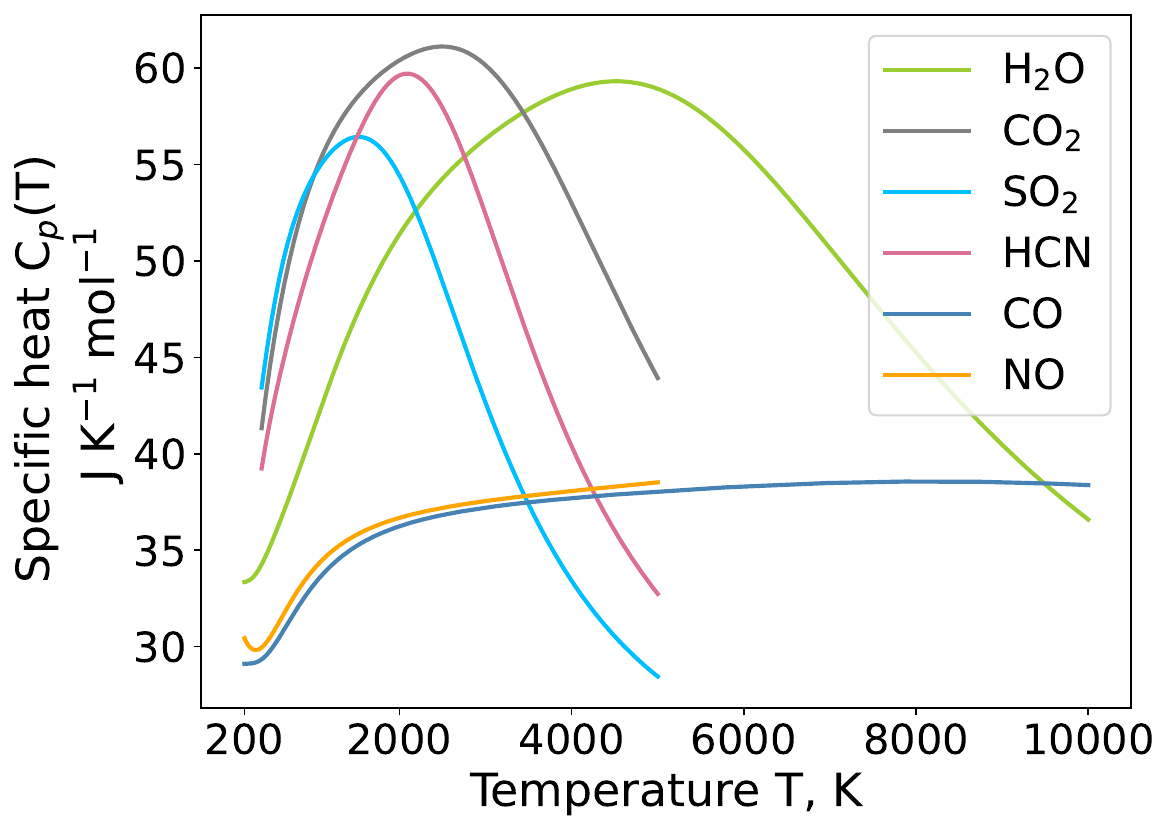}
    \caption{Specific heats of $^1$H$_2^{16}$O, $^{12}$C$^{16}$O$_2$, $^{32}$S$^{16}$O$_2$, $^1$H$^{12}$C$^{14}$N, $^{12}$C$^{16}$O, and $^{14}$N$^{16}$O with temperatures from 200 to 10000 K or $T_{\rm max}$ for the given line list, using the ExoMol line lists.}
    \label{fig:specificheat}
\end{figure}

The molar specific heat (J K${}^{-1}$ mol${}^{-1}$) at constant pressure $\textsl C_p$ for a given temperature using  the molecule partition function as given by \citet{92MaFrGi}: is defined  calculated on a grid of 
\begin{equation} 
    \textsl{C}_p(T) = R\left [\frac{\textsl{Q}''}{\textsl{Q}}-\left (\frac{\textsl{Q}'}{\textsl{Q}} \right )^2 \right ]+\frac{5R}{2}.
    \label{eq:specificheat}
\end{equation}
where $R$ is the molar gas constant (= 8.31446261815324 J K${}^{-1}$ mol${}^{-1}$). The partition function $\textsl{Q}(T)$ is defined in Eq.~(\ref{eq:partition}). The first two moments of the partition function can be evaluated directly from the energy term values :
\begin{equation}
\begin{split}
    \textsl{Q}'(T) = T\frac{\mathrm{d} \textsl{Q}}{\mathrm{d} T} = \sum_n g_n^{\mbox{\scriptsize{tot}}}\left (\frac{c_2 \tilde{E}_n}{T}\right ) \exp \left (-\frac{c_2 \tilde{E}_n}{T}\right ), \\
    \textsl{Q}''(T) = T^2\frac{\mathrm{d}^2 \textsl{Q}}{\mathrm{d} T^2}+2\textsl{Q}' = \sum_n g_n^{\mbox{\scriptsize{tot}}} \left (\frac{c_2 \tilde{E}_n}{T}\right )^2 \exp \left (-\frac{c_2 \tilde{E}_n}{T}\right ). 
    \label{eq:delta q}
\end{split}
\end{equation}
\textsc{PyExoCross} uses these equation in connection with the molecular States file to compute specific heats on a grid of temperatures. Table~\ref{tab:cp} lists the specific heat file format and Table ~\ref{tab:specificheat} gives an example of the specific heat file of AlH. Figure~\ref{fig:specificheat} shows the specific heats of NO, MgH and AlH at $T\in [0,5000]$ K.
\begin{table}
\centering
\setlength{\tabcolsep}{3mm}{
\begin{threeparttable}
\caption{Specification of the \texttt{.cp} specific heat file format.}
\label{tab:cp}
\begin{tabular}{llll}
\toprule
Field & Fortran format & C format & Description\\
\midrule
$\textsl{T}$ & \texttt{F8.1} & \texttt{\%8.1f} & Temperature in K\\
$\textsl{C}_p(T)$ & \texttt{F15.4} & \texttt{\%15.4f} & Specific heat (dimensionless)\\
\bottomrule
\end{tabular}
Fortran format: \texttt{(F8.1,1x,F15.4)}.
\end{threeparttable}
}
\end{table}
\begin{table}
\centering
\setlength{\tabcolsep}{19.5mm}{
\begin{threeparttable}
\caption{Extract from the \texttt{.cp} specific heat file generated from the $_{}^{27}$Al$_{}^{1}$H line lists \citep{jt732}.}
\label{tab:specificheat}
\begin{tabular}{cc} 
\toprule
$T$ & $\textsl{C}_p(T)$ \\
\midrule
200.0 & 29.1538 \\
201.0 & 29.1545 \\
202.0 & 29.1553 \\
203.0 & 29.1560 \\
204.0 & 29.1567 \\
205.0 & 29.1575 \\
206.0 & 29.1583 \\
207.0 & 29.1591 \\ 
208.0 & 29.1955 \\
209.0 & 29.1973 \\
\bottomrule
\end{tabular}
$T$: Temperature in K; \\
$\textsl{C}_p(T)$: Specific heat (J K${}^{-1}$ mol${}^{-1}$).
\end{threeparttable}
}
\end{table}

We note that following the work of \citet{jt899} a near complete set of specific heats are now available on the ExoMol website.

\subsection{Cooling functions}
\label{sec:cooling}

\begin{figure}
    \centering
    \includegraphics[width=0.4\textwidth]{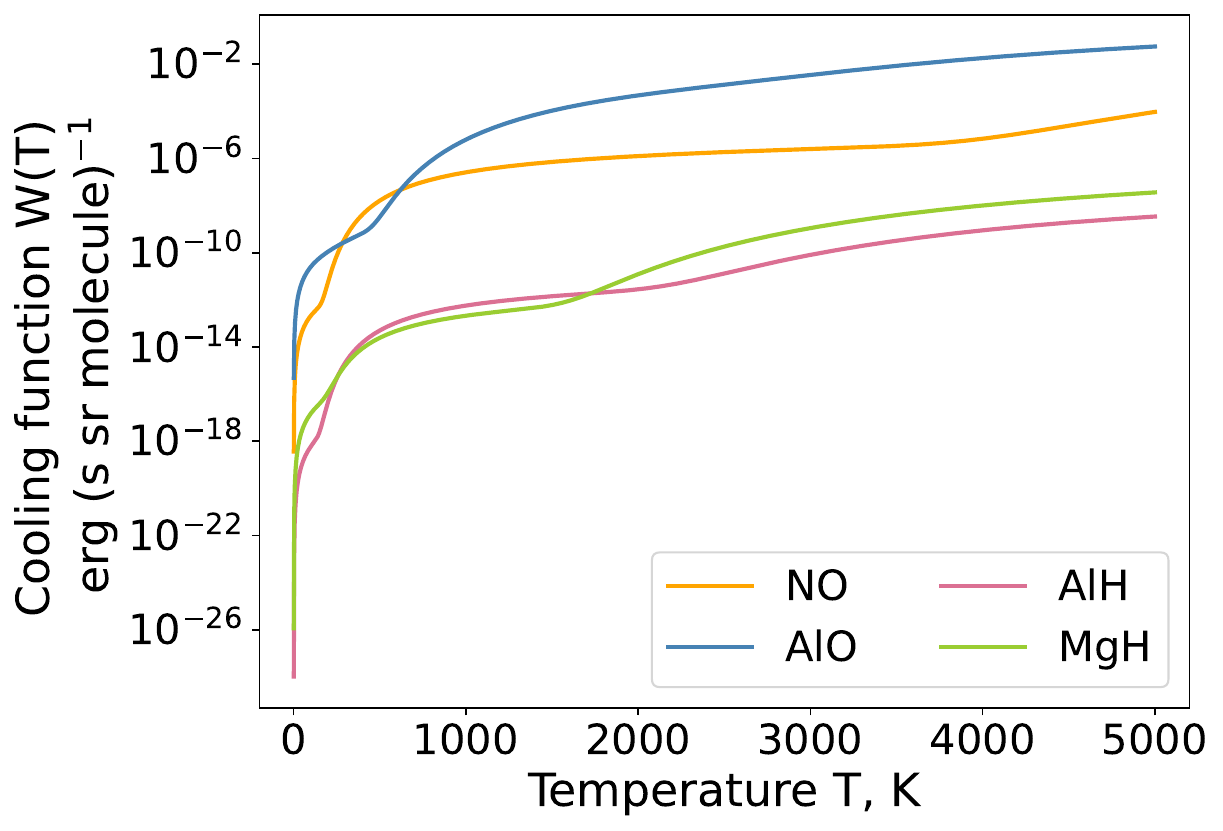}
    \caption{The temperature-dependent cooling functions of $^{14}$N$^{16}$O, $^{27}$Al$^{16}$O, $^{27}$Al$^{1}$H, and $^{24}$Mg$^{16}$O with temperatures from 0 K to 5000 K using the ExoMol line lists.}
    \label{fig:cooling}
\end{figure}

Cooling functions play an important role in a whole range of astronomical environments ranging from the early Universe to the interstellar medium, star formation, and issues with planetary stability \citep{jt506,jt551,wang2014radiative,nguyen2022impact,gerard2023h3+}. The temperature-dependent cooling function $\textsl{W}(T)$ [units erg (s molecule sr)$^{-1}$ $=10^{-7}$ Watts (molecule sr)$^{-1}$] is the total energy per unit solid angle (in steradians) emitted by a molecule at temperature $T$ \citep{jt660}. Table~\ref{tab:cooling} and \ref{tab:coolingfile} give the file format and an example of the cooling function file. Figure~\ref{fig:cooling} shows the temperature-dependent cooling function of NO, MgH, and AlH from the ExoMol database.The cooling function is produced by the emissivity \citep{neale1996spectroscopic}:
\begin{equation} 
    \textsl{W}(T) = \frac{1}{4 \pi \textsl{Q}(T)} \sum_{f,i} A_{fi} h c \tilde{\nu}_{fi} g'_f e^{-c_2 \tilde{E}'_f / T}.
    \label{eq:cooling}
\end{equation}
In Eq.~(\ref{eq:cooling}), $\textsl{Q}(T)$ is the partition function at temperature $T$; $g'_f$ is the upper state degeneracy and $\tilde{E}'_f$ is the upper state energy which can be read from the \texttt{.states} States file; $A_{fi}$ and $\tilde{\nu}_{fi}$ are the Einstein $A$-coefficient (s$^{-1}$) and transition wavenumber (cm$^{-1}$) in the \texttt{.trans} Transitions file (see Table~\ref{tab:trans} and \ref{tab:transfile}). \textsc{PyExoCross} requires the States files, Transitions files and partition function file to compute  $\textsl{W}(T)$ on a grid of temperatures.  Although some smaller Transitions files provide the wavenumbers as column 4 (see next), it is important that they evaluated  using the upper and lower state term values from the States files as given by 
\begin{equation} 
    \tilde{\nu}_{fi}=\tilde{E}'_f-\tilde{E}''_i. 
    \label{eq:v}
\end{equation}

\begin{table}
\centering
\setlength{\tabcolsep}{2.8mm}{
\begin{threeparttable}
\caption{Specification of the \texttt{.cf} cooling function file format.}
\label{tab:cooling}
\begin{tabular}{llll}
\toprule
Field & Fortran format & C format & Description\\
\midrule
$\textsl{T}$ & \texttt{F8.1} & \texttt{\%8.1f} & Temperature in K\\
$\textsl{W}(T)$ & \texttt{ES12.4} & \texttt{\%12.4E} & Cooling function in erg (s molecule sr)${}^{-1}$\\
\bottomrule
\end{tabular}
Fortran format: \texttt{(F8.1,1x,ES12.4)}.
\end{threeparttable}
}
\end{table}
\begin{table}
\centering
\setlength{\tabcolsep}{20mm}{
\begin{threeparttable}
\caption{Extract from the \texttt{.cf} cooling function file generated from the $_{}^{27}$Al$_{}^{1}$H line lists \citep{jt732}.}
\label{tab:coolingfile}
\begin{tabular}{cc} 
\toprule
$T$ & $\textsl{W}(T)$ \\
\midrule
1.0 & 1.04219598E-28 \\
2.0 & 8.95038739E-25 \\
3.0 & 1.82111455E-23 \\
4.0 & 8.06932723E-23 \\
5.0 & 1.94455237E-22 \\
6.0 & 3.52589825E-22 \\
7.0 & 5.56853398E-22\\
8.0 & 8.18964607E-22 \\ 
9.0 & 1.15275411E-21 \\
10.0 & 1.56910825E-21 \\
\bottomrule
\end{tabular}
$T$: Temperature in K; \\
$\textsl{W}(T)$: Cooling function in erg (s molecule sr)${}^{-1}$.
\end{threeparttable}
}
\end{table}  

\subsection{Radiative lifetimes}
\label{sec:lifetime}

Radiative lifetimes are important for determining the criticial density and as they represent a competing process to collisional thermalisation in
environments including plasmas such as the interstellar medium. The radiative lifetime $\tau_i$ (s) of state $i$ can be calculated by inverting the sum of  Einstein $A$-coefficients $A_{fi}$ (s$^{-1}$) with $\tilde{E}_j<\tilde{E}_i$ from the (\texttt{.trans}) Transitions file(s) (see Table~\ref{tab:transfile}) \citep{jt624}:
\begin{equation} 
    \tau_i = \frac{1}{{\textstyle \sum_{f} A_{fi}}}.
    \label{lifetime}
\end{equation} 
To calculate the state-dependent lifetime, the process has been divided into three steps. Initially,  transitions are reorganized with different upper state categories by \texttt{pandas.DataFrame.groupby} \citep{mckinney2010data}. Each upper state $f$ corresponds to one or more than one Einstein $A$-coefficient $A_{fi}$. Secondly, the sum over of the Einstein $A$-coefficients per upper state is computed. 
%The number of sums is equal to the number of states in the \texttt{.states} states file. 
Finally, the inverse of each sum is calculated. 
In principle, $\tau_i$  can be calculated for each state $i$ in the States file (see Table~\ref{tab:states}). In practice, however, the \texttt{.trans} files are not always complete. In this case the lifetime is considered to be undefined with \texttt{NaN}. 
%the blank. 
Some lowest states, particularly for high symmetry molecules  such methane or H$_3^+$, have infinite lifetimes, for which `inf' is used. The inclusion of lifetimes is being
made a standard part of the (recommended) States files in ExoMol as part of the generalization of the ExoMol data format to allow for predissociation lifetimes \citep{jt898}.
%Therefore, in Eq.~(\ref{lifetime}), $i$ in $\tau_i$ means state $i$ in the states file, not the lower state in the \texttt{.trans} transitions file and $f$ in $\sum_f$ is the upper state in transitions file.

\begin{table}
\centering
\setlength{\tabcolsep}{5.3mm}{
\begin{threeparttable}
\caption{Specification of the \texttt{.trans} transitions file format.}
\label{tab:trans}
\begin{tabular}{llll} 
\toprule
Field & Fortran format & C format & Description\\ 
\midrule
$f$ & \texttt{I12} & \texttt{\%12d} & Upper state ID\\
$i$ & \texttt{I12} & \texttt{\%12d} & Lower state ID\\
$A$ & \texttt{ES10.4} & \texttt{\%10.4E} & Einstein $A$-coefficient in
$\mathrm{s^{-1}}$ \\
$\tilde{\nu}_{fi}$&\texttt{F15.6} & \texttt{\%15.6f} & Transition wavenumber in cm$^{-1}$ (optional) \\
\bottomrule
\end{tabular}
Fortran format: \texttt{(I12,1x,I12,1x,ES10.4,1x,F15.6)} 
\end{threeparttable}
}
\end{table}

\begin{figure}
  \centering
\includegraphics[width=8.3cm]{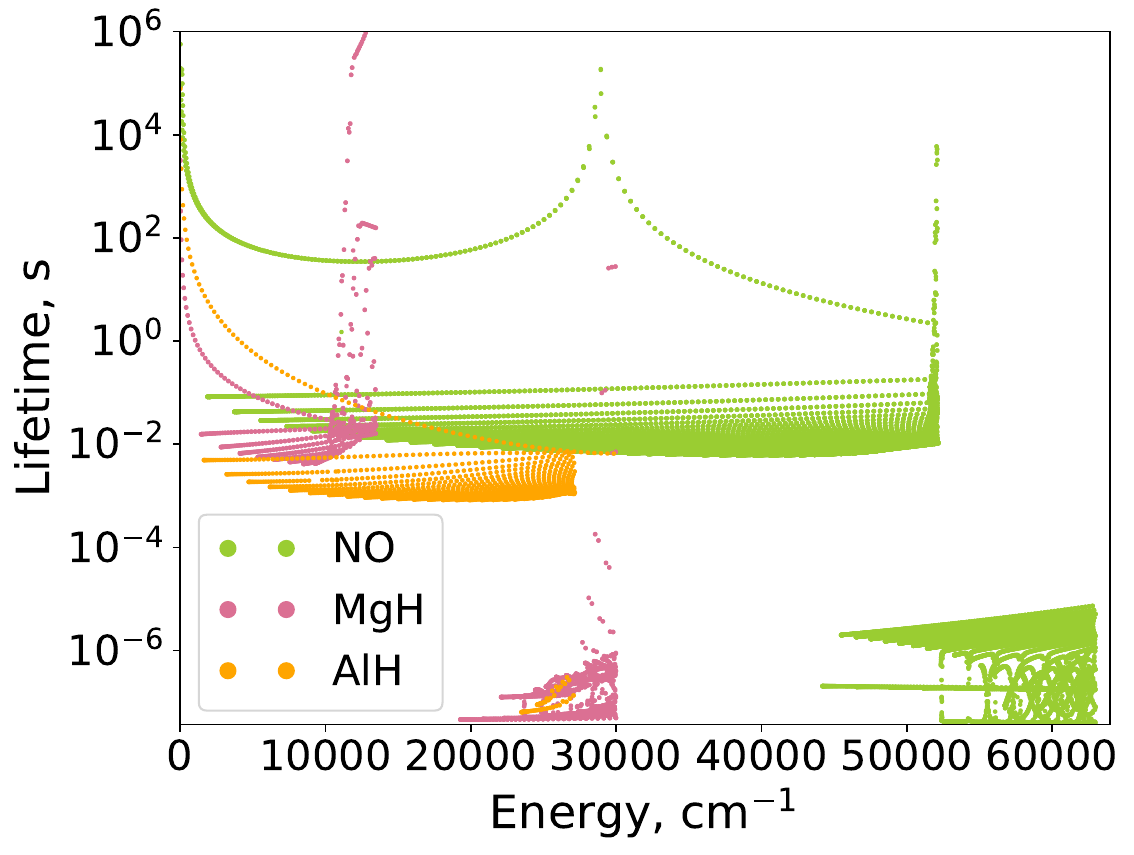}
  \includegraphics[width=9.25cm]{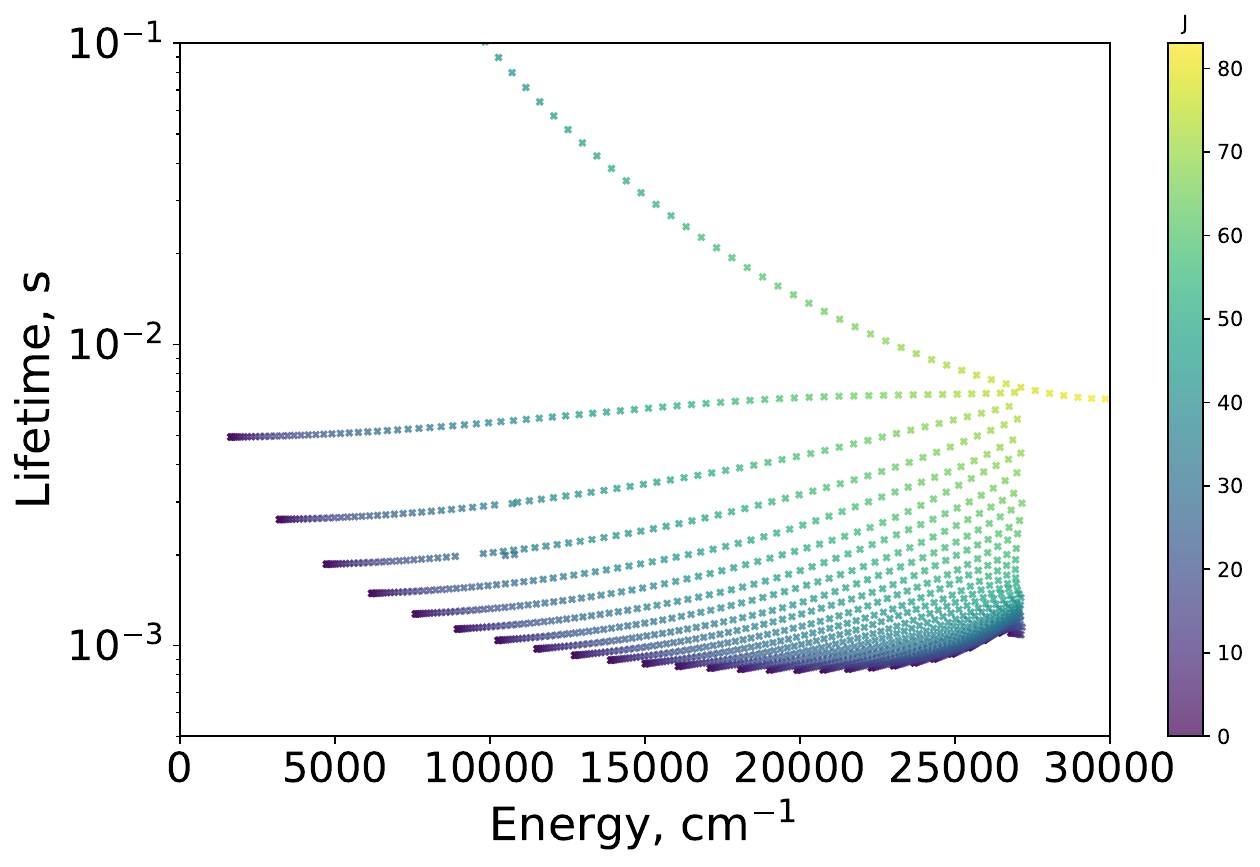}
  \caption{Left-hand panel: Lifetime of $^{14}$N$^{16}$O, $^{24}$Mg$^{1}$H, and $^{27}$Al$^{1}$H using the ExoMol line lists. Right-hand panel: State-dependent lifetime in range of $10^{-4}-10^{-1}$ s of $^{27}$Al$^{1}$H from the ExoMol database. The right-hand scale gives the $J$ values for which lifetimes are plotted.}\label{fig:lifetime}
\end{figure}

\textsc{PyExoCross} adds the lifetimes to the States file in the ExoMol data format. \textsc{PyExoCross} also provides a user option to save either an uncompressed \texttt{.states} file or a compressed  \texttt{.states.bz2} file.  Figure~\ref{fig:lifetime} (left-hand panel) shows states-dependent lifetimes for different molecules while the right-hand panel shows the lifetimes computed from AlH line list. These lifetimes are fully state specific; recently \citet{jt904} used ExoMol data to construct a lifetime database (LiDB)  for plasma modelling which stores lifetimes and decay routes for vibronic states by lumping them over rotational and fine structure states.

\begin{table}
\centering
\setlength{\tabcolsep}{4.1mm}{
\begin{threeparttable}
\caption{Specification of the mandatory part of the \texttt{.states} states file with extra data options unc, $\tau$ and $g$.}
\label{tab:states}
\begin{tabular}{llll} 
\toprule
Field & Fortran format & C format & Description\\ 
\midrule
$i$ & \texttt{I12} & \texttt{\%12d} & State ID\\
$E$ & \texttt{F12.6} & \texttt{\%12.6f} & State energy in $\mathrm{cm^{-1}}$\\
$g_\mathrm{tot}$ & \texttt{I6} & \texttt{\%6d} & State degeneracy\\
$J$ & \texttt{I7/F7.1} & \texttt{\%7d/\%7.1f} & $J$-quantum number (integer/half-integer)\\
(Unc) & \texttt{F12.6} & \texttt{\%12.6f} & Uncertainty in state energy in $\mathrm{cm^{-1}}$ (optional) \\
($\tau$) & \texttt{ES12.4} & \texttt{\%12.4E} & Lifetime in s (optional)\\
($g$) & \texttt{F10.6} & \texttt{\%10.6f} & Land\'e $g$-factor  (optional)\\
(Extra) & & & Extra quantum numbers, any format (optional)\\
\bottomrule
\end{tabular}
ID: State identifier: a non-negative integer index, starting at 1; \\
$J$: Total angular momentum quantum, excluding nuclear spin; \\
Fortran format, $J$ integer: \texttt{(I12,1x,F12.6,1x,I6,I7,1x,ES12.4,1x,F10.6)} \\
or $J$ half-integer: \texttt{(I12,1x,F12.6,1x,I6,F7.1,1x,ES12.4,1x,F10.6)}.
\end{threeparttable}
}
\end{table}

\subsection{Oscillator strengths}
\label{sec:oscillatorstrength}

The oscillator strength is a dimensionless quantity used in spectroscopy to indicate the probability of electromagnetic radiation being emitted or absorbed in transitions between energy levels of an atom or molecule \citep{demtroder1981laser}. The fractional amount of effective electrons participating in the transition is represented by the oscillator strength $f$ \citep{orloski2001weighted}.  
%The statistical weight of an energy level, denoted by $g$, is a measure of the probability of a transition between two energy levels in an atom or ion. 
The weighted oscillator strength $gf$ is defined as:
\begin{equation}
\label{eq:oscillatorstrength}
    gf=\frac{g'_\textrm{tot}A_{fi}}{(c\tilde{v}_{fi})^2},
\end{equation}
where $g'_{\textrm{tot}}=g_{\textrm{ns}}(2J'+1)$ is the degeneracy  of the upper states with upper angular momentum $J'$; $A_{fi}$ and $\tilde{v}_{fi}$ are the Einstein $A$-coefficient (s$^{-1}$) and transition wavenumber (cm$^{-1}$), respectively; and $c$ is the velocity of light (cm s$^{-1}$). 

\begin{figure}
    \centering
    \includegraphics[scale=0.45]{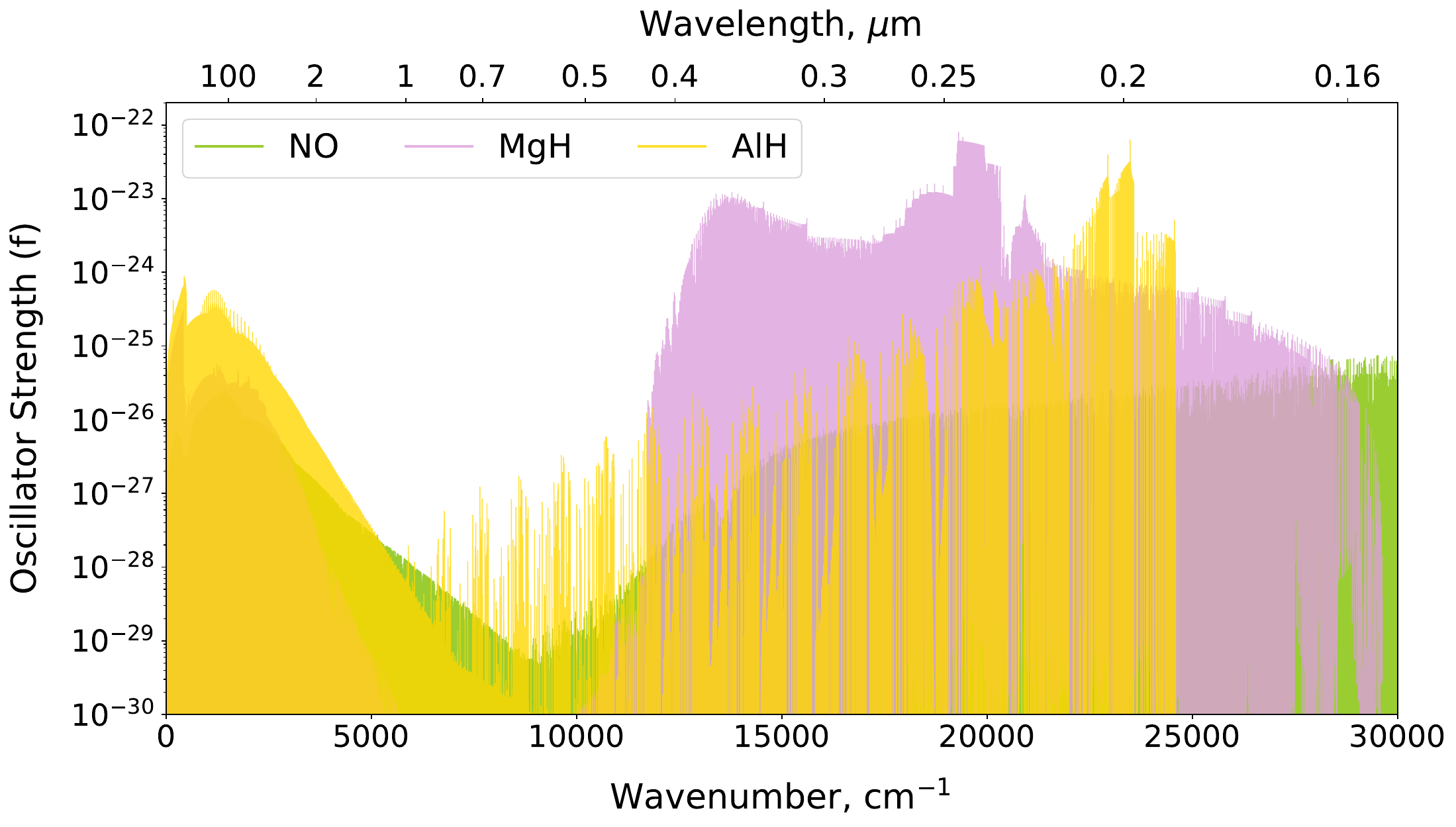}
    \caption{Oscillator strength of $^{14}$N$^{16}$O, $^{24}$Mg$^{16}$O, and $^{27}$Al$^{1}$H from the ExoMol line lists.}
    \label{fig:oscillator}
\end{figure}

The actual oscillator strength is defined by $f=gf/g''_{\textrm{tot}}$,  where $g''_{\textrm{tot}}$ is the degeneracy  of the lower state.
Table~\ref{tab:oscillator} and ~\ref{tab:oscillatorfile} give the file format and an example of the oscillator strength file. Figure~\ref{fig:oscillator} shows the oscillator strength of NO, MgH, and AlH from the ExoMol database. 
\begin{table}
\centering
\setlength{\tabcolsep}{5.2mm}{
\begin{threeparttable}
\caption{Specification of the \texttt{.os} oscillator strength file format.}
\label{tab:oscillator}
\begin{tabular}{llll}
\toprule
Field & Fortran format & C format & Description\\
\midrule
$f$ & \texttt{I12} & \texttt{\%12d} & Upper state degeneracy \\
$i$ & \texttt{I12} & \texttt{\%12d} & Lower state degeneracy \\
$gf$ & \texttt{ES10.4} & \texttt{\%10.4E} & Oscillator strength \\
$\tilde{\nu}_{i}$ & \texttt{F15.6} & \texttt{\%15.6f} & Central bin wavenumber, cm${}^{-1}$ (optional)\\
\bottomrule
\end{tabular}
Fortran format: \texttt{(I12,1x,I12,1x,ES10.4,1x,F15.6)} 
\end{threeparttable}
}
\end{table} 
\begin{table}
\centering
\setlength{\tabcolsep}{9.8mm}{
\begin{threeparttable}
\caption{Extract from the \texttt{.os} oscillator strength file generated from the $^{14}$N$^{16}$O line lists \citep{jt831}.}
\label{tab:oscillatorfile}
\begin{tabular}{rrcc} 
\toprule
$\thead{f}$ & $\thead{i}$ & $gf$ & $\tilde{\nu}_{i}$ \\
\midrule
        2237 &        2604 &      2.3938E-29 &   32406.776335 \\
       16945 &       16990 &      2.6529E-30 &   32406.778880 \\ 
       11519 &       11563 &      9.3397E-30 &   32406.843732 \\
       15241 &       14758 &      4.6102E-26 &   32406.846700 \\
       15986 &       15778 &      1.0011E-25 &   32406.866875 \\
       13921 &       13689 &      9.0563E-31 &   32406.888211 \\
        7437 &        7184 &      1.1891E-30 &   32406.897702 \\
       11504 &       11847 &      1.0124E-28 &   32406.925712 \\
       16602 &       16639 &      4.0684E-30 &   32406.926676 \\
       16353 &       16394 &      5.5179E-31 &   32406.942107 \\
\bottomrule
\end{tabular}
$f$: Upper state ID; \\
$i$: Lower state ID; \\
$gf$: Oscillator strength;\\
$\tilde{\nu}_{i}$: Central bin wavenumber, cm${}^{-1}$ (optional). 
\end{threeparttable}
}
\end{table} 

\subsection{Line intensities and stick spectra}
\label{sec:stick spectra}

\subsubsection{Absorption coefficient}

The absorption coefficient,  also known as the absorption line intensity ${I}_{fi}$ (cm molecule$^{-1}$),  can be calculated at a general temperature $T$  using  \citep{96JoLaIw.CH,jt708}:
\begin{equation} 
\label{eq:absintensity}
    I_{f \gets i} = \frac{g'_f {A}_{fi}}{8 \pi c \tilde{\nu}^2_{fi}} \frac{e^{-c_2 \tilde{E}''_i / T} (1 - e^{-c_2 \tilde{\nu}_{fi} / T })}{\textsl{Q}(T)},
\end{equation}
where 
$g'_f$ is the upper state degeneracy;  $\tilde{E}''_i$ is the lower state energy;  $A_{fi}$ is the Einstein $A$-coefficient (s$^{-1}$); $\tilde{\nu}_{fi}$ is the transition wavenumber (cm$^{-1}$); $c_2= hc/k_B$ is the second radiation constant (cm K), where $h$ is the Planck constant (erg s), $c$ is the speed of light (cm s$^{-1}$), and $k_B$ is the Boltzmann constant (erg K$^{-1}$); and $T$ is the temperature in K. The temperature-dependent partition function $\textsl{Q}(T)$ can be accessed from the partition function file (\texttt{.pf}). When computing  $I_{f \gets i}$ with \textsc{PyExoCross}, the energy term values are read from  the ExoMol States (\texttt{.states}) file, which are also used to evaluate $\tilde{\nu}^2_{fi}$, the  Einstein $A$-coefficient are read from  the ExoMol  Transitions (\texttt{.trans}) file, and the partition function is either taken from the partition function file or calculated on the fly using the energy term values and Eq.~\eqref{eq:partition}, as in the case for the cooling functions. 

We note that the definitions of line intensities given in Eqs.~(\ref{eq:absintensity} and \ref{eq:emiintensity}) are for 100 per cent abundance and, unlike HITRAN and HITEMP, are not scaled to allow for isotopic fractional abundances. The results should therefore be multiplied by the appropriate fractional abundance for many applications.

\subsubsection{Emission coefficient}

The emission coefficient is also called the emission line intensity and emissivity $\varepsilon$ [units erg (s molecule sr)$^{-1}=10^{-7}$ Watts (molecule sr)$^{-1}$] \citep{jt708}. 
\begin{equation} 
\label{eq:emiintensity}
    \varepsilon(i \gets f) = \frac{g'_f hc {A}_{fi} \tilde{\nu}_{fi}} {4 \pi} \frac{e^{-c_2 \tilde{E}'_f / T}}{\textsl{Q}(T)},
\end{equation}
where $g'_f$ is the upper state degeneracy and $\tilde{E}'_f$ is the upper state energy term value. As in the case of the absorption coefficients, the energy term values $\tilde{E}_f$ are read from the ExoMol States file, which are also used for $\tilde{\nu}_{fi}$, ${A}_{fi}$ are taken from the Transition files, and $Q(T)$ is from the partition function file or computed using Eq.~\eqref{eq:partition}. When calculating the emission coefficient for the HITRAN database, the HITRAN format line lists file (\texttt{.par}) only has the lower state energy $E_{\textrm{lower}}$ ($E''_i$). Then the upper state energy  must be evaluated using Eq.~(\ref{eq:v}) and the partition function is taken from the partition function file.

\subsubsection{Stick spectra}

\begin{figure}
\centering    
\includegraphics[scale=0.45]{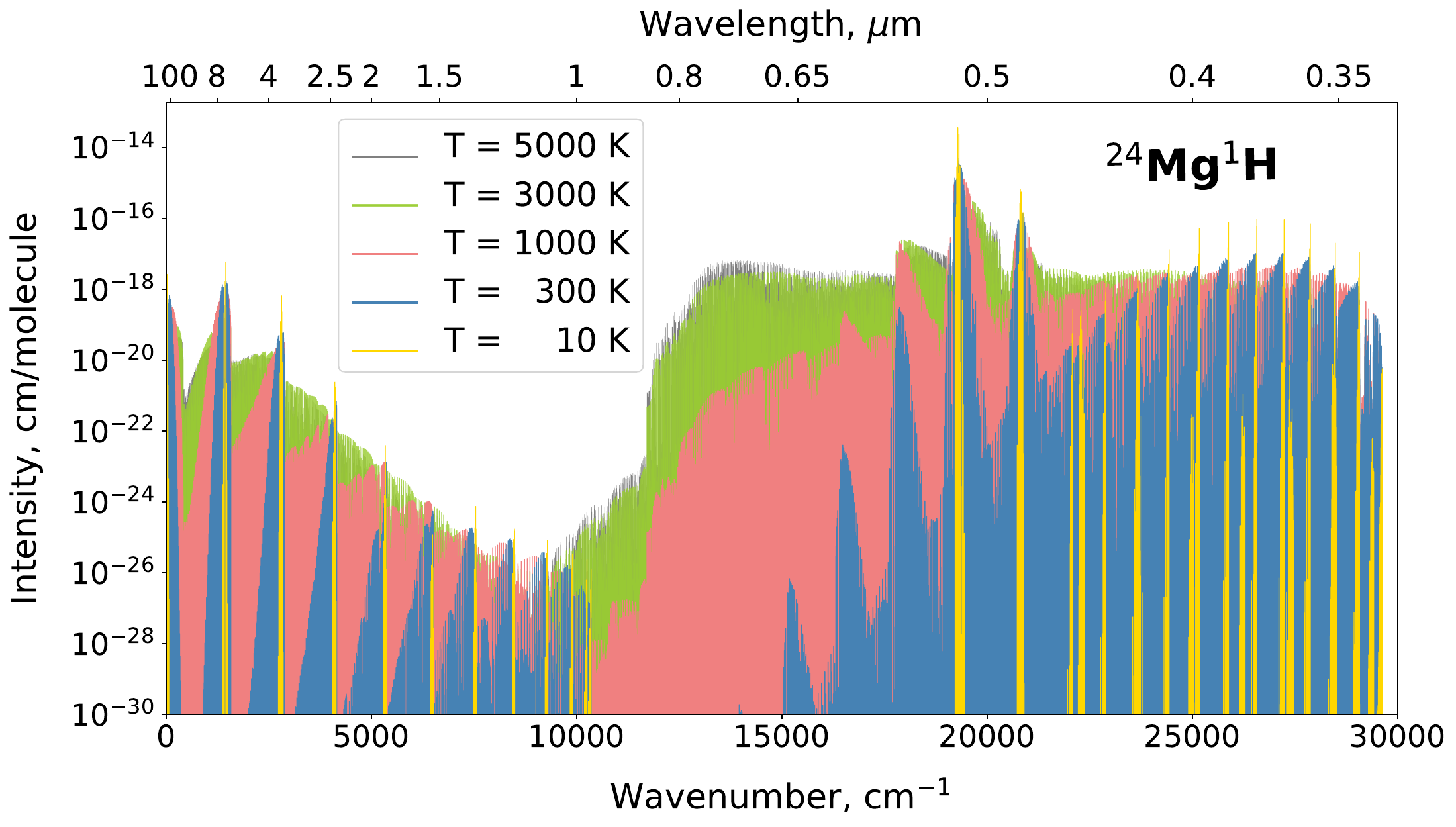} 
\quad
\includegraphics[scale=0.45]{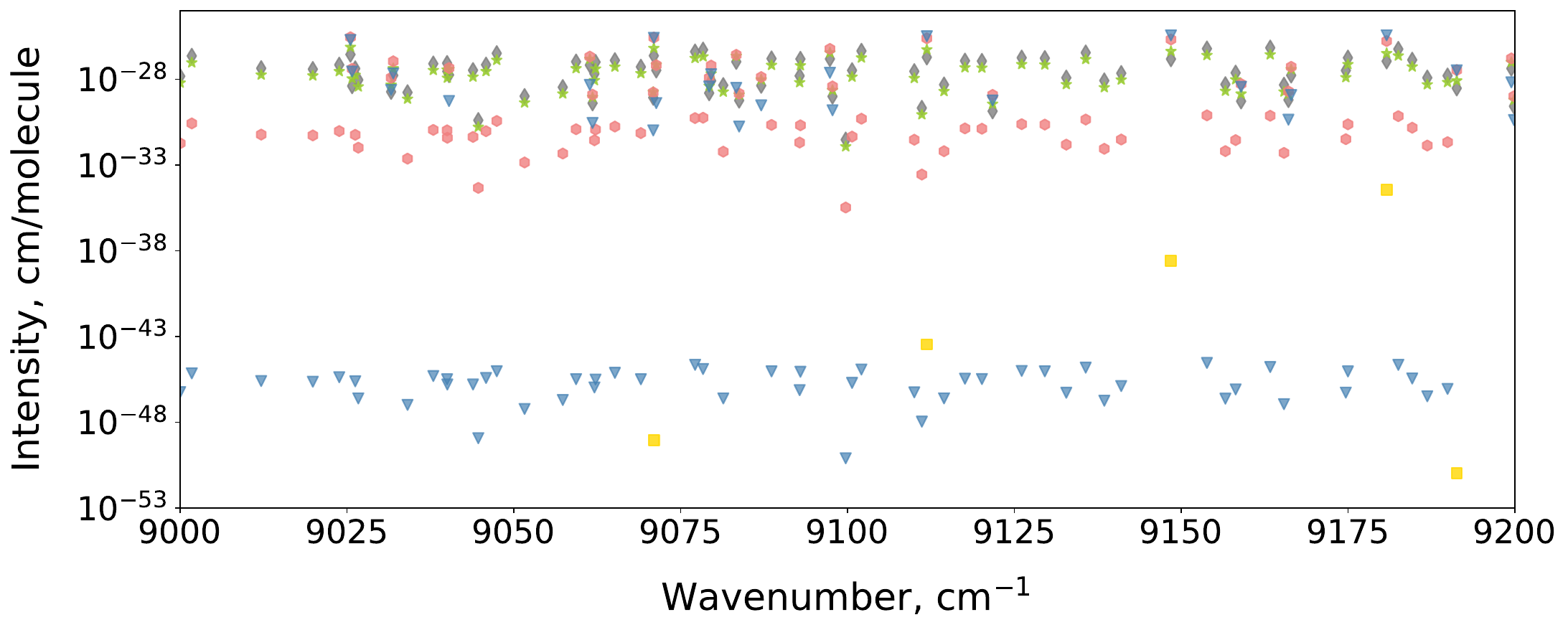}
\caption{ Stick spectra: Absorption intensities of $^{24}$Mg$^1$H at $T=$ 10, 300, 1000, 3000, and 5000 K simulated using the ExoMol line list of dataset XAB \citep{jt858}. Upper panel: Intensities in wavenumber range {$[0,30000]$} cm$^{-1}$. Lower panel: Intensities zoom at {$[9000, 9200]$} cm$^{-1}$.}    
\label{fig:stickspectra}    
\end{figure}

Stick spectra are lists of frequencies and line intensities, absorption or emission, calculated using Eqs.~(\ref{eq:absintensity}) and (\ref{eq:emiintensity}), respectively, along with their descriptions (required quantum numbers) for the upper and lower states \citep{jt708}.  \textsc{PyExoCross} can apply  filters of uncertainties, thresholds, and quantum numbers when producing stick spectra. Table~\ref{tab:stick} gives the mandatory part of the  stick (\texttt{.stick}) spectra file. Table~\ref{tab:MgHstick} shows an example of a stick spectrum output. Lines are plotted as `sticks', with their height given by their intensity (see Figure~\ref{fig:stickspectra}). %\red{Can we show a zoom-in where the sticks can be better seen?}

\subsubsection{Intensity thresholds}

\textsc{PyExoCross} provides an intensity threshold filter to skip weak lines. Using an intensity threshold can speed up the cross-section calculations or reduce the output of stick spectra calculations. In \textsc{PyExoCross}, the default value of the intensity threshold is $10^{-30}$ cm molecule$^{-1}$ but this value can be changed in the input file.

\begin{table}
\centering
\setlength{\tabcolsep}{9.15mm}{
\begin{threeparttable}
\caption{Specification of the mandatory part of the \texttt{.stick} stick spectra file with the extra columns for the  optional quantum numbers.}
\label{tab:stick}
\begin{tabular}{llll} 
\toprule
Field & Fortran format & C format & Description\\ 
\midrule
$\tilde{\nu}$ & \texttt{F12.6} & \texttt{\%12.6f} & Frequency (wavenumber) in $\mathrm{cm^{-1}}$\\
$I$ & \texttt{ES14.8} & \texttt{\%14.8E} & Intensity in cm molecule$^{-1}$\\
$J'$ & \texttt{I7/F7.1} & \texttt{\%7d/\%7.1f} & $J$-quantum number (integer/half-integer) for upper state\\
$E'$ & \texttt{F12.4} & \texttt{\%12.4f} & Upper state energy in $\mathrm{cm^{-1}}$\\
$J''$ & \texttt{I7/F7.1} & \texttt{\%7d/\%7.1f} & $J$-quantum number (integer/half-integer) for lower state\\
$E''$ & \texttt{F12.4} & \texttt{\%12.4f} & lower state energy in $\mathrm{cm^{-1}}$\\
(QN$'$) & & & Extra quantum numbers for upper state, any format (optional)\\
(QN$''$) & & & Extra quantum numbers for lower state, any format (optional)\\
\bottomrule
\end{tabular}
$J$: Total angular momentum quantum, excluding nuclear spin.
\end{threeparttable}
%\item Fortran format, $J$ integer: \texttt{(I12,1x,F12.6,1x,I6,I7,1x,ES12.4,1x,F10.6)}\\
%\item or $J$ half-integer: \texttt{(I12,1x,F12.6,1x,I6,F7.1,1x,ES12.4,1x,F10.6)}.\\
}
\end{table}
\begin{table}
\centering
\setlength{\tabcolsep}{0.75mm}{
\begin{threeparttable}
\caption{Extract from the \texttt{.stick} stick spectra file generated from the absorption $^{24}$Mg$^{1}$H line list \citep{jt858}.}
\label{tab:MgHstick}
\begin{tabular}{ccrcrccccrccccccrccc}
\toprule
$\tilde{v_{fi}}$ & $I_{fi}$ & \thead{$J'$} & $\tilde{E}'$ & \thead{$J''$} & $\tilde{E}''$ & par$'$ & e$/$f $'$ & State$'$ & \thead{$\varv'$} & $|\upLambda|'$ & $|\upSigma|'$ & $|\upOmega|'$ & par$''$ & e$/$f $''$ & State$''$ & \thead{$\varv''$} & $|\upLambda|''$ & $|\upSigma|''$ & $|\upOmega|''$ \\
\midrule
1.242966 & 7.42574603E-49 & 20.5 & 10749.7776 & 20.5 & 10748.5346 & - & f & X(2Sigma+) & 8  & 0 & -0.5 & -0.5 & + & e & X(2Sigma+) & 9  & 0 & 0.5 & 0.5 \\
1.252056 & 7.53243618E-46 & 20.5 & 10749.7776 & 19.5 & 10748.5256 & - & f & X(2Sigma+) & 8  & 0 & -0.5 & -0.5 & + & f & X(2Sigma+) & 9  & 0 & 0.5 & 0.5 \\
1.337756 & 9.08537842E-46 & 21.5 & 10749.8724 & 20.5 & 10748.5346 & - & e & X(2Sigma+) & 8  & 0 & -0.5 & -0.5 & + & e & X(2Sigma+) & 9  & 0 & 0.5 & 0.5 \\
1.569618 & 2.65779920E-52 & 36.5 & 12150.5019 & 36.5 & 12148.9323 & - & f & X(2Sigma+) & 7  & 0 & -0.5 & -0.5 & + & e & X(2Sigma+) & 8  & 0 & 0.5 & 0.5 \\
1.597018 & 1.38115517E-48 & 36.5 & 12150.5019 & 35.5 & 12148.9049 & - & f & X(2Sigma+) & 7  & 0 & -0.5 & -0.5 & + & f & X(2Sigma+) & 8  & 0 & 0.5 & 0.5 \\
1.816673 & 1.86393524E-48 & 37.5 & 12150.7490 & 36.5 & 12148.9323 & - & e & X(2Sigma+) & 7  & 0 & -0.5 & -0.5 & + & e & X(2Sigma+) & 8  & 0 & 0.5 & 0.5 \\
2.859016 & 2.89098064E-45 & 0.5  & 10353.9808 & 0.5  & 10352.2738 & - & f & X(2Sigma+) & 11 & 0 & -0.5 & -0.5 & + & e & X(2Sigma+) & 11 & 0 & 0.5 & 0.5 \\
2.862028 & 5.84278951E-45 & 1.5  & 10354.0094 & 0.5  & 10352.2738 & - & e & X(2Sigma+) & 11 & 0 & -0.5 & -0.5 & + & e & X(2Sigma+) & 11 & 0 & 0.5 & 0.5 \\
3.589218 & 8.68228140E-47 & 33.5 & 11929.6911 & 34.5 & 11926.1019 & - & e & X(2Sigma+) & 9  & 0 & -0.5 & -0.5 & + & e & X(2Sigma+) & 8  & 0 & 0.5 & 0.5 \\
3.606156 & 8.58388055E-47 & 32.5 & 11929.6121 & 33.5 & 11926.0060 & - & f & X(2Sigma+) & 9  & 0 & -0.5 & -0.5 & + & f & X(2Sigma+) & 8  & 0 & 0.5 & 0.5 \\
\bottomrule
\end{tabular}
$\tilde{\nu}_{fi}$: Absorption frequency (wavenumber) (in cm$^{-1}$); \\
$I_{fi}$: Absorption intensity (in cm molecule$^{-1}$); \\
$J'$ and $J''$: Total angular momentum quantum number for upper and lower state; \\
$\tilde{E}'$ and $\tilde{E}''$: Term value (in cm$^{-1}$) for upper and lower state; \\
par$'$ and par$''$: Total parity for upper and lower state; \\
e$/$f $'$ and e$/$f $''$: Rotationless parity for upper and lower state; \\
State$'$ and State$''$: Electronic term value for upper and lower state; \\
$\varv'$ and $\varv''$: Vibrational quantum number for upper and lower state; \\
$|\upLambda|'$ and $|\upLambda|''$: Absolute value of the projection of electronic angular momentum for upper and lower state; \\
$|\upSigma|'$ and $|\upSigma|''$: Absolute value of the projection of the electronic spin for upper and lower state; \\
$|\upOmega|'$ and $|\upOmega|''$: Absolute value of the projection of the total angular momentum for upper and lower state.
\end{threeparttable}
}
\end{table}

\subsection{Cross-sections}
\label{sec:crosssection}

\begin{figure}
\centering    
\includegraphics[scale=0.45]{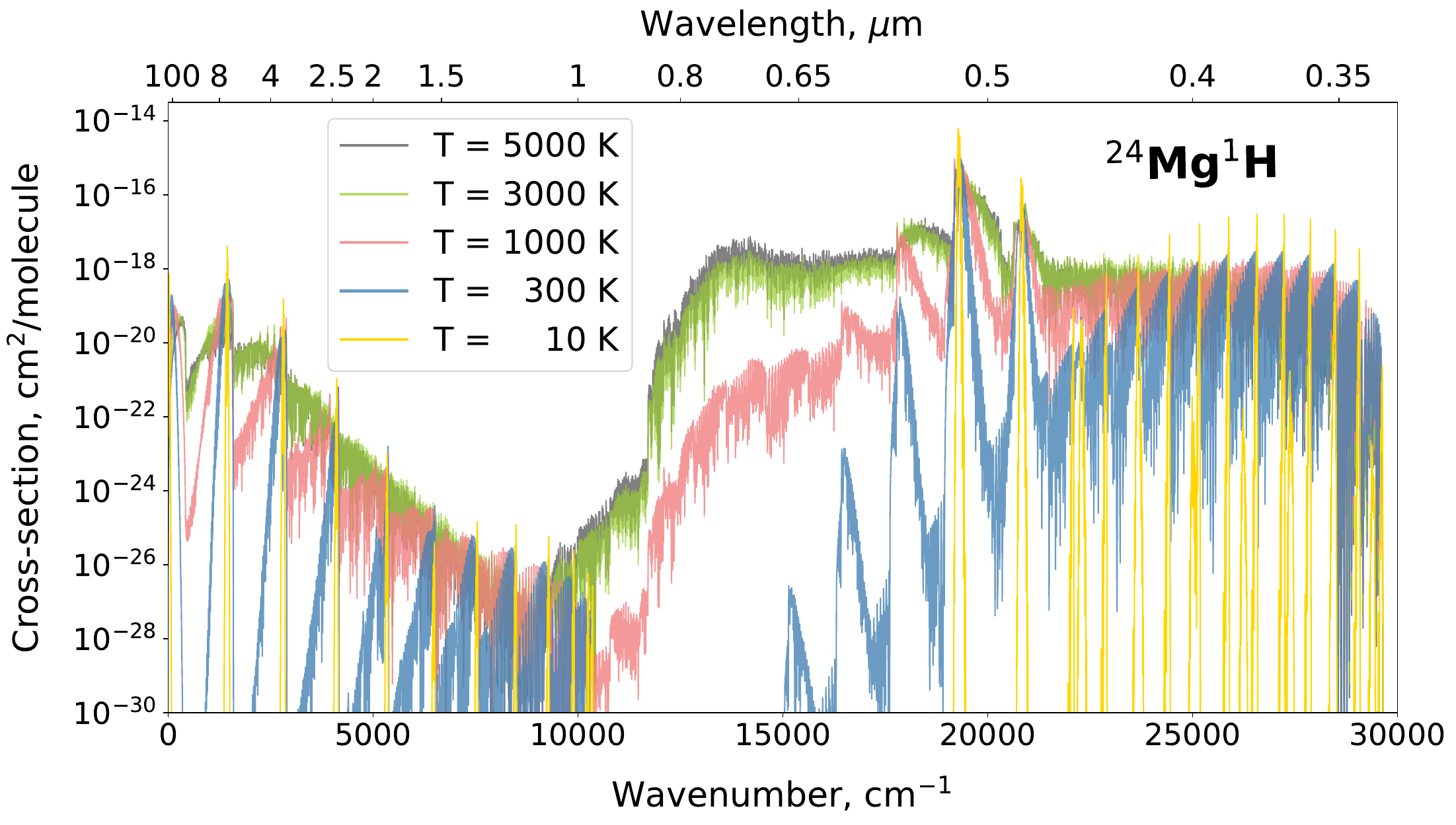} 
\quad
\includegraphics[scale=0.45]{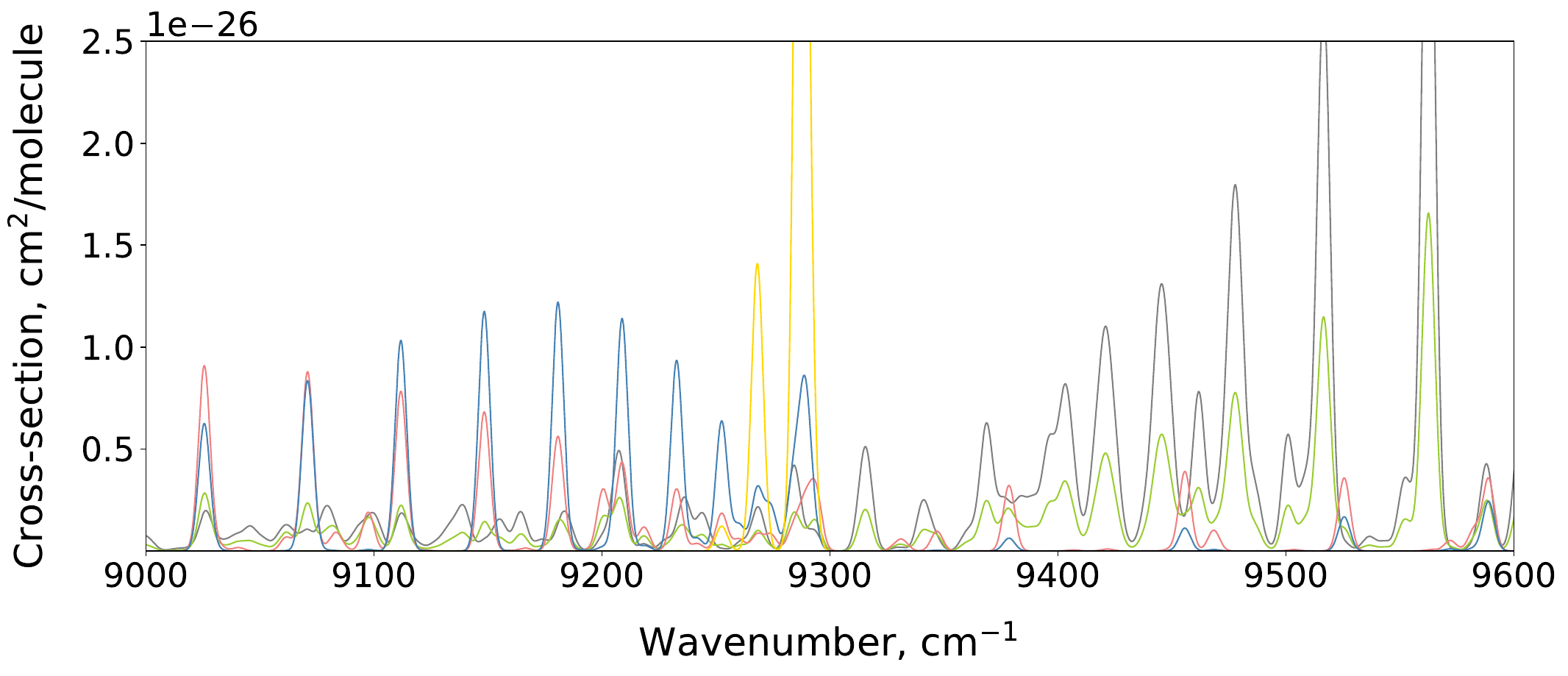}
\caption{Cross-section: Absorption spectrum of $^{24}$Mg$^1$H at $T=$ 10, 300, 1000, 3000, and 5000 K simulated using the ExoMol line list of the dataset XAB \citep{jt858} broadened using a Gaussian line profile with its HWHM $\alpha_{\rm D}=3$ cm$^{-1}$. Upper panel: Cross-sections in wavenumber range {$[0,30000]$} cm$^{-1}$. Lower panel: Cross-sections zoom for {$[9000, 9600]$} cm$^{-1}$.}    
\label{fig:cross sections}    
\end{figure}

\begin{figure}
\centering      
\includegraphics[scale=0.45]{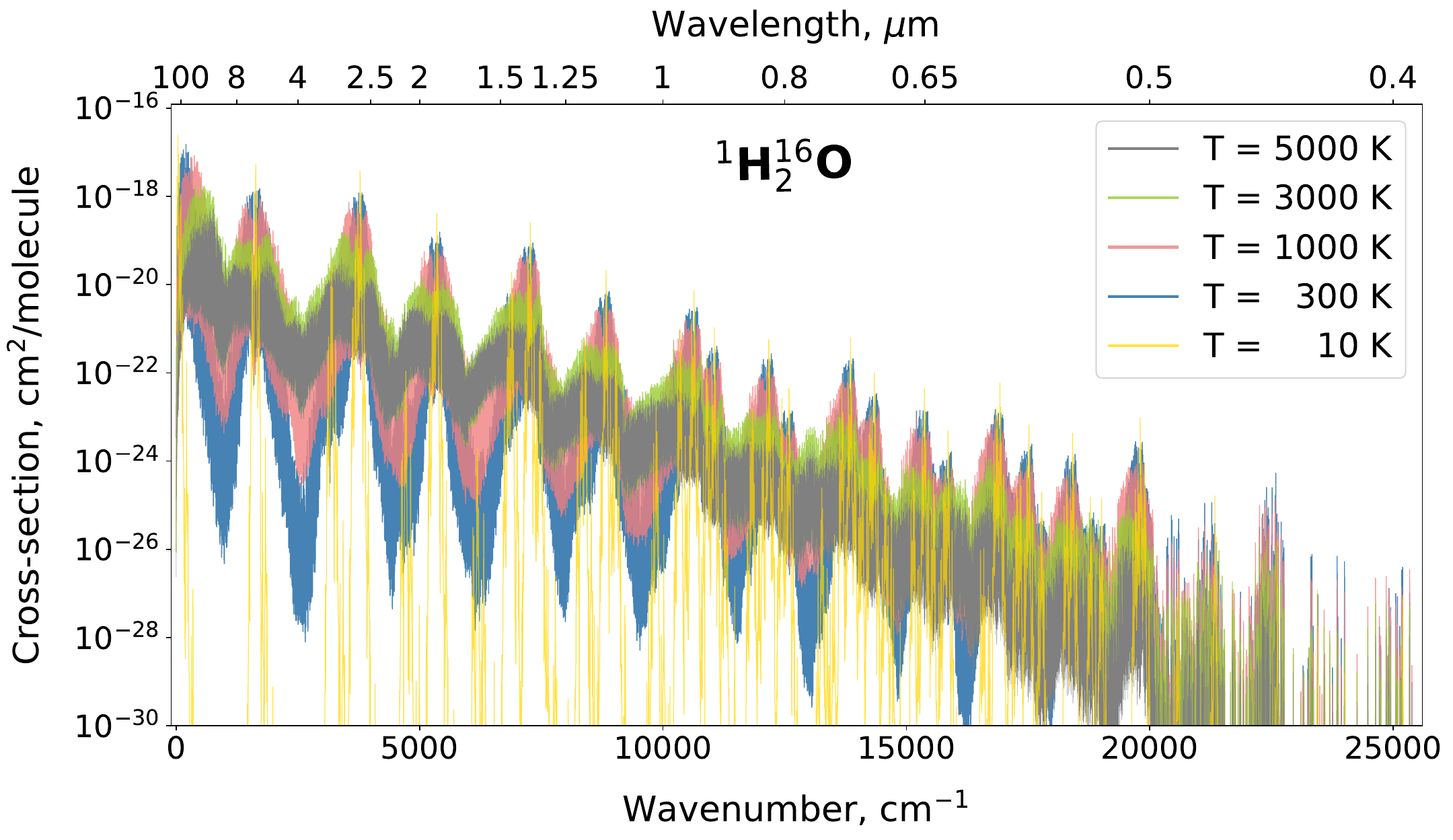} 
\quad
\includegraphics[scale=0.45]{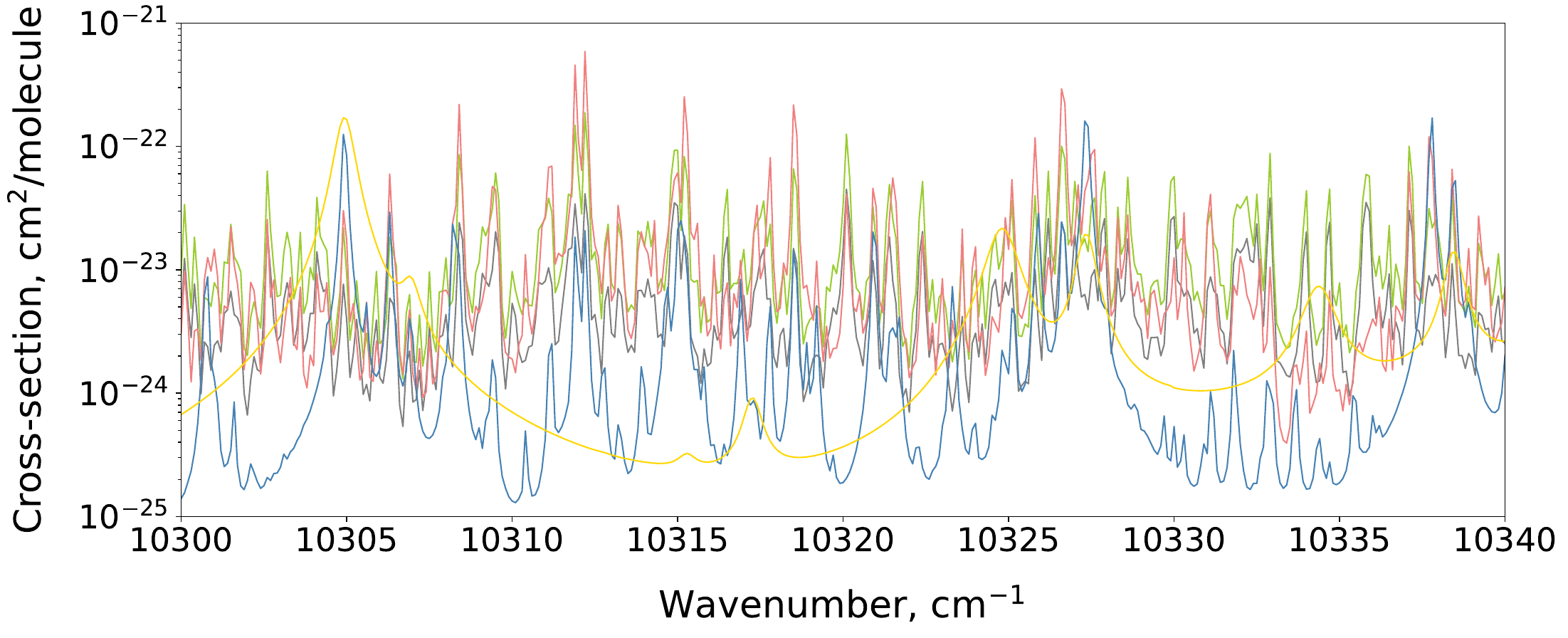}
\caption{Cross-section: Absorption spectrum of $^1$H$_2^{16}$O at $T=$ 10, 300, 1000, 3000, and 5000 K simulated using the ExoMol line list of the dataset POKAZATEL \citep{jt734} broadened using a SciPy Voigt line profile with the bin size 0.1 cm$^{-1}$ at $P=1$ bar, HWHM mixed with H$_2$ and He broadeners and also applied uncertainty and threshold filters (0.01 cm$^{-1}$ and $10^{-30}$ cm molecule$^{-1}$). Upper panel: Cross-sections in wavenumber range {$[0,26000]$} cm$^{-1}$. Lower panel: Cross-sections zoom for {$[10300, 10340]$} cm$^{-1}$.}    
\label{fig:cross sections H2O T}    
\end{figure}

\begin{figure}
\centering       
\includegraphics[scale=0.45]{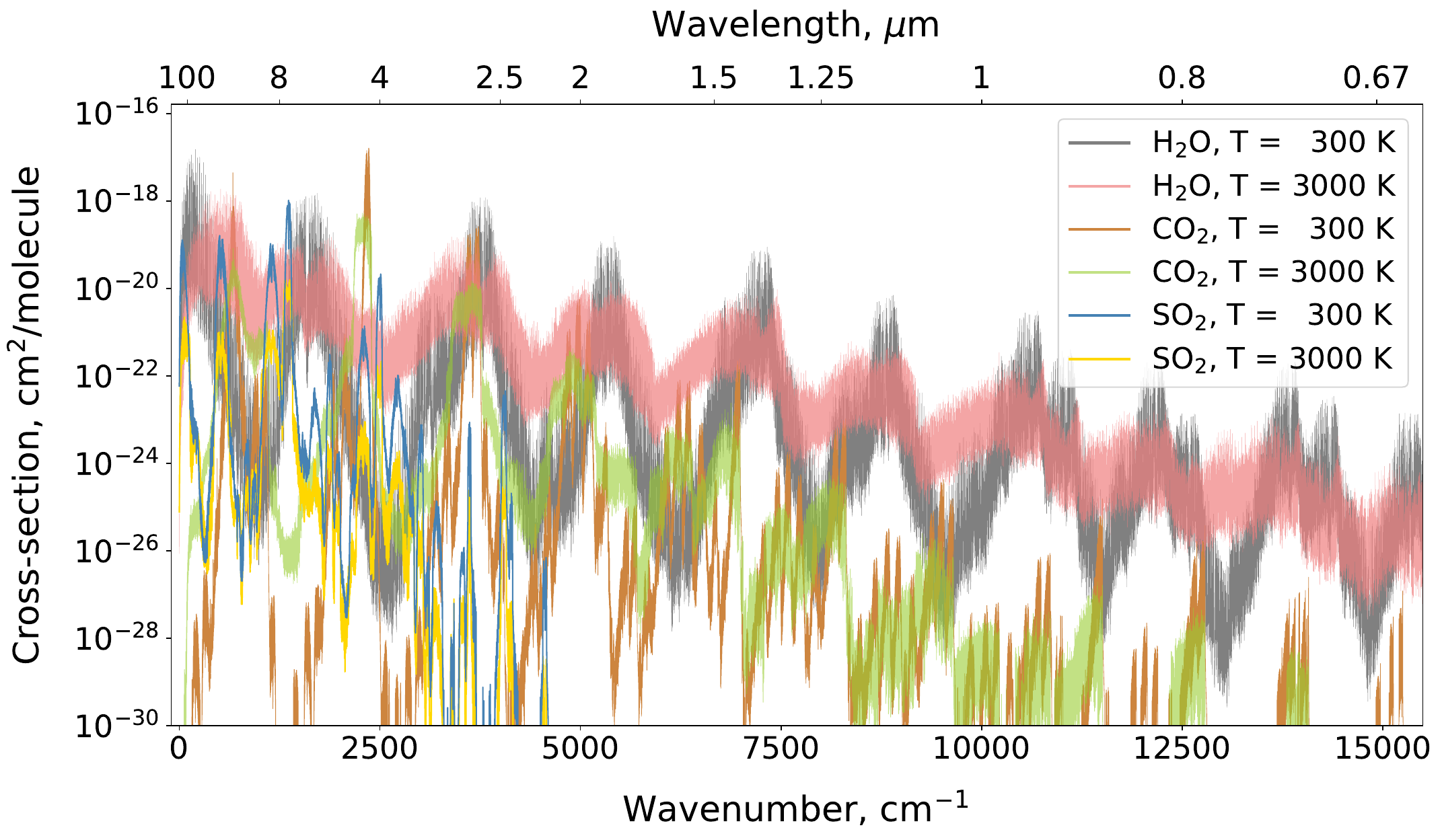} 
\quad
\includegraphics[scale=0.45]{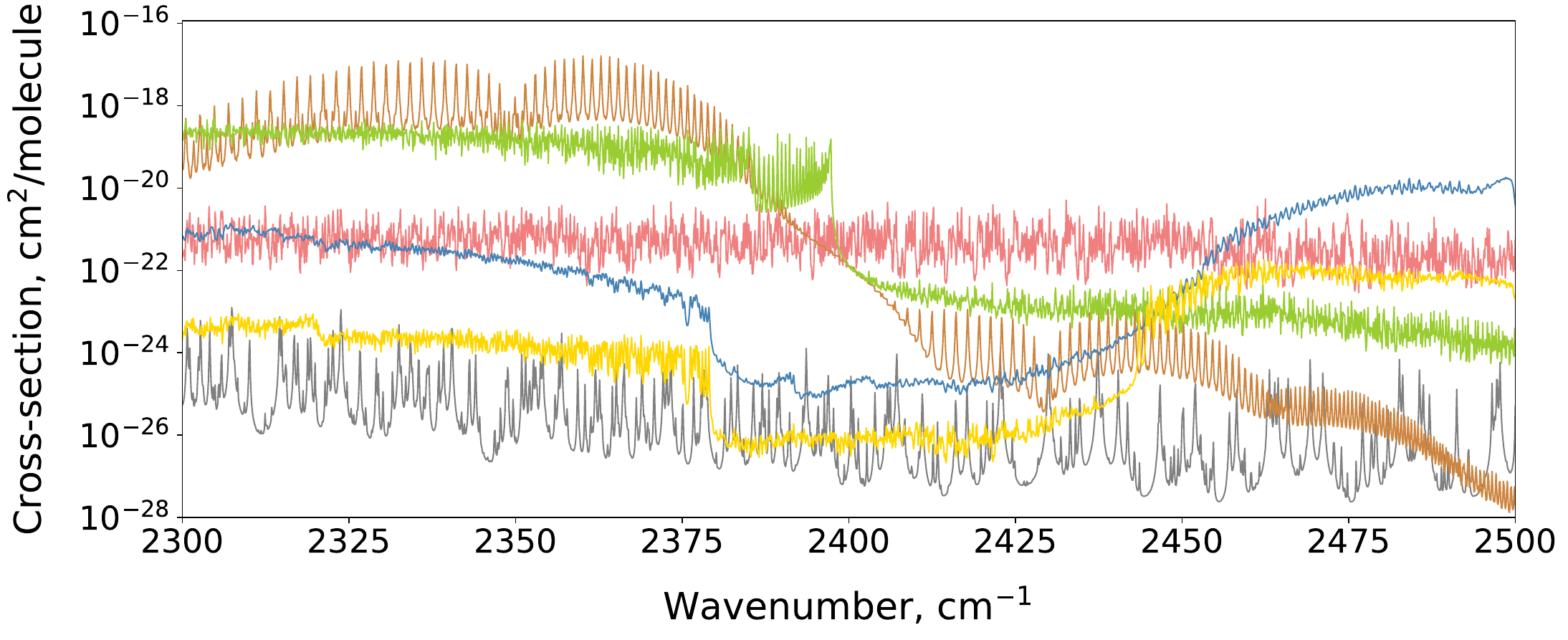}
\caption{Cross-section: Absorption spectrum of $^1$H$_2^{16}$O, $^{12}$C$^{16}$O$_2$, and $^{32}$S$^{16}$O$_2$, at $T=$ 300 and 3000 K simulated using the ExoMol line list of the dataset POKAZATEL \citep{jt734}, UCL-4000 \citep{jt804}, and ExoAmes \citep{jt635} broadened using a SciPy Voigt line profile with bin size 0.1 cm$^{-1}$ at $P=1$ bar, HWHM mixed with H$_2$ and He broadeners and also applied uncertainty and threshold filters (0.01 cm$^{-1}$ and $10^{-30}$ cm molecule$^{-1}$). Upper panel: Cross-sections in wavenumber range {$[0,15500]$} cm$^{-1}$. Lower panel: Cross-sections zoom for {$[2300,2500]$} cm$^{-1}$.}    
\label{fig:cross sections H2O CO2 SO2}    
\end{figure}

\textsc{PyExoCross} computes the cross-sections ($\sigma_{\rm ab}$ or $\sigma_{\rm em}$) for the given wavenumber range, temperature, and pressure, line intensity (absorption coefficient $I_{fi}$ in Eq.~(\ref{eq:absintensity}) or emission coefficient $\epsilon_{if}$ in Eq.~(\ref{eq:emiintensity})) and line profile ($f_{\tilde{\nu}}(\tilde{\nu})$, more details in Section~\ref{sec:profile}). \textsc{PyExoCross} provides uncertainty, threshold, and quantum number filters for calculating cross-sections (see Section~\ref{sec:filter}). Figures~\ref{fig:cross sections} and \ref{fig:cross sections H2O T} show examples of cross-sections of MgH and H$_2$O at $P=1$ bar and $T=$ 10, 300, 1000, 3000, and 5000 K simulated using the ExoMol line list XAB \citep{jt858} and POKAZATEL \citep{jt734}. Figure~\ref{fig:cross sections H2O CO2 SO2} compares cross-sections of H$_2$O \citep{jt734}, CO$_2$ \citep{jt804}, and SO$_2$ \citep{jt635} at $T=$ 300 and 3000 K.
Table~\ref{tab:cross} shows the cross-section file format and Table~\ref{tab:crossfile} specifies a part of the cross-section file.

\begin{table}
\centering
\setlength{\tabcolsep}{3mm}{
\begin{threeparttable}
\caption{Specification of the \texttt{.cross} cross-section file format.}
\label{tab:cross}
\begin{tabular}{llll}
\toprule
Field & Fortran format & C format & Description   \\
\midrule
$\tilde{\nu}_{i}$ & F12.6 & \%12.6f & Central bin wavenumber, cm${}^{-1}$\\
$\sigma _i$ & ES14.8 & \%14.8E & Absorption cross section, cm${}^2$ molecule${}^{-1}$\\
\bottomrule
\end{tabular}
Fortran format: (F12.6,1x,ES14.8).
\end{threeparttable}
}
\end{table}
\begin{table}
\centering
\setlength{\tabcolsep}{16.2mm}{
\begin{threeparttable}
\caption{Extract from the \texttt{.cross} cross-section file generated from the $^1$H$_2^{16}$O POKAZATEL ExoMol line lists \citep{jt734}.}
\label{tab:crossfile}
\begin{tabular}{cc} 
\toprule
$\tilde{\nu}$ (cm${}^{-1}$) & $\sigma _i$ (cm${}^2$ molecule${}^{-1}$) \\
\midrule
 5000.000000 & 8.03246668E-25 \\
 5000.100000 & 1.84420150E-24 \\
 5000.200000 & 2.00318098E-23 \\
 5000.300000 & 3.84642766E-24 \\
 5000.400000 & 8.77939177E-25 \\
 5000.500000 & 4.33350935E-25 \\
 5000.600000 & 2.91607795E-25 \\
 5000.700000 & 2.30487176E-25 \\
 5000.800000 & 2.01726401E-25 \\
 5000.900000 & 2.32892097E-25 \\
\bottomrule
\end{tabular}
$\tilde{\nu}$: Central bin wavenumber, cm${}^{-1}$; \\
$\sigma _i$: Absorption cross section, cm${}^2$ molecule${}^{-1}$.
\end{threeparttable}
}
\end{table}

\subsubsection{Wavenumber grids}

\textsc{PyExoCross} uses uniform wavenumber grids  when calculating cross-sections. The grids can be defined using two methods for a given  wavenumber  range $[\tilde{\nu}_\textrm{min},\tilde{\nu}_\textrm{max}]$: (i) using  the number of grid points $N_{\textrm{points}}$ with $N_{\textrm{points}} - 1$ as the number of grid intervals or (ii) the size of the grid bin $\Delta\tilde{\nu}$. The wavenumber step size is defined as
\begin{equation} 
\label{grids}
    \Delta\tilde{\nu} = \frac{\tilde{\nu}_\textrm{max} - \tilde{\nu}_\textrm{min}}{N_{\textrm {points}} - 1}.
\end{equation}
In order to let $\Delta\tilde{\nu}$ be a `round' value, the number of grid points is usually set to an odd number \citep{jt708}.

\subsubsection{Absorption and emission cross-sections}

The absorption cross-section $\sigma_{\rm ab}$ at  wavenumber $\tilde{\nu}$ for a transition line $f \leftarrow i$ with intensity $I_{fi}$ (Eq.~(\ref{eq:absintensity})) and line profile ($f_{\tilde{\nu}_{fi}}$) is given by
\begin{equation} 
\label{eq:absxsec}
    I_{fi}=\int_{-\infty}^{\infty}\sigma_{fi}(\tilde{\nu})\mathrm{d} \tilde{\nu} \Rightarrow \sigma_{\rm ab}(\tilde{\nu}) = I_{fi} f_{\tilde{\nu}_{fi}}(\tilde{\nu}).
\end{equation}
Similarly, the emission cross-section $\sigma_{\rm em}$ at  wavenumber $\tilde{\nu}$ for a transition line $i \rightarrow f$ is calculated with the emission coefficient $\epsilon_{if}$ (Eq.~(\ref{eq:emiintensity})) and line profile ($f_{\tilde{\nu}_{fi}}$):
\begin{equation} 
\label{eq:emixsec}
    \epsilon_{if}=\int_{-\infty}^{\infty}\epsilon_{if}(\tilde{\nu})\mathrm{d}\tilde{\nu}\Rightarrow \sigma_{\rm em}(\tilde{\nu}) = \epsilon_{if} f_{\tilde{\nu}_{if}}(\tilde{\nu}).
\end{equation}   

\subsection{Filters}
\label{sec:filter}

\textsc{PyExoCross} can apply filters including uncertainty, threshold, quantum number labels, formats, and values based on the information given in the \texttt{.states} file if users require the filters.

\subsubsection{Uncertainty filter}

An uncertainty filter can be applied when computing stick spectra, cross-sections, or performing format conversion. The uncertainty refers to the accuracy of each individual transition wavenumbers. \textsc{PyExoCross} applies an uncertainty filter with a given value which is usually taken to be the upper bound of the selected uncertainties. Laboratory spectroscopy uses uncertainties in cm$^{-1}$ and the uncertainties less than 0.01 cm$^{-1}$ are usually assumed to be accurate, but even uncertainties of 0.05 cm$^{-1}$ are probably good.
Astronomers use the resolving power $R=\frac{\lambda}{\Delta \lambda}$ in wavelength as a measure of the spectral resolution of an instrument and this is approximately the same as $R=\frac{\tilde v}{\Delta \tilde v}$ in wavenumbers. $R>100\,000$ indicates high resolution, but for some astronomy fields, $R>50\,000$ could be seen as high enough resolution as well.

\subsubsection{Intensity threshold filter}

An intensity threshold filter can be applied when computing stick spectra, cross-sections, or performing format conversion. A large intensity filter means only stronger intensities are processed in the task in question. \textsc{PyExoCross} extracts and process  lines, stronger than intensity threshold requested. The default intensity threshold value is  $I>10^{-30}$ cm molecule$^{-1}$. Users are recommended to use the threshold filter when processing the molecules with large datasets to reduce run times; however, we should note that including weak lines with $I<10^{-30}$ cm molecule$^{-1}$ can signifcantly alter the
results \citep{jt572}.

\subsubsection{Quantum number filter}

\begin{figure}
    \centering
    \includegraphics[scale=0.45]{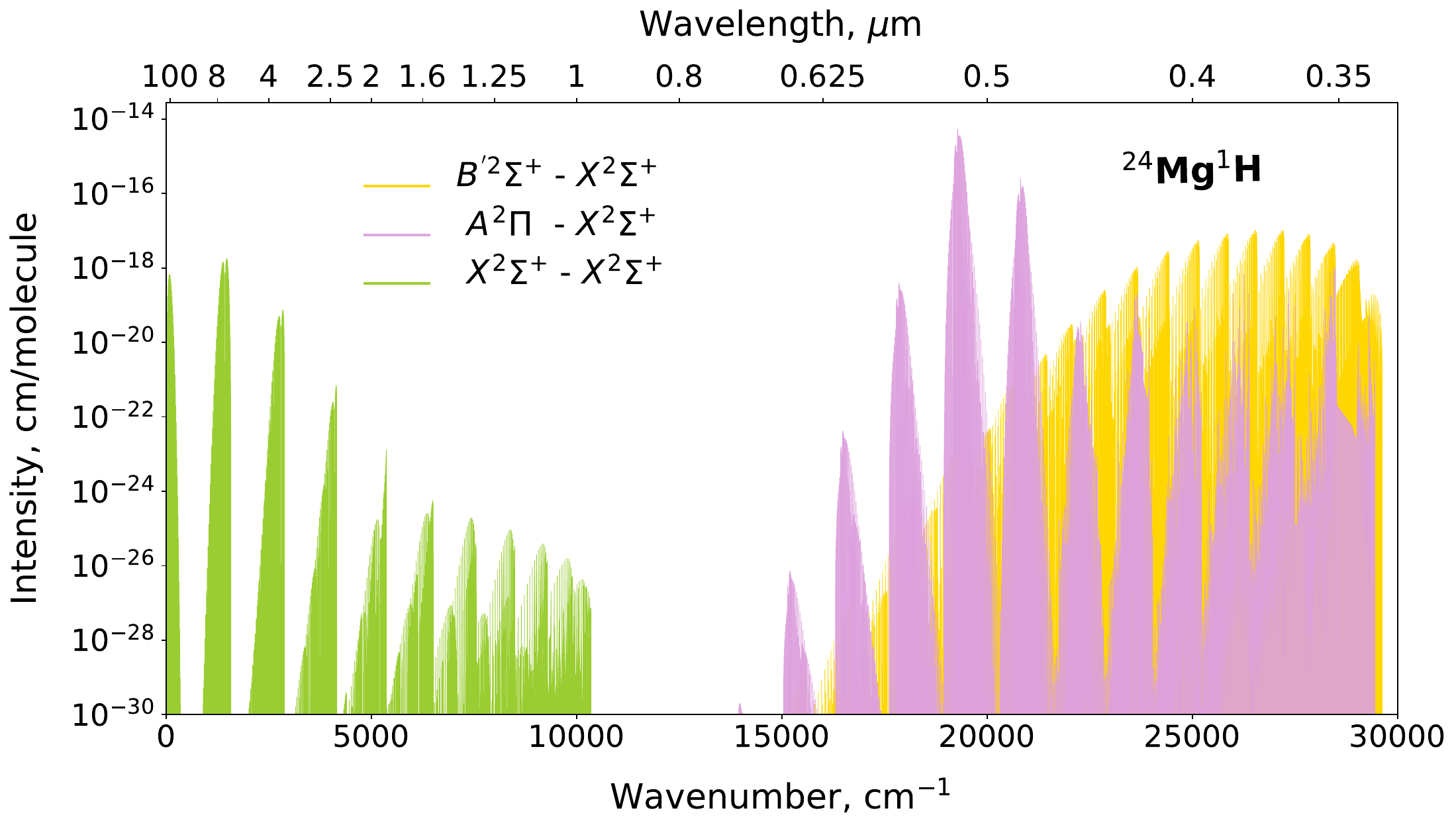}
    \caption{Absolute absorption line intensities of different electronic states X–X, A–X, and B–X bands of the ExoMol $^{24}$Mg$^{1}$H simulated at $T=300$ K.}
    \label{fig:eS}
\end{figure}

A quantum number filter can be applied in calculations of  stick spectra and cross-sections. This filter can be used to select transitions for a given set of the upper/lower  quantum numbers. The user can also select the  quantum numbers to be printed out as well as the printing format. Figure~\ref{fig:eS} shows the stick spectra of MgH using the electronic states as the quantum number filter.
%In addition, the user has the option of extracting some entries from the States file by specifying which quantum number values should be saved. 
%It is possible to filter by quantum number label and specify upper and lower quantum numbers. 
%The quantum number filters can help analyse the spectra as a function of different quantum labels.

\subsection{Plotting options}

\textsc{PyExoCross} provides the functionality to optionally plot figures for stick spectra and cross-sections. Users also can set the bottom $y$-axis view limits of data coordinates on plotting stick spectra and cross-sections. The default value of the bottom $y$-axis limit is $10^{-30}$ cm molecule$^{-1}$.

\subsection{Format conversion} 
\label{sec:conversion}

\subsubsection{Converting data format from ExoMol to HITRAN}

\textsc{PyExoCross} can convert data from the ExoMol format to the HITRAN format (see Tables~\ref{tab:HITRAN format} and \ref{tab:HITRAN data format}). This involves the following steps: (i) reading the States file (\texttt{.states}) from the ExoMol database,  (ii) reading the ExoMol Transition files (\texttt{.trans}) in chunks, (iii) reading the ExoMol partition function file, (iv) for each line, calculating the line intensity and applying the uncertainty and intensity thresholds, and (v) finally, printing out the transition descriptors in the HITRAN format, with the corresponding  error codes for the given uncertainties, etc.
%After the calculations, extracting high-resolution line lists from the results which have low uncertainties. Finally, recording results in the HITRAN format and saving the error codes instead of the exact values of the uncertainties. 

For the format conversion, the user has to provide the HITRAN molecule and isotopologue IDs. If the molecule is included in the HITRAN database, \textsc{PyExoCross} finds the necessary isotopic abundance from HITRANOnline, otherwise the programme uses 1 as the abundance to calculate the intensities.
\textsc{PyExoCross} can extract Einstein $A$-coefficients from the ExoMol Transition files, lower state energy; upper and lower total degeneracy from the ExoMol States file. \textsc{PyExoCross} also supports extracting $\gamma_{\rm self}$ and $n_{\rm air}$ from \texttt{.self} and \texttt{.air} broadening files, separately, if they exist. Otherwise, the default values $\gamma_{\rm self}=0.07$ cm$^{-1}$ per atm and $n_{\rm air}=0.5$ cm$^{-1}$ per atm are used.  \textsc{PyExoCross} only considers the first two HITRAN error codes  which are error codes of the line position and intensity. The remaining four integers are set to zero. The line position uncertainty error codes are described in Table~\ref{tab:uncertainty}. For the intensity error codes, the record `0' is used for all states since they are undefined in the ExoMol database. Table~\ref{tab:qn value} gives some information on  how the quantum number formats are converted.

\textsc{PyExoCross} provides more flexibility to help users choose which quantum number labels they prefer to save in the HITRAN format as well as their formats. Users can provide global and local quantum number labels and their corresponding formats in the input file so that they can arbitrarily change the quantum number formats to satisfy the 60-character limit of the HITRAN database format. 
%To make sure the total number of characters of quantum numbers is 60, \textsc{PyExoCross} needs users to provide which labels they want and their formats. 
If the total number of  characters provided is less than 60, \textsc{PyExoCross} will truncate them to be exactly 60 characters.

%\red{What about the quantum numbers? How do you format them in order to fit into HITRAN 60 characters?} \green{jingxin: \textsc{PyExoCross} provides more flexibility to help users choose which quantum number labels they prefer to save in the HITRAN format results and in what kind of formats. Users can provide global and local quantum number labels and their corresponding formats in the input file so that they can arbitrarily change the quantum number formats to satisfy the 60-character limit of the HITRAN database format.} \red{I think for HITRAN format it is important to follow it exactly and limit  the QN output to exactly 60 characters. Is it what you do here but let the users decide which QN to write? } \green{Yes, \textsc{PyExoCross} needs users to provide which labels they want and their formats. The number of total characters provided by users can be less than 60, \textsc{PyExoCross} can fix them to be exactly 60 characters.}

\textsc{PyExoCross} currently cannot provide air pressure-induced line shifts ($\delta_{\textrm{air}}$), reference indices (I$_\textrm{ref}$) and an associated uncertainty  flag. 

\subsubsection{Converting data format from  HITRAN to ExoMol}

In principle
\textsc{PyExoCross} can convert HITRAN \texttt{.par} files into the ExoMol format (\texttt{.states}, \texttt{.trans}, \texttt{.pf}), but this is not straightforward since the HITRAN data model does not ensure that the tabulated lower state energies and transitions are are self-consistent. The main challenge is thereforre to construct the ExoMol States file and define the associated state IDs. 
To this end, the following steps are used.

\textbf{(A) Construct the ExoMol States file}\\
(i) Read HITRAN line list \texttt{.par} file; 
(ii) for each line, calculate the upper state energy $\tilde{E}'_{i}$ with the wavenumber $\tilde{\nu}_{j}$ and lower state energy $\tilde{E}''_j$ by Eq.~(\ref{eq:v});
(iii) convert the uncertainty error code into the upper bound of the absolute uncertainty range;
(iv) process quantum numbers and extract angular momentum $F'$ and $F''$ from the HITRAN quantum numbers;
(v) split HITRAN data into the upper and lower  datasets, called DataFrames in Python, which include the Einstein $A$-coefficients ($A$), the statistical weights ($g$), uncertainties, state energies $\tilde{E}$, global quantum number and local quantum numbers, and then concatenate these two DataFrames with the same columns; 
(vi) create a new DataFrame  which merges levels with the same statistical weights $g_{\rm tot}$, state energies $\tilde{E}$, uncertainties, and quantum numbers, in this process, we use the lines with the lowest uncertainty when they have the same quantum numbers and statistical weights; 
(vii) sort with increasing state energies and use the continuum-increasing integers as the state IDs; and 
(viii) create the ExoMol States file in the ExoMol data format with the state ID $i$, state energy $\tilde{E}$, total degeneracy $g_{\rm tot}$, angular momentum $J$ or $F$, uncertainties $Unc$, and quantum numbers (see Table~\ref{tab:states}). 

\textbf{(B) Construct the ExoMol Transitions file}\\
(i) For each line in the HITRAN DataFrame, reconstruct the upper and lower state IDs; 
(ii) calculate new transitions frequencies with upper and lower state energies; and 
(iii) write an ExoMol Transitions file with the upper ID, lower state ID, Einstein A-coefficient $A$, and frequency wavenumber $\tilde{\nu}$ (see Table~\ref{tab:trans}).

\textbf{(C) Construct an ExoMol air-broadening file (\texttt{.air})}\\
(i) Concatenate the initial HITRAN DataFrame with the $\gamma_{\rm air}$, $n_{\rm air}$ and $J''$ columns, and then drop duplicated rows and 
(ii) fill `a0' in the first column called `code' and save code, $\gamma_{\rm air}$, $n_{\rm air}$, and $J''$ to the ExoMol air broadening file with the ExoMol data format (see Table~\ref{tab:broad}). Note this procedure assumes the relatively simple a0 $J$-diet for line broadening \citep{jt684}.

\textbf{(D) Construct ExoMol self-broadening files (\texttt{.self})}\\
(i) Concatenate the initial HITRAN DataFrame with the $\gamma_{\rm self}$, $n_{\rm air}$, and $J''$ columns, and then drop duplicated rows and 
(ii) fill `a0' in the first column called `code' and save code, $\gamma_{\rm air}$, $n_{\rm air}$, and $J''$ to the ExoMol self broadening file with the ExoMol data format (see Table~\ref{tab:broad}, see next for more details on the line broadening codes). 

\textsc{PyExoCross} can only convert the standard 160-character HITRAN format data to the ExoMol format and for reasons given above the  results
need to be review carefully. As  the HITRAN database has added extra H$_2$, He, CO$_2$, and H$_2$O broadening, users can replace air and self broadening columns with these extra broadening data. The quantum numbers are formatted following Tables S1 and S2 in the HITRAN 2020 supplementary material \citep{jt841}. The quantum numbers are written using the same format as used in the HITRAN .par file, therefore there is no need for the format specification of the quantum numbers.

\begin{table}
\centering
\setlength{\tabcolsep}{0.85mm}{
\begin{threeparttable}
\caption{HITRAN  line-by-line format in 2004 edition \citep{rothman2005hitran} (160-character record).}
\label{tab:HITRAN format}
\begin{tabular}{lccccccccccccccccccc}
\toprule
Parameter & M & I & $\nu$ & $S$ & $A$ & $\gamma_{\textrm{air}}$ & $\gamma_{\textrm{self}}$ & $E''$ & $n_{\textrm{air}}$ & $\delta_{\textrm{air}}$ & $V'$ & $V''$ & $\textsl{Q}'$ & $\textsl{Q}''$ & I$_{\textrm{err}}$ & I$_{\textrm{ref}}$ & * & $g'$ & $g''$ \\
\midrule
Field length & 2 & 1 & 12 & 10 & 10 & 5 & 5 & 10 & 4 & 8 & 15 & 15 & 15 & 15 & 6 & 12 & 1 & 7 & 7 \\
Fortran format & I2 & I1 & F12.6 & E10.3 & E10.3 & F5.4 & F5.4 & F10.4 & F4.2 & F8.6 & A15 & A15 & A15 & A15 & 6I1 & 6I2 & A1 & F7.1 & F7.1 \\
C format & \%2d & \%1d & \%12.6f & \%10.3e & \%10.3e & \%5.4f & \%5.4f & \%10.4f & \%4.2f & \%8.6f & \%15s & \%15s & \%15s & \%15s & \%6d & \%12d & \%1s & \%7.1f & \%7.1f \\
\bottomrule
\end{tabular}
\end{threeparttable}
}
\end{table}
\begin{table}
\centering
\setlength{\tabcolsep}{5.5mm}{
\begin{threeparttable}
\caption{Description of the parameters presented in the 160-character records in the line-by-line section of the HITRAN database \citep{rothman2005hitran}.}
\label{tab:HITRAN data format}
\begin{tabular}{llll}
\toprule
Parameter & Meaning & Field length & Type \\
\midrule
M & Molecule number & 2 & Integer \\
I & Isotopologue number & 1 & Integer \\
$\nu$ & Vacuum wavenumber & 12 & Real \\
$S$ & Intensity & 10 & Real \\
$A$ & Einstein $A$-coefficient &10 & Real \\
$\gamma_{\textrm{air}}$ & Air-broadened half-width & 5 & Real \\
$\gamma_{\textrm{self}}$ & Self-broadened half-width & 5 & Real \\
$E''$ & Lower state energy & 10 & Real \\
$n_{\textrm{air}}$ & Temperature-dependent exponent for $\gamma_{\textrm{air}}$ & 4  & Real \\
$\delta_{\textrm{air}}$ & Air pressure-induced line shift & 8  & Real \\
$V'$ & Upper state global `quanta' & 15 & Hollerith \\
$V''$ & Lower state global `quanta' & 15 & Hollerith \\
$\textsl{Q}'$ & Upper state local `quanta' & 15 & Hollerith \\
$\textsl{Q}''$ & Upper state local `quanta' & 15 & Hollerith \\
I$_{\textrm{err}}$ & Uncertainty indices & 6 & Integer \\
I$_{\textrm{ref}}$ & Reference indices & 12 & Integer \\
* & Flag & 1 & Character \\
$g'$ & The statistical weight of the upper state & 7 & Real \\
$g''$ & The statistical weight of the lower state & 7 & Real \\
\bottomrule
\end{tabular}
\end{threeparttable}
}
\end{table}
\begin{table}
\centering
\setlength{\tabcolsep}{13mm}{
\begin{threeparttable}
\caption{The uncertainty codes used by the HITRAN database \citep{jt857}.}
\label{tab:uncertainty}
\begin{tabular}{cc}
\toprule
Code & Absolute uncertainty (unc) range  \\
\midrule
0 & $1 \le \textrm{unc}$ or Unreported \\
1 & $0.1 \le \textrm{unc} < 1$ \\
2 & $0.01 \le \textrm{unc} < 0.1$ \\
3 & $0.001 \le \textrm{unc} < 0.01$ \\
4 & $0.0001 \le \textrm{unc} < 0.001$ \\
5 & $0.00001 \le \textrm{unc} < 0.0001$ \\
6 & $0.000001 \le \textrm{unc} < 0.00001$ \\
7 & $0.0000001 \le \textrm{unc} < 0.000001$ \\
8 & $0.00000001 \le \textrm{unc} < 0.0000001$ \\
9 & $0.000000001 \le \textrm{unc} < 0.00000001$ \\
\bottomrule
\end{tabular}
\end{threeparttable}
}
\end{table}
\begin{table}
\centering
\setlength{\tabcolsep}{4.4mm}{
\begin{threeparttable}
\caption{Conversion of the quantum number values between the ExoMol and HITRAN database.}
\label{tab:qn value}
\begin{tabular}{c|cccc}
\hline
Quantum number value in ExoMol & \multicolumn{4}{c}{Quantum value in HITRAN} \\
Description & Total symmetry & Vibrational symmetry & Rotational symmetry & Rotational parity (0 or 1) \\
Label & G$_{\rm tot}$ & G$_{\rm vib}$ & G$_{\rm rot}$ & $\tau_i$ \\
\hline
0 &        &        &        & s \\
1 & A1$'$  & A1$'$  & A1$'$  & a \\
2 & A2$'$  & A2$'$  & A2$'$  &   \\
3 & E$'$   & E$'$   & E$'$   &   \\
4 & A1$''$ & A1$''$ & A1$''$ &   \\
5 & A2$''$ & A2$''$ & A2$''$ &   \\ 
6 & E$''$  & E$''$  & E$''$  &   \\
\hline
\end{tabular}
\end{threeparttable}
}
\end{table}

\section{Line broadening and line profiles}
\label{sec:profile} 

An individual spectral line at high resolution is characterized by (i) its transition frequency, (ii)  transition integrated intensity, and (iii) line profile. The line profile is governed by line broadening effects which are functions of temperature and pressures, and whose inclusion is essential for most practical applications. The temperature-dependent Doppler broadening is represented by a Doppler line profile in the form of a Gaussian funciton. The pressure broadening can be modelled by  a Lorentzian line profile, while the
Voigt line profiles provide a convolution of the Doppler and Lorentzian profiles  \citep{17Schreier}.  It may also be necessary to include lifetime effects
but these also produce a Lorentzian so do not alter the discussion below \citep{jt898}.

Line profiles $f_{\tilde{\nu}}(\tilde{\nu})$ are integrable functions whose areas are normalised to unity:
\begin{equation} 
\label{eq:Fv}
    \int_{-\infty}^{\infty}f_{\tilde{\nu}}(\tilde{\nu})\mathrm{d}v=1.
\end{equation} 
 \textsc{PyExoCross} offers a choice of four profiles namely Doppler, generalized Gaussian, Lorentzian and Voigt.
 Of course the Doppler profile is a Gaussian but with the line width  defined based on the mass  of the species concerned,  temperature and the line position, while Gaussian allows the user to specify the half-width at half-maximum (HWHM; $\alpha_{\rm G}$). Section \ref{sec:scipyVoigt} explores
 seven different methods of calculating Voigt profiles. 

\subsection{Wing cutoff}

As line profiles have in principle infinite spread, frequency (or wavelength) cutoffs are often used in practical calculations to limit the calculation region to the line centre only and to allow for the fact that Voigt/Lorentzian functions do not generally give a good representation of the far wings \citep{jt708}. 
Aside from influencing computation time and the accuracy of the cross-sections, a wing cutoff is also necessary if functions representing, e.g. the water continuum are to be used, as was recently advocated for exoplanet studies by \citet{jt850}. 
\citet{jt909} discuss this issue in detail. 
\textsc{PyExoCross} uses a default cutoff of 25 cm${}^{-1}$ in line with HITRAN, HITEMP, and most representations of the
water continuum. This choice is a matter of convention and can be changed by the user; however, we note that the default
value is in line with the recent recommendations of \citet{jt909}.

\subsection{Broadening parameters}
\label{sec:broadening}

Pressure broadening in the ExoMol database is currently based on the ExoMol diet \citep{jt684}; the database stores  line broadening parameters in 
  \texttt{.broad} broadening files; broadeners available for a given species are specified in the  \texttt{.def} definition file.
Users can provide as many broadening files with appropriate file suffixes, e.g. \texttt{.H2O.broad} and \texttt{.CO2.broad}. \textsc{PyExoCross} can then use suffixes to read the corresponding broadening files and, if necessary, process mixtures of broadeners (see Section~\ref{s:mixtures}). \textsc{PyExoCross} can process user-supplied broadening files  provided they use  the ExoMol \texttt{.broad} file format.  The format of the broadening file is expressed in Table~\ref{tab:broad} and the specification of the broadening file is given in Table~\ref{tab:broadfile}. Currently, the ExoMol database contains   \texttt{.broad} files for air,  Ar, CH$_4$, CO, CO$_2$, H$_2$, H$_2$O, He, N$_2$, NH$_3$, NO, O$_2$ and the rapidly increasing number of broadeners being included in the ExoMol database. The HITRAN database has air and  self broadening for all species as well as parameters for the broadening by He, H$_2$, H$_2$O, and CO$_2$ \citep{ngo2013isolated,16WiGoKoHi.broad,19TaKoRo.broad,22TaSkSa.broad,21WcThSt.broad} for a small number of key species. 

\begin{table}
\centering
\setlength{\tabcolsep}{7.7mm}{
\begin{threeparttable}
\caption{Specification of the mandatory part of the \texttt{.broad} broadening file format.}
\label{tab:broad}
\begin{tabular}{llll}
\toprule
Field & Fortran format & C format & Description   \\
\midrule
code & A2 & \%2s & Code identifying quantum number set following $J$\\
$\gamma _\textrm{ref}$ & F6.4 & \%6.4f & Lorentzian half-width at reference temperature and pressure in cm${}^{-1}/\textrm{bar}$\\
$n_\textrm{L}$ & F6.3 & \%6.3f & Temperature exponent\\
$J"$ & I7/F7.1 & \%7d/\%7.1f & Lower $J$-quantum number integer/half-integer\\
\bottomrule
\end{tabular}
Fortran format: $J$ integer: (A2,1x,F6.4,1x,F6.3,1x,I7) or $J$ half-integer: (A2,1x,F6.4,1x,F6.3,1x,F7.1).
\end{threeparttable}
}
\end{table}
\begin{table}
\centering
\setlength{\tabcolsep}{8.5mm}{
\begin{threeparttable}
\caption{Extract from the \texttt{.broad} broadening file of the H$_2$ broadening file \texttt{14N-1H3\_\_H2.broad} \citep{jt684}.}
\label{tab:broadfile}
\begin{tabular}{cccc}
\toprule
Code & $\gamma_\textrm{ref}$ & $n_\textrm{L}$ & $J''$ \\
\midrule
a0 & 0.0908 & 0.583       & 0 \\
a0 & 0.0935 & 0.559       & 1 \\
a0 & 0.0926 & 0.527       & 2 \\
a0 & 0.0850 & 0.508       & 3 \\
a0 & 0.0800 & 0.529       & 4 \\
a0 & 0.0768 & 0.506       & 5 \\
a0 & 0.0728 & 0.501       & 6 \\
a0 & 0.0698 & 0.506       & 7 \\
a0 & 0.0671 & 0.514       & 8 \\
a0 & 0.0655 & 0.500       & 9 \\
\bottomrule
\end{tabular}
Code: Code identifying quantum number set following $J$; \\
$\gamma_\textrm{ref}$: Lorentzian half-width at reference temperature and pressure in cm${}^{-1}/\textrm{bar}$; \\
$n_\textrm{L}$: Temperature exponent; \\
$J''$: Lower $J$-quantum number integer/half-integer.
\end{threeparttable}
}
\end{table}

\subsection{Doppler and Gaussian profile}
\label{sec:doppler gaussian}

The Gaussian line profile is defined as 
\begin{equation} 
\label{eq:gaussian}
    f_{\tilde{\nu}_{fi},\alpha_{\rm G}}(\tilde{\nu})=\sqrt{\frac{\ln{2}}
    {\pi}}
    \frac{1}{\alpha_{\rm G}}\exp\left(-\frac{(\tilde{\nu}-\tilde{\nu}_{fi})^{2}\ln{2}}{\alpha_{\rm G}^{2}}\right), 
\end{equation}
where $\alpha_{\rm G}$ is the HWHM and $\tilde{\nu}_{fi}$ is the line centre. 
$\alpha_{\rm G}$ is  related to the  Gaussian standard deviation $\sigma_{\rm G}$ as \citep{squires2001practical}:
\begin{equation} 
\label{eq:sigma}
    \sigma_{\rm G}=\frac{\alpha_{\rm G}}{\sqrt{2\ln{2}}}
\end{equation}
and can be specified in the \textsc{PyExoCross} input.

The Doppler profile is based on the Gaussian shape specified in Eq.~(\ref{eq:gaussian}) with HWHM $\alpha_{\rm D}$ given by \citep{squires2001practical}
\begin{equation} 
\label{eq:alpha}
    \alpha_{\rm D}=\sqrt{\frac{2N_Ak_BT\ln{2}}{M}}\frac{\tilde{\nu}_{fi}}{c},
\end{equation}
which depends on the temperature, mass of the molecule, and the line position.  Here, $k_B$ (erg K) is the Boltzmann constant, $c$ (cm s$^{-1}$) is the speed of light, $T$ (K) is the temperature, $M$ (Da) (g mol$^{-1}$) is the molar mass of the isotopologue in grams \citep{ngo2013isolated} which can be taken from the HITRANOnline website or from the \texttt{.def} definition file of the ExoMol database, and  
$N_A$ (mol$^{-1}$) is Avogadro's number. $N_A/M$ can be replaced by $1/1000m$ where isotopologue mass $m$ (kg) is available from the ExoMol definition file. For Doppler as well as for the generalized  Gaussian, the HWHMs do not depend on the (upper/lower) molecular states.

\subsection{Lorentzian profiles}
\label{sec:lorentzian}

The Lorentzian profile is the Cauchy distribution whose probability density function is  \citep{feller1971introduction,johnson1994continuous}
\begin{equation} 
\label{eq:lorentzian}
    f_{\tilde{\nu}_{fi},\gamma_{L}}^{L}(\tilde{\nu})=\frac{1}{\pi}\frac{\gamma_{L}}{\left(\tilde{\nu}-\tilde{\nu}_{fi}\right)^{2}+\gamma_{L}^{2}},
\end{equation}
where $\tilde{\nu}_{fi}$ is the centroid in wavenumbers identifying the position of the peak of the distribution and  $\gamma_{L}$ is Lorentzian HWHM, which in \textsc{PyExoCross}  conventionally assumes the following temperature and pressure-dependent form:
\begin{equation} 
\label{eq:gamma}
    \gamma_{\rm L}=\gamma_\textrm{ref}\left(\frac{T_{\textrm {ref}}}{T}\right)^{n_{\rm L}}\frac{P}{P_{\textrm {ref}}},
\end{equation}
where $\tilde{\nu}_{fi}$ is the line centre. As in HITRAN, $T_{\textrm {ref}}=296$~K is the reference temperature and $P_{\textrm {ref}}=1$ bar is the reference pressure \citep{schreier2021computational}, and 
$\gamma_\textrm{ref}$ and $n_L$ are the reference HWHM (see Sections~\ref{sec:broadening} and \ref{sec:extract broad}) and the temperature exponent, which are the broadening parameters for the given broadeners in the ExoMol database. In the HITRAN database, $\gamma_{\textrm{air}}$ and $\gamma_{\textrm{self}}$ are used to calculate $\gamma_{\textrm{ref}}$ with mixing ratios  provided by users, see Section~\ref{s:mixtures}. $n_{\rm L}$ is as $n_{\textrm{air}}$ in the HITRAN line lists. In general, $\gamma_\textrm{ref}$ and  $n_{\rm L}$ are (upper/lower) state-dependent.  The specification of the broadeners and the broadener mixture ratio must be set in the \textsc{PyExoCross}  \texttt{.inp} input file. 

Users can choose to input the HWHM ($\gamma_{\rm L}$) of the Lorentzian profile directly as a constant for calculations instead of calculating with the parameters from the line-broadening files of the ExoMol database.

\subsubsection{Extracting broadening parameters from \texttt{.broad}}
\label{sec:extract broad}

Calculation of the Lorentzian/Voigt broadening requires information from the  States (\texttt{.states}) file and the parameters from the broadening file (\texttt{.broad}). If the value of $J''$ in the States file is larger than $J''_\textrm{max}$ in the broadening file, then the broadening parameters $\gamma_{\rm L}$  and $n_{\rm L}$ corresponding to $J''_\textrm{max}$ are assumed. If $J''$ is half-integer, the parameters corresponding to the index value $J'' - \frac{1}{2}$ in the broadening file are used. This process of extracting values based on their indices is especially efficient in Python. A part of the Python code for extracting broadening parameters from the \texttt{.broad} file is shown in Listing~\ref{listing:broad}.

\begin{python}[caption={A sample of the Python code for extracting broadeners from the \texttt{.broad} broadening file. \textsc{PyExoCross} reads the \texttt{.broad} broadening file and saves the data as a DataFrame \texttt{broad\_{df}}. \texttt{states\_trans\_{df}} is a DataFrame storing the line lists of the \texttt{.states} States file merged by upper and lower states IDs from the \texttt{.trans} Transitions file. \texttt{Jpp} and \texttt{J"} means lower $J$ and \texttt{id\_{broad}} indicates the index of the broad DataFrame. \texttt{gamma\_L} and \texttt{n\_{air}} are $\gamma_\textrm{ref}$ and $n_\textrm{air}$ denoting the reference HWHM and the temperature exponents.}, label={listing:broad}]
def extract_broad(broad_df, states_trans_df):
    max_broad_J = max(broad_df['Jpp'])
    Jpp = states_trans_df['J"'].values
    Jpp[Jpp > max_broad_J] = max_broad_J
    id_broad = (states_trans_df['J"']-0.1).round(0).astype('int32').values
    gamma_L = broad_df['gamma_L'][id_broad]
    n_air = broad_df['n_air'][id_broad]
    return(gamma_L, n_air)
\end{python}
% \end{lstlisting}

% \begin{listing}[!ht]
% \centering
% \begin{minted}{python}
% def extract_broad(broad_df, states_trans_df):
%     max_broad_J = max(broad_df['Jpp'])
%     Jpp = states_trans_df['J"'].values
%     Jpp[Jpp > max_broad_J] = max_broad_J
%     id_broad = (states_trans_df['J"']-0.1).round(0).astype('int32').values
%     gamma_L = broad_df['gamma_L'][id_broad]
%     n_air = broad_df['n_air'][id_broad]
%     return(gamma_L, n_air)
% \end{minted}
% \caption{A sample of the Python code for extracting broadeners from the \texttt{.broad} broadening file. \textsc{PyExoCross} reads the \texttt{.broad} broadening file and saves the data as a DataFrame \texttt{broad\_{df}}. \texttt{states\_trans\_{df}} is a DataFrame storing the line lists of the \texttt{.states} states file merged by upper and lower states IDs from the \texttt{.trans} transitions file. \texttt{Jpp} and \texttt{J"} means lower $J$ and \texttt{id\_{broad}} indicates the index of the broad DataFrame. \texttt{gamma\_L} and \texttt{n\_{air}} are $\gamma_\textrm{ref}$ and $n_\textrm{air}$ denote the reference HWHM and the temperature exponents.}
% \label{listing:broad}
% \end{listing}

\subsubsection{Mixtures of broadeners}
\label{s:mixtures}

\textsc{PyExoCross} allows for inclusion of different broadeners or their mixtures. For mixtures, in the input file, users can specify the fractional abundance  ($X_i$) of the $i$th broadener. Then, the overall value of $\gamma_{\rm L}$ is the weighted sum of the values of $\gamma_i^{\rm L}$ from each broadener \citep{jt708}:
\begin{equation} 
\label{eq:total gamma}
        \gamma_{\rm L}=\sum_{i}\gamma_i^{\rm L}X_i \quad \textrm{and} \quad \sum_{i}X_i=1.
\end{equation}

\subsubsection{Lifetime broadening}

In principle, the natural lifetime contributes to the line profile alongside the effects  of the temperature-dependent Doppler broadening and the pressure-dependent collision broadening. In practice, the natural lifetime generally contributes very little to the line profile and is consequently ignored. However, for predissociating states, the effect of the lifetime broadening can be important, see \citet{jt874} and \citet{jt922} for example; these can easily be included in the Voigt profile \citep{jt898}. Like the pressure broadening, lifetime broadening $\gamma_{\tau}$ produces a Lorentzian line shape with the HWHM  given by
\begin{equation}
\label{eq:liftime broadener}
    \gamma_\tau = \frac{\hbar}{2\tau^{'} h c},   
\end{equation}
where $h$ is the Planck constant (erg s) and $\hbar=h/(2\pi)$ is the reduced Planck constant. $c$ is the speed of light (cm s$^{-1}$). $\tau^{'}$ is the upper state lifetime (s). 
The sum of the pressure broadening and lifetime broadening $\gamma_{\rm tot}=\gamma_p+\gamma_{\tau}$ can be then used as the total Lorentzian component of the line profile. Since a Voigt profile is already being utilized, this has no computational impact on a calculation, implying that routine use of $\gamma_{\rm tot}$ for the half-width on a routine basis would eliminate the need to worry about whether or not predissociation has to be considered.

\subsection{Voigt profiles}
\label{sec:voigt}
The Voigt profile results from the convolution of the Gaussian and Lorentzian profiles and is given by  \citep{armstrong1967spectrum,herbert1974spectrum}
\begin{equation} 
\label{eq:voigt}
    f_{\tilde{\nu}_{fi},\sigma_{\rm G},\gamma_{\rm L}}^{V}(\tilde{\nu})
    =f_{\tilde{\nu}_{fi},\sigma_{\rm G}}^G(\tilde{\nu})\otimes f_{\tilde{\nu}_{fi},\gamma_{L}}^{\rm L}(\tilde{\nu})
    =\int_{-\infty}^{+\infty}\mathrm{d}\tilde{\nu}'f^G(\tilde{\nu}-\tilde{\nu}',\sigma_{\rm G})\times f^{\rm L}(\tilde{\nu}'-\tilde{\nu}_{fi},\gamma_{\rm L})
    =\frac{y}{\pi^{3/2}\sqrt{2}\sigma_{\rm G}}\int_{-\infty}^{+\infty}\frac{e^{-t^{2}}}{\left(x-t\right)^{2}+y^{2}}\mathrm{d}t,
\end{equation}
where 
\begin{equation} 
\label{eq:xy}
    x=\frac{\tilde{\nu}-\tilde{\nu}_{fi}}{\sqrt{2}\sigma_{\rm G}} \quad \quad \textrm{and} \quad \quad y=\frac{\gamma_{\rm L}}{\sqrt{2}\sigma_{\rm G}}.
\end{equation}
Here $x$ gives the wavenumber scale in units of the Gaussian standard deviation, $y$ is the ratio of Lorentzian widths to the Gaussian standard deviation,  $\sigma_{\rm G}$ is the Gaussian standard deviation, and $\gamma_{\rm L}$ is HWHM of the Lorentzian profile. The central wavenumber or frequency, $\tilde{\nu}_{fi}$ is the energy difference of the transition \citep{schreier2018voigt}.

\textsc{PyExoCross} offers nine methods for calculating the Voigt profile with five of them as pseudo-Voigt profiles as described in the following.

\subsubsection{\textsc{SciPy} Voigt profile}
\label{sec:scipyVoigt}

Use of the Python open-source library \textsc{SciPy} \citep{2020SciPy-NMeth} gives the Voigt profile directly, see Listing~\ref{listing:scipyvoigt}. 
% \begin{listing}[!ht]
% \begin{minted}{python}
% from scipy.special import voigt_profile
% def scipy_Voigt_profile(dv, sigma, gamma):
%     return voigt_profile(dv, sigma, gamma)
% \end{minted}
% \caption{Sample of the Python code used to calculate Voigt Profile by SciPy package \citep{2020SciPy-NMeth}. $\mathrm{d}v=\tilde{\nu}-\tilde{\nu}_{fi}$ and $\tilde{\nu}_{fi}$ is the line centre. \texttt{sigma} is the standard deviation $\sigma_{\rm G}$ of the Gaussian profile. \texttt{gamma} is the half-width for pressure broadening $\gamma_{\rm L}$.}
% \label{listing:scipyvoigt}
% \end{listing}

\begin{python}[caption={Sample of the Python code used to calculate Voigt Profile by \textsc{SciPy} package \citep{2020SciPy-NMeth}. $\mathrm{d}v=\tilde{\nu}-\tilde{\nu}_{fi}$ and $\tilde{\nu}_{fi}$ is the line centre. \texttt{sigma} is the standard deviation $\sigma_{\rm G}$ of the Gaussian profile. \texttt{gamma} is the half-width for pressure broadening $\gamma_{\rm L}$.}, label={listing:scipyvoigt}]
from scipy.special import voigt_profile
def scipy_Voigt_profile(dv, sigma, gamma):
    return voigt_profile(dv, sigma, gamma)
\end{python}
% \end{lstlisting}

\texttt{scipy.special.voigt\_profile} uses  \texttt{scipy.signal.convolve} to compute the Voigt profile with the convolution of a one-dimensional (1D) Normal distribution (Gaussian profile) with the standard deviation $\sigma_{\rm G}$ and a 1D Cauchy distribution (Lorentzian profile) with HWHM $\gamma_{\rm L}$ \citep{2020SciPy-NMeth}, see Eq.~(\ref{eq:voigt}). The method of \texttt{scipy.signal.convolve} is given in Listing~\ref{listing:convolve}. 

% \begin{listing}[!ht]
% \begin{minted}{python}
% from scipy.signal import convolve
% convolved=dx*convolve(Lorentzian_profile, Gaussian_profile, mode="same")
% \end{minted}
% \caption{Method for computing Voigt profile by the convolution of the Gaussian and Lorentzian profile using the \texttt{SciPy} Python package as used by the \texttt{scipy.special.voigt\_function}.}
% \label{listing:convolve}
% \end{listing}

\begin{python}[caption={Method for computing Voigt profile by the convolution of the Gaussian and Lorentzian profile using the \textsc{SciPy} Python package as used by the \texttt{scipy.special.voigt\_function}.}, label={listing:convolve}]
from scipy.signal import convolve
convolved = dx*convolve(Lorentzian_profile, Gaussian_profile, mode="same")
\end{python}
% \end{lstlisting}

\subsubsection{\textsc{SciPy} wofz Voigt profile}
\label{sec:scipywofzVoigt}

The Python open-source library \textsc{SciPy} can be used to calculate the Voigt profile, see Listing~\ref{listing:scipywofzvoigt}. The \textsc{SciPy} function  \texttt{scipy.special.wofz} calculates the Voigt profile for $\sigma_{\rm G}$ and $\gamma_{\rm L}$ as given  by 
\begin{equation} 
\label{eq:voigtwofz}
    f_{\tilde{\nu}_{fi},\sigma_{\rm G},\gamma_{\rm L}}^{V}(\tilde{\nu})
    =\frac{1}{\sqrt{2\pi}\sigma_{\rm G}}K(x,y),
\end{equation}
where $x$ and $y$ are given in Eq.~(\ref{eq:xy}), $K(x,y)$ is the Voigt function
\begin{equation} 
\label{eq:k}
    K(x,y)=\frac{y}{\pi}\int_{-\infty}^{+\infty}\frac{e^{-t^{2}}}{\left(x-t\right)^{2}+y^{2}}\mathrm{d}t
\end{equation}
satisfying $\int K \mathrm{d}x=\sqrt{\pi}$.

$K(x,y)$ comprises the real part of the complex error function (Faddeeva function) $\omega(z)$ \citep{Faddeeva1961TablesOV}:
\begin{equation}
\label{eq:w}
    \omega(z)\equiv K(x,y)+iL(x,y)=\frac{i}{\pi}\int_{-\infty}^{+\infty}\frac{e^{-t^{2}}}{z-t}\mathrm{d}t.
\end{equation}
where 
\begin{equation} 
\label{eq:z}
    z=x+iy=\frac{\left(v-\tilde{\nu}_{fi}+i\gamma_{\rm L}\right)}{\sqrt{2}\sigma_{\rm G}}
\end{equation}
and
\begin{equation} 
\label{eq:wofz voigt}
    f_{\tilde{\nu}_{fi},\sigma_{\rm G},\gamma_{\rm L}}^{V}(\tilde{\nu})=\frac{{\bf Re}\left[\omega\left(z\right)\right]}{\sigma_{\rm G}\sqrt{2\pi}}. 
\end{equation}
In Eq.~(\ref{eq:z}),  in the limiting cases of $\sigma_{\rm G}=0$ and $\gamma_{\rm L}=0$, Eq.~(\ref{eq:wofz voigt}) is simplified  to the Lorentzian profile and Gaussian profile, respectively.

% \begin{listing}[!ht]
% \begin{minted}{python}
% import numpy as np
% from scipy.special import wofz
% def scipy_wofz_Voigt_profile(dv, sigma, gamma):
%     z = (dv + 1j * gamma) / sigma / np.sqrt(2)
%     voigt = (np.real(wofz(z)) / sigma / np.sqrt(2 * np.pi))
%     return voigt
% \end{minted}
% \caption{A sample of the Python code used to calculate the Voigt profile \citep{2020SciPy-NMeth,hill2020learning} with the real part of the Faddeeva function. \texttt{wofz} in \texttt{scipy.special.wofz} can be computed with error function $\textrm{erf}(-iz)$ given by \citet{pierluissi1977fast}: $\omega(z)=e^{-z^2}(1-\textrm{erf}(-iz))$.}
% \label{listing:scipywofzvoigt}
% \end{listing}

\begin{python}[caption={A sample of the Python code used to calculate the Voigt profile \citep{hill2020learning,2020SciPy-NMeth} with the real part of the Faddeeva function. \texttt{wofz} in \texttt{scipy.special.wofz} can be computed with error function $\textrm{erf}(-iz)$ given by \citet{pierluissi1977fast}: $\omega(z)=e^{-z^2}(1-\textrm{erf}(-iz))$.}, label={listing:scipywofzvoigt}]
import numpy as np
from scipy.special import wofz
def scipy_wofz_Voigt_profile(dv, sigma, gamma):
    z = (dv + 1j * gamma) / sigma / np.sqrt(2)
    voigt = (np.real(wofz(z)) / sigma / np.sqrt(2 * np.pi))
    return voigt
\end{python}
% \end{lstlisting}

\subsubsection{Huml\'i\v{c}ek Voigt profile}
\label{sec:humlicek}

Huml\'i\v{c}ek’s algorithm \citep{humlivcek1979efficient} for approximating the Voigt profile function is used extensively. Huml\'i\v{c}ek's Voigt profile gives a method for computing the complex probability function $\omega(z)$ where $z=x+iy$ (see Eqs.~(\ref{eq:voigt}, \ref{eq:w}, and  \ref{eq:z})). $x=\frac{\left(\tilde{\nu}-\tilde{\nu}_{fi}\right)\sqrt{\ln{2}}}{\alpha_{\rm D}}$ is the wavenumber scale in units of the Doppler HWHM $\alpha_{\rm D}$ and $y=\frac{\gamma_{\rm L}\sqrt{\ln{2}}}{\alpha_{\rm D}}$ is the ratio of the Lorentzian HWHM $\gamma_{\rm L}$ to the Doppler HWHM $\alpha_{\rm D}$.
A rational approximation is used in Huml\'i\v{c}ek’s w4 algorithm, in which the imaginary part on the real axis is minimized in order to minimize its relative error \citep{humlivcek1982optimized}. The following procedure can evaluate both the real and imaginary parts of $\omega(z)$ with a high degree of relative accuracy \citep{humlivcek1979efficient,humlivcek1982optimized,kuntz1997new}. 
\begin{equation}
    \omega(z)= 
    \begin{cases}
        \frac{t}{(0.5+t^2)\sqrt{\pi}}, & \text{ if } s \ge 15, \\ \\
        \frac{t\left(1.4104739589+\frac{u}{\sqrt{\pi}}\right)}{0.75+u(3+u)}, & \text{ if } 5.5 \le s < 15, \\ \\
        \frac{16.4955+t(20.20933+t(11.96482+t(3.778987+0.5642236t)))}{16.4955+t(38.82363+t(39.27121+t(21.69274+t(6.699398+t)))}, & \text{ if } s < 5.5 \text{ and } y \ge 0.195\left|x\right|-0.176, \\ \\
        e^u-\frac{t(36183.31-u(3321.99-u(1540.787-u(219.031-u(35.7668-u(1.320522-u0.56419))))))}{32066.6-u(24322.8-u(9022.23-u(2186.18-u(364.219-u(61.5704-u(1.84144-u))))))}, & \text{ if } s < 5.5 \text{ and } y < 0.195\left|x\right|-0.176,
    \end{cases}
\end{equation}
where complex function $t=y-xi$, $s=\left|x\right|+y$, and $u=t^2$. \citet{17Zaghol} and \citet{78HuArWr} defined a simpler, more efficient, and relatively accurate approximation for the evaluation of the Faddeyeva function with six regions of the proposed partitioning.

The Huml\'i\v{c}ek Voigt line profile is then given by \citep{humlivcek1979efficient,humlivcek1982optimized}
\begin{equation}
    f_{\tilde{\nu}_{fi},\alpha_{\rm D},\gamma_{\rm L}}^{V}(\tilde{\nu}) = \frac{{\bf Re}\left[\omega(z)\right]\sqrt{\ln{2}}}{\pi\alpha_{\rm D}}.
\end{equation}

\subsubsection{Pseudo-Voigt profiles}

\textsc{PyExoCross} provides five pseudo-Voigt line profiles. They are not the recommended algorithms and we do not consider pseudo-Voigt profiles in comparison Section~\ref{sec:compare profile} because they have lower accuracy and speed of run time than other Voigt profiles.

\textbf{Thompson pseudo-Voigt profile}
\label{sec:pseudo}

The pseudo-Voigt function $V(\mathrm{d}v,\gamma_{\rm V})$ is frequently used for calculating experimental spectral line shapes \citep{ida2000extended}. As an approximation to the Voigt profile $V(x)$, the pseudo-Voigt profile combines a Gaussian curve $G(x)$ and a Lorentzian curve $L(x)$ linearly, rather than by convolution \citep{ida2000extended}. The following equation provides the mathematical definition of the normalized pseudo-Voigt profile $V(\mathrm{d}v,\gamma_{\rm V})$ \citep{thompson1987rietveld,sanchez1997use}: 
\begin{equation} 
\label{eq:V}
    V(\mathrm{d}v,\gamma_{\rm V})=\eta L(\mathrm{d}v,\alpha_{\rm D})+(1-\eta) G(\mathrm{d}v,\gamma_{\rm L}) \quad \textrm{with}\quad 0<\eta <1,
\end{equation}
where $\mathrm{d}v=\tilde{\nu}-\tilde{\nu}_{fi}$; $\alpha_{\rm D}$ and $\gamma_{\rm L}$ are the HWHM of the Gaussian and Lorentzian profiles. 

The accuracy of the pseudo-Voigt profile depends on several possible choices for calculating the parameter $\eta$ and HWHM of the Voigt profile $\gamma_{\rm V}$.  For the Thompson pseudo-Voigt profile  $\eta$ is given in Eq.~(\ref{eq:etakielkopf}) \citep{thompson1987rietveld}:
\begin{equation} 
\label{eq:etakielkopf}
    \eta \approx 1.36603(\gamma_{\rm L}/\gamma_{\rm V})-0.47719(\gamma_{\rm L}/\gamma_{\rm V})^2+0.11116(\gamma_{\rm L}/\gamma_{\rm V})^3.
\end{equation}
From the  associated   Gaussian and Lorentzian HWHMs, the value of $\gamma_{\rm V}$ of the Thompson pseudo-Voigt profile  can be determined using \citep{thompson1987rietveld}
\begin{equation} 
\label{eq:hv}
\gamma_{\rm V}=\left[\alpha_{\rm D}^5+2.69269\alpha_{\rm D}^4\gamma_{\rm L}+2.42843\alpha_{\rm D}^3\gamma_{\rm L}^2+4.47163\alpha_{\rm D}^2\gamma_{\rm L}^3+0.07842\alpha_{\rm D}\gamma_{\rm L}^4+\gamma_{\rm L}^5\right]^{1/5}.
\end{equation}

\textbf{Kielkopf pseudo-Voigt profile}
\label{sec:pseudokielkopf}

\citet{kielkopf1973new} proposed a weighted-sum method to 
 approximate the value of $\gamma_{\rm V}$ of the pseudo-Voigt profile via Eqs.~(\ref{eq:V} and \ref{eq:etakielkopf}) as given by (see also \citet{schreier2011optimized})
\begin{equation} 
\label{eq:hVKielkopf}
    \gamma_{\rm V}\approx \frac{1}{4}\left[(1+0.099\ln{2})\gamma_{\rm L}+\sqrt{(1-0.099\ln{2})^2\gamma_{\rm L}^2+4\ln{2}\alpha_{\rm D}^2}\right] \approx 0.5346\gamma_{\rm L}+\sqrt{0.2166f\gamma_{\rm L}^2+\alpha_{\rm D}^2}. 
\end{equation}

\textbf{Olivero pseudo-Voigt profile}
\label{sec:pseudoOlivero}

\citet{olivero1977empirical} provided another alternative to Eqs.~(\ref{eq:V} and \ref{eq:etakielkopf}) with the value of HWHM  $\gamma_{\rm V}$ approximated by
\begin{equation}
\label{eq:hvOlivero}
    \gamma_{\rm V}\approx \left[1-0.18121(1-d^2)-(0.023665e^{0.6d}+0.00418e^{-1.9d})\sin{(\pi d)}\right](\alpha_{\rm D}+\gamma_{\rm L}),
\end{equation}
where $d$ is a dimensionless parameter related to $\alpha_{\rm D}$ and $\gamma_{\rm L}$  \citep{olivero1977empirical} as
\begin{equation}
\label{eq:d}
    d=\frac{\gamma_{\rm L}-\alpha_{\rm D}}{\gamma_{\rm L}+\alpha_{\rm D}}.
\end{equation}

\textbf{Liu-Lin pseudo-Voigt profile}
\label{sec:pseudoliulin}

\citet{liu2001simple} used $d$ related to the Gaussian and Lorentzian half widths ($\alpha_{\rm D}$ and $\gamma_{\rm L}$)  defined in Eq.~(\ref{eq:d}) to provide a new conversion of $\eta$ from \citet{kielkopf1973new}, see Eq.~(\ref{eq:etakielkopf}). The half-width of the Voigt profile is the same as Eq.~(\ref{eq:hvOlivero}). The Liu-Lin pseudo-Voigt profile is calculated from Eqs.~(\ref{eq:V}, \ref{eq:hvOlivero}, \ref{eq:d}, and \ref{etaliulin}) using
\begin{equation}
\label{etaliulin}
    \eta \approx 0.68188+0.61293d-0.18384d^2-0.11568d^3.
\end{equation}

\textbf{Rocco pseudo-Voigt profile}
\label{sec:pseudorocco}

\citet{di2012voigt} provided a theoretically justified method for calculating $\gamma_{\rm V}$ and $\eta$ in the pseudo-Voigt methodology and defined them as follows:
\begin{equation}
\label{eq:etaRocco}
\begin{split}
     & \gamma_{\rm V} = \omega_G b_{1/2}(y), \\
     & \eta \approx \frac{\sqrt{\pi}V_y(0)\gamma_{\rm V}-\sqrt{\ln{2}}}{\sqrt{\pi}V_y(0)\gamma_{\rm V}(1-\sqrt{\pi\ln{2}})}, \\
\end{split}
\end{equation}
where 
\begin{equation}
 \omega_G = \frac{\alpha_{\rm D}}{\sqrt{\ln{2}}},
\end{equation}
$V_y(0)$ is the Voigt profile in Eq.~(\ref{eq:voigt}) at $x=0$ 
\begin{equation}
V_y(0)=\frac{e^{y^2}(1-\textrm{erf}(y))}{\sqrt{\pi}\omega_G}. 
\end{equation}
Here,
\begin{equation}
y = \frac{\gamma_{\rm L}}{\omega_G}
\quad \rm{and} \quad 
 b_{1/2}(y) = y+\sqrt{\ln{2}}e^{-0.6055y+0.0718y^2-0.0049y^3+0.000136y^4}. 
\end{equation}

%and $y=\frac{\gamma_{\rm L}}{\sqrt{2}\sigma_{\rm G}}=\frac{\gamma_{\rm L}%\sqrt{\ln{2}}}{\alpha_{\rm D}}$ given by:

%\textsc{PyExoCross} uses Eqs.~(\ref{eq:V}, \ref{eq:hvRocco}, \ref{eq:etaRocco}) to calculate Rocco pseudo-Voigt profile.

\subsection{Binned line profiles}

All the above methods use a sampling algorithm to evaluate the Voigt profile on a set of frequency wavenumbers $\tilde{\nu}_i$. The disadvantage of these methods is that they can lead to undersampling of the opacity (absorption or emission) at low sampling resolution, see e.g. \citet{jt542}. In order to avoid this problem and guarantee that  the full  opacity is preserved, the following binned approach can be used by defining the analytically integrated (averaged) cross-sections within a bin $[\tilde{\nu}_k-\Delta{\tilde{\nu}}/2,\tilde{\nu}_k+\Delta{\tilde{\nu}}/2]$ as a  sum over the contributions from each line \citep{jt542} as given by  
\begin{equation}
\label{eq:binned}
\begin{split}
    &\bar{\sigma}_{k}^{{\rm ab},fi}=\frac{I_{fi}}{\Delta\tilde{\nu}}\int_{\tilde{\nu}_k-\Delta\tilde{\nu}/2}^{\tilde{\nu}_k+\Delta\tilde{\nu}/2}f_{\tilde{\nu}_{fi}}(\tilde{\nu})\mathrm{d}\tilde{\nu}, \\
    &\bar{\sigma}_{k}^{{\rm em},if}=\frac{\epsilon_{if}}{\Delta\tilde{\nu}}\int_{\tilde{\nu}_k-\Delta\tilde{\nu}/2}^{\tilde{\nu}_k+\Delta\tilde{\nu}/2}f_{\tilde{\nu}_{if}}(\tilde{\nu})\mathrm{d}\tilde{\nu},
\end{split}
\end{equation}
where $\tilde{\nu}_{fi}$ is the line centre, $I_{fi}$ and $\epsilon_{if}$ are the line intensities (absorption and  emission) from Eqs.~(\ref{eq:absintensity} and \ref{eq:emiintensity}).

The related functions for the binned cross-sections for the Doppler, Gaussian, Lorentzian and Voigt profiles are specified below. 

\subsubsection{Binned Doppler and Gaussian profile}
\label{sec:binned doppler gaussian}

Both the Doppler and the Gaussian profiles use the Gaussian function. Taking advantage of the fact that the integral of the Gaussian function has an analytic solution, the error function erf can be used:
\begin{equation}
\label{eq:erf}
    \int e^{-x^2}\mathrm{d}x=\frac{\sqrt{\pi}}{2}\textrm{erf}(x). 
\end{equation}
Therefore, we apply the analytical integral to the Gaussian profile \citep{jt542}:
\begin{equation}
\label{eq:binnedg}
    \bar{\sigma}_{k}^{fi}=\frac{I_{fi}}{\Delta\tilde{\nu}}\int_{\tilde{\nu}_k-\Delta\tilde{\nu}/2}^{\tilde{\nu}_k+\Delta\tilde{\nu}/2}f_{\tilde{\nu}_{fi}}^G(\tilde{\nu})\mathrm{d}\tilde{\nu} =\frac{I_{fi}}{2\Delta\tilde{\nu}}\left[\textrm{erf}\left(x_{k,fi}^+\right)-\textrm{erf}\left(x_{k,fi}^-\right)\right],
\end{equation}
where $x_{k,fi}^{\pm}$ are the scaled limits of the frequency bin $k$ centred on $\tilde{\nu}_k$ related to the line centre $\tilde{\nu}_{fi}$ given by \citet{jt542}:
\begin{equation}
\label{eq:erfx}
    x_{k,fi}^{\pm}=\frac{\sqrt{\ln{2}}}{\alpha_{\rm D}}\left[\tilde{\nu}_k-\tilde{\nu}_{fi}\pm\frac{\Delta\tilde{\nu}}{2}\right].
\end{equation}

\subsubsection{Binned Lorentzian profile}
\label{sec:binned lorentzian}

The integral of the Lorentzian function has an analytical solution in the format
\begin{equation}
\label{eq:arctan}
    \int \frac{\textrm{d}x}{x^2+\gamma^2}=\frac{1}{\gamma}\arctan\left(\frac{x}{\gamma}\right) +C.
\end{equation}

If we calculate profiles in the range of the wings which have large distances from the line centre, then the normalization correction factor is unity. To solve the far wings in calculating profiles and save an excessive amount of computing time, we truncate the profiles at some distance $\Delta\tilde{\nu}/2$ from the line centre \citep{07ShBuxx.VO}. Therefore, \textsc{PyExoCross} provides two methods for calculating binned Lorentzian and Voigt profiles with or without a cutoff filter for users.  

The same approach of an analytical binned integral can be applied to the Lorentzian profile  $f_{\tilde{\nu}_{fi}}^{\rm L}(\tilde{\nu})$ as follows \citep{jt708}:
\begin{equation}
\label{eq:binnedlnob}
    \bar{\sigma}_{k}^{fi}=\frac{I_{fi}}{\Delta\tilde{\nu}}\int_{\tilde{\nu}_k-\Delta\tilde{\nu}/2}^{\tilde{\nu}_k+\Delta\tilde{\nu}/2}f_{\tilde{\nu}_{fi}}^{\rm L}(\tilde{\nu})\mathrm{d}\tilde{\nu}
 =\frac{I_{fi}}{\pi\Delta\tilde{\nu}}\left[\arctan\left(y_{k,fi}^+\right)-\arctan\left(y_{k,fi}^-\right)\right],
\end{equation}
where $\bar{\sigma}_{k}^{fi}$ is an averaged cross-section for the  bin $k$ centred at $\tilde{\nu}_k$ over the interval $\Delta \tilde{\nu}$, $\tilde{\nu}_{fi}$ is the centre of the line, and $y_{k,fi}^{\pm}$  are  given by
\begin{equation}
\label{eq:yk}
    y_{k,fi}^{\pm}=\frac{\tilde{\nu}_k-\tilde{\nu}_{fi}\pm\Delta\tilde{\nu}/2}{\gamma_{\rm L}}.
\end{equation}

Because the line profiles are usually cut off at a distance $\Delta\tilde{\nu}/2$ from the line centre, it is common to preserve the total strength of the line  by re-normalizing its  area to unity  \citep{07ShBuxx.VO}.  Here we multiply Eq.~\eqref{eq:binnedlnob} by  the  correction factor $b_{\rm corr}\ge 1$ given by
\begin{equation}
\label{eq:bnorm}
    b_{\rm corr}=\frac{\pi}{{\arctan\left(y_{\textrm{end},fi}\right)-\arctan\left(y_{\textrm{start},fi}\right)}},
\end{equation}
where $y_{\textrm{start},fi}$ and $y_{\textrm{end},fi}$ are defined as
\begin{equation}
\label{eq:ystartend}
y_{\textrm{start},fi}=\frac{\tilde{\nu}_\textrm{start}-\tilde{\nu}_{fi}}{\gamma_{\rm L}} \quad {\rm and} \quad y_{\textrm{end},fi}=\frac{\tilde{\nu}_\textrm{end}-\tilde{\nu}_{fi}}{\gamma_{\rm L}}.
\end{equation}

\subsubsection{Binned Voigt profile}
\label{sec:binned voigt}

Gauss-Hermite quadrature is a form of the Gaussian quadrature used in numerical analysis to approximate the value of integrals \citep{abramowitz1972handbook}:
\begin{equation}
\label{eq:g-h}
    \int_{-\infty}^{+\infty}e^{-t^2}f(t)\approx \sum_{m=1}^n\omega_mf(t_m),
\end{equation}
where $n$ is the number of the used sample points and $t_m$ are the roots of the Hermite polynomial $H_n(t)$ $(m=1,2,...,n)$ (physicists' version) and the Gauss-Hermite quadrature weights $\omega_m$ are given by \citet{abramowitz1972handbook}:
\begin{equation}
\label{eq:wm}
    \omega_m=\frac{2^{n-1}n!\sqrt{\pi}}{n^2[H_{n-1}(t_m)]^2}.
\end{equation}
Combined with the Gaussian and Lorentzian profiles, we use Huml\'i\v{c}ek’s algorithm \citep{humlivcek1979efficient} as the Gauss-Hermite quadratures to create a formulation of a similar integral method for the Voigt profile \citep{jt708} with HWHM of Doppler (and Gaussian) profile $\alpha_{\rm D}$ and the HWHM of Lorentzian profile $\gamma_{\rm L}$ (see Eq.~(\ref{eq:voigt})):
\begin{equation}
\label{eq:voigtalpha}
    f_{\tilde{\nu}_{fi},\alpha_{\rm D},\gamma_{\rm L}}^{V}(\tilde{\nu})
    =\frac{y\ln{2}}{\pi^{3/2}\alpha_{\rm D}^2}\int_{-\infty}^{+\infty}\frac{e^{-t^{2}}}{\left(x-t\right)^{2}+y^{2}}\mathrm{d}t
    =\frac{y\ln{2}}{\pi^{3/2}\alpha_{\rm D}^2}\sum_{m=1}^{N_{\textrm{G-H}}}\frac{\omega_m^{\textrm{G-H}}}{(x-t_m)^2+y^2},
\end{equation}
where $t_m$ are the Gauss–Hermite quadrature points (roots) and $\omega_m^{\textrm{G-H}}$ are the Gauss–Hermite quadrature weights. Using Eq.~(\ref{eq:binnedlnob}) to form the area-conserved integrals, for the binned Voigt profile we obtain \citep{jt708}
\begin{equation}
\label{eq:binnedvoigtnob}
    \bar{\sigma}_{k}^{fi}=\frac{I_{fi}}{\pi^{3/2}\Delta\tilde{\nu}}\sum_{m=1}^{N_{\textrm{G-H}}}\omega_m^{\textrm{G-H}}\left[\arctan\left(z_{k,fi,m}^+\right)-\arctan\left(z_{k,fi,m}^-\right)\right],
\end{equation}
where the scaled limits $z_{k,fi,m}^{\pm}$ of the bin $k$ centred at the grid point $\tilde{\nu}_k$ is given by
$$
 z_{k,fi,m}^{\pm}=\frac{\tilde{\nu}_k-\tilde{\nu}_{fi}-t_m\, \sigma_{\rm G}\pm\Delta\tilde{\nu}/2}{\gamma_{\rm L}}.
$$
We commonly take $N_{\textrm{G-H}}=20$ as the number of Gauss-Hermite points used.

This expression can be further corrected for the missing area due to the wing's cutoff similarly to the Lorentzian example above. 
Here, we multiply $\bar{\sigma}_{k}^{fi}$ in Eq.~(\ref{eq:binnedvoigtnob}) with the normalization correction factor $b_{\rm corr}$ given by
\begin{equation}
\label{eq:bnorm:Voigt}
    b_{\rm corr}=\frac{\pi}{{\arctan\left(z_{\textrm{end},fi,m}\right)-\arctan\left(z_{\textrm{start},fi,m}\right)}},
\end{equation}
where 
\begin{equation}
\label{eq:zstartend}
     z_{\textrm{start},fi,m}=\frac{\tilde{\nu}_\textrm{start}-\tilde{\nu}_{fi}-t_m\, \sigma_{\rm G}}{\gamma_{\rm L}} \quad  {\rm and}  \quad 
     z_{\textrm{end},fi,m}=\frac{\tilde{\nu}_\textrm{end}-\tilde{\nu}_{fi}-t_m\,\sigma_{\rm G}}{\gamma_{\rm L}}.
\end{equation}

\subsection{Line profile comparisons}
\label{sec:compare profile}

\subsubsection{Comparison of the sampling and binned line profile methods}

When the cross-sections resolution (i.e. the number of points per wavenumber interval) is low, the result of cross-sections calculations becomes very sensitive to the way how profiles are evaluated, using sampling or binning.  While the sampling line profile methods  are constructed to accurately reproduce  the cross-sections at the given sampling points, the binned methods guarantee the integrated cross-section within a given bin, regardless of the number of grid points or integration intervals. Accordingly, the binned methods are especially advantageous when the conservation of the total flux, absorption, or emission coefficients, is required. The sampling method are preferred  when the accuracy of the cross-sections at the given frequencies is critical. 

\begin{figure}
\centering  
%\subfigure[Doppler profile]{    
\includegraphics[scale=0.25]{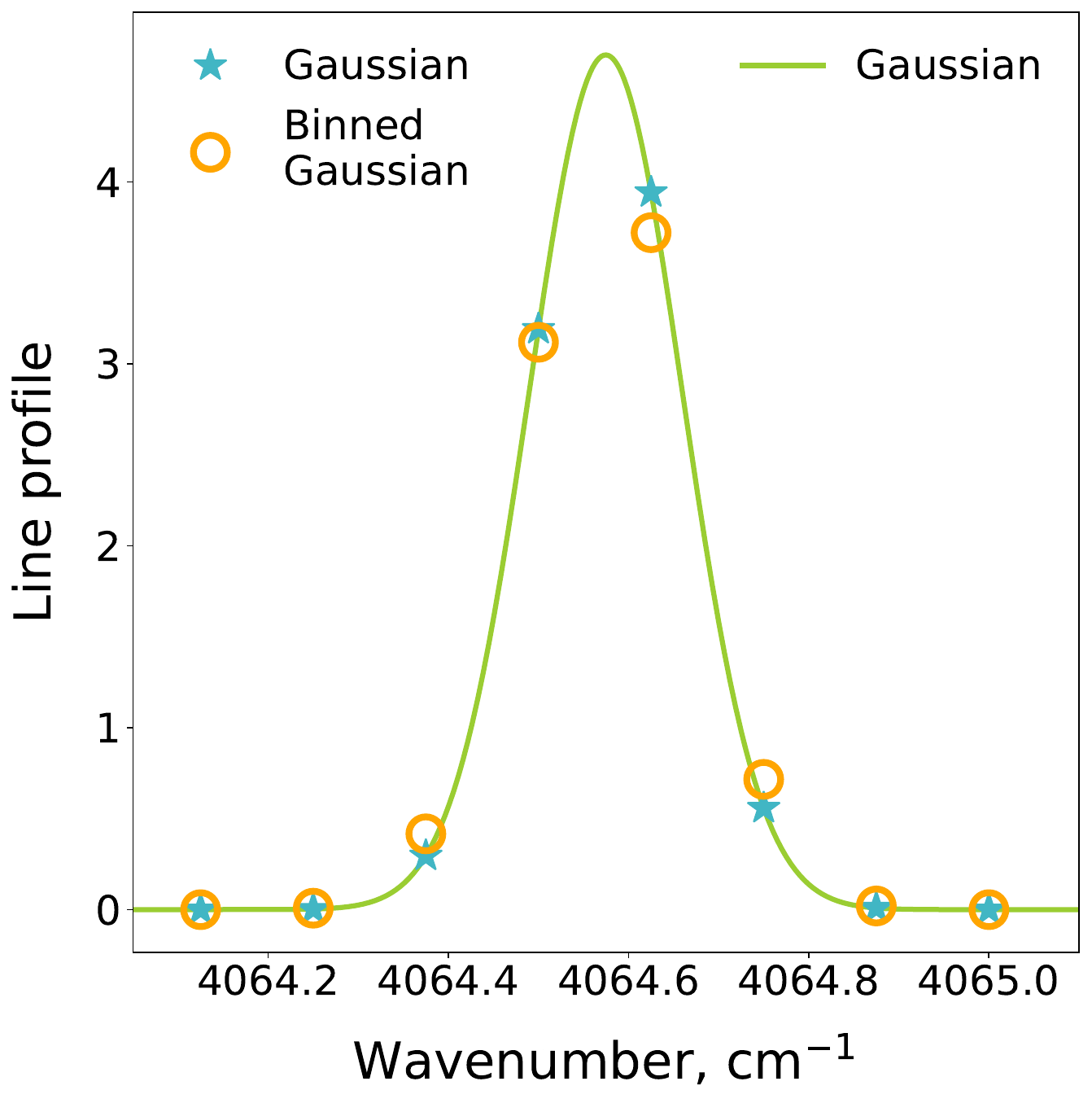} 
%}
%\subfigure[Lorentzian profile]{  
\includegraphics[scale=0.25]{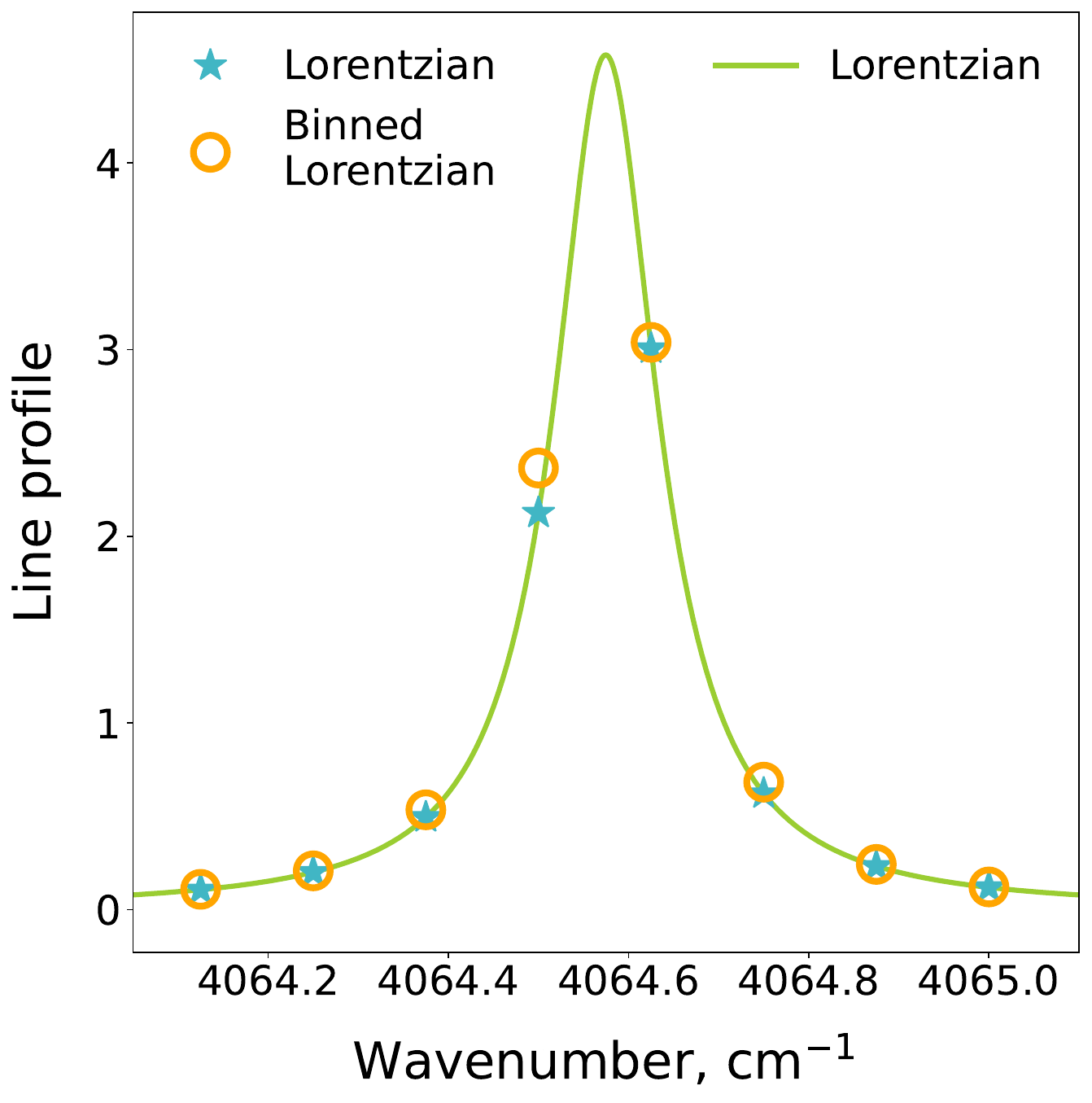}
%}
%\subfigure[Voigt profile]{  
\includegraphics[scale=0.25]{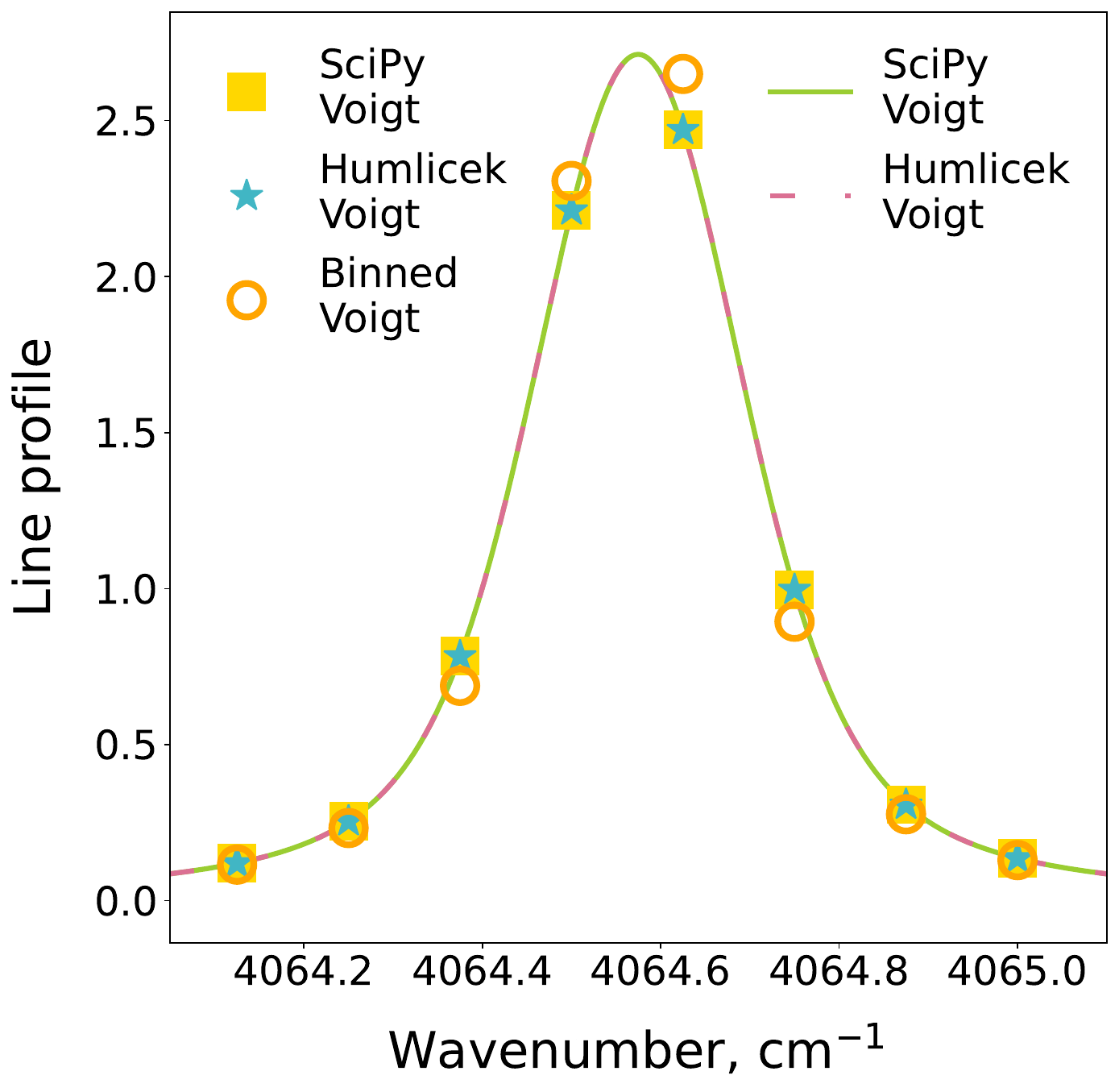}
%}
\caption{Comparison of the sampling and binned methods with different line profiles for a single normalized transition with a cutoff $=100$ cm$^{-1}$ and the bin size of 0.125~cm$^{-1}$ (8 points per 1 cm$^{-1}$) at the same temperature $T=300$ K and pressure $P=1$ bar. Left-hand panel: Doppler profile; middle panel: Lorentzian profile; and right-hand panel: Voigt profile. }    
\label{fig:profile plot}    
\end{figure}

Figure~\ref{fig:profile plot} gives examples of the Doppler, Lorentzian, and Voigt line profiles calculated using the sampling and binned methods for a single normalized transition in the wavenumber range [4064, 4065] cm$^{-1}$ at the temperature $T=300$ K and pressure $P=1$ bar using the bin size of 0.125~cm$^{-1}$ (8 points per cm$^{-1}$) with a cutoff of $100$ cm$^{-1}$. The lines show the correct line profiles modelled, while points represent the corresponding calculated values. For the sampling method, the profile lines go exactly through the calculated cross-section points, while for the binned method, the points are shifted away (up or down) from the curve in order to maintain the integrated cross-section (unity in this case) as follows:
\begin{equation}
    \int_{-\infty}^{\infty}f_{\tilde{v}_{fi}}(\tilde{v})dv \equiv  \sum_i f_{\tilde{v}_{fi}}(\tilde{v_i}) \Delta \nu, 
    \label{eq:lineprofile1}
\end{equation}
where the rectangular integration rule is used. 

%\subsubsection{Comparison  of line profile errors}

%We use bin size $\Delta \tilde v=0.1$ cm$^{-1}$ to calculate $\sum f_{\tilde{v}_{fi}}(\tilde{v})\Delta \tilde v$. A smaller line profile area error occurs when the sum approaches 1. 

The profile normalization error given as the difference of the integrated area from unity of a (normalized) line profile is given by definition   
\begin{equation}
\label{e:eps:area}
    \epsilon_{\rm area} =1-\sum f_{\tilde{\nu}_{fi}}(\tilde{\nu})\Delta \tilde \nu.
\end{equation}
For the binned methods, $\epsilon_{\rm area}$ is zero by construction, while for the sampling methods, it depends on the number of sampling points. Choosing an adequate number of sampling points is critical in radiative transfer applications \citep{jt801}. 
Undersampling can lead to underestimation of opacity, while too many points can make the calculations too slow. The profile sampling error can be defined a root-mean-squares error (RMSE) deviation of the calculated from the exact profiles at the sampling points as given by 
\begin{equation}
\label{e:eps:sampl}
    \epsilon_{\rm sampling} = \sqrt{\frac{1}{N} \sum_{i=1}^N \left( f(\tilde{\nu_i}) - f_{\rm calc}(\tilde{\nu_i}) \right)^2 }, 
\end{equation}
where $N$ is the number of sampling points. The sampling error $\epsilon_{\rm sampling}$ should be  zero for the sampling methods by construction (within the numerical accuracy). 

\begin{figure}
\centering  
{\includegraphics[scale=0.34]{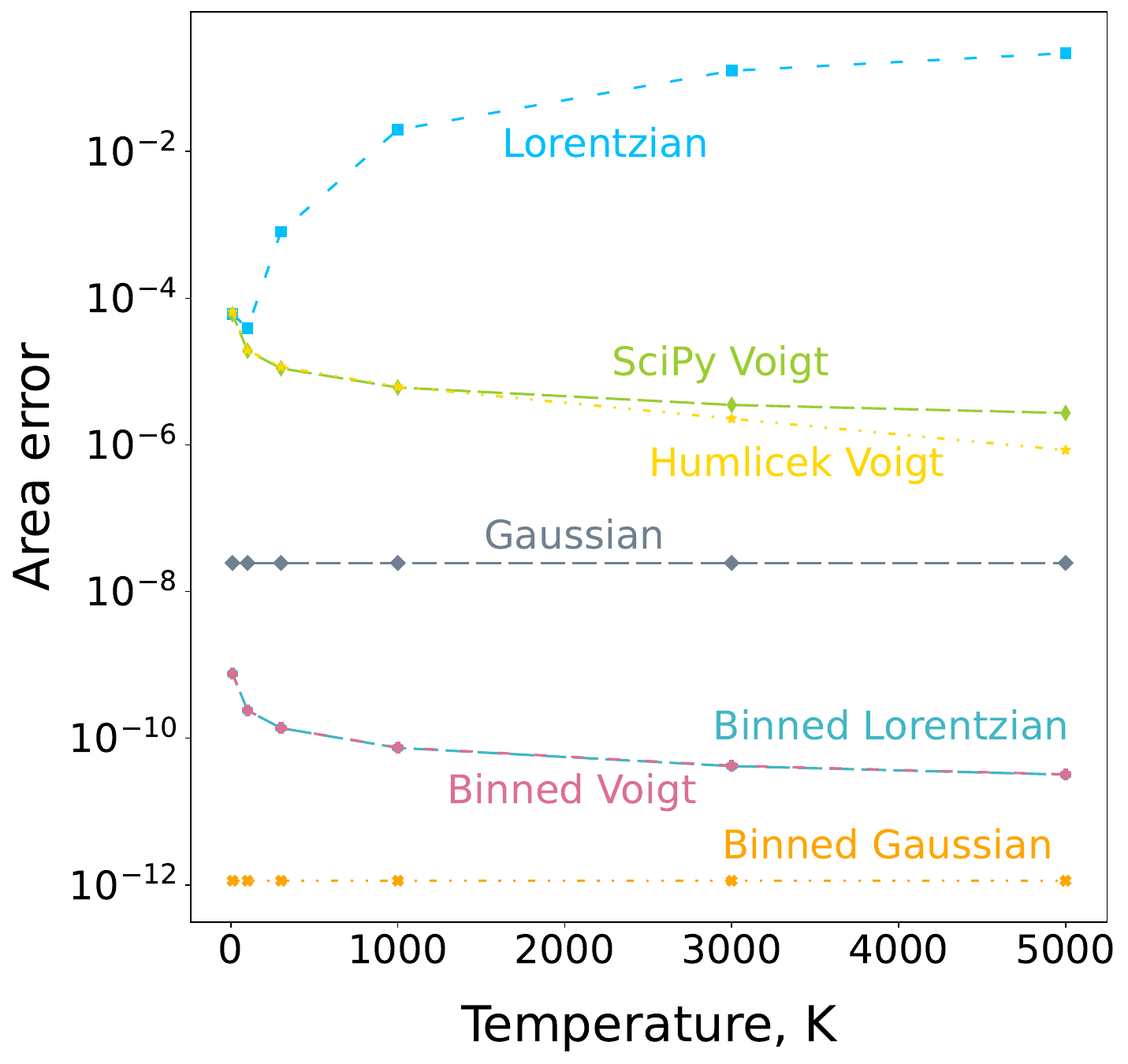}}
{\includegraphics[scale=0.34]{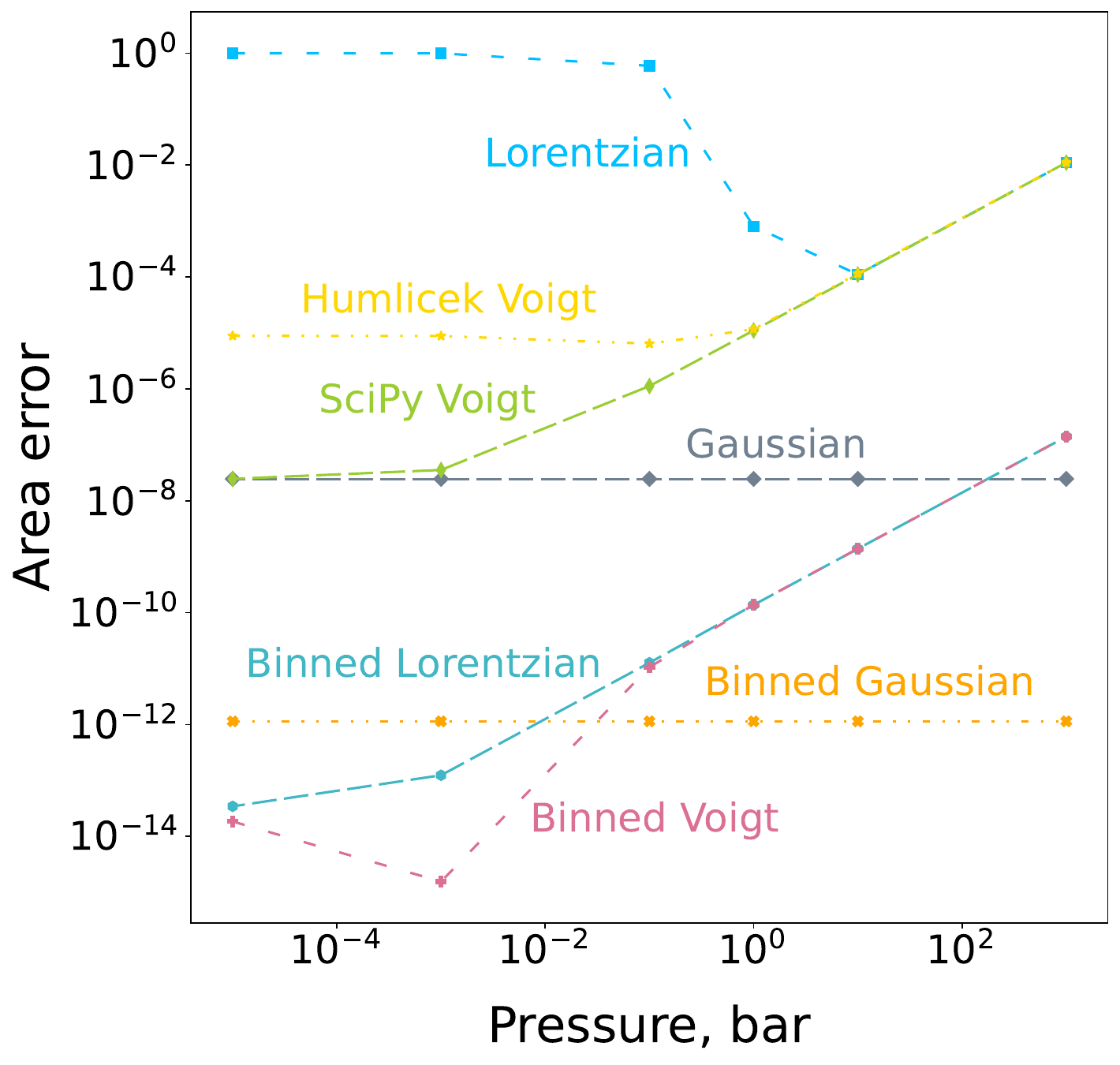}}
{\includegraphics[scale=0.34]{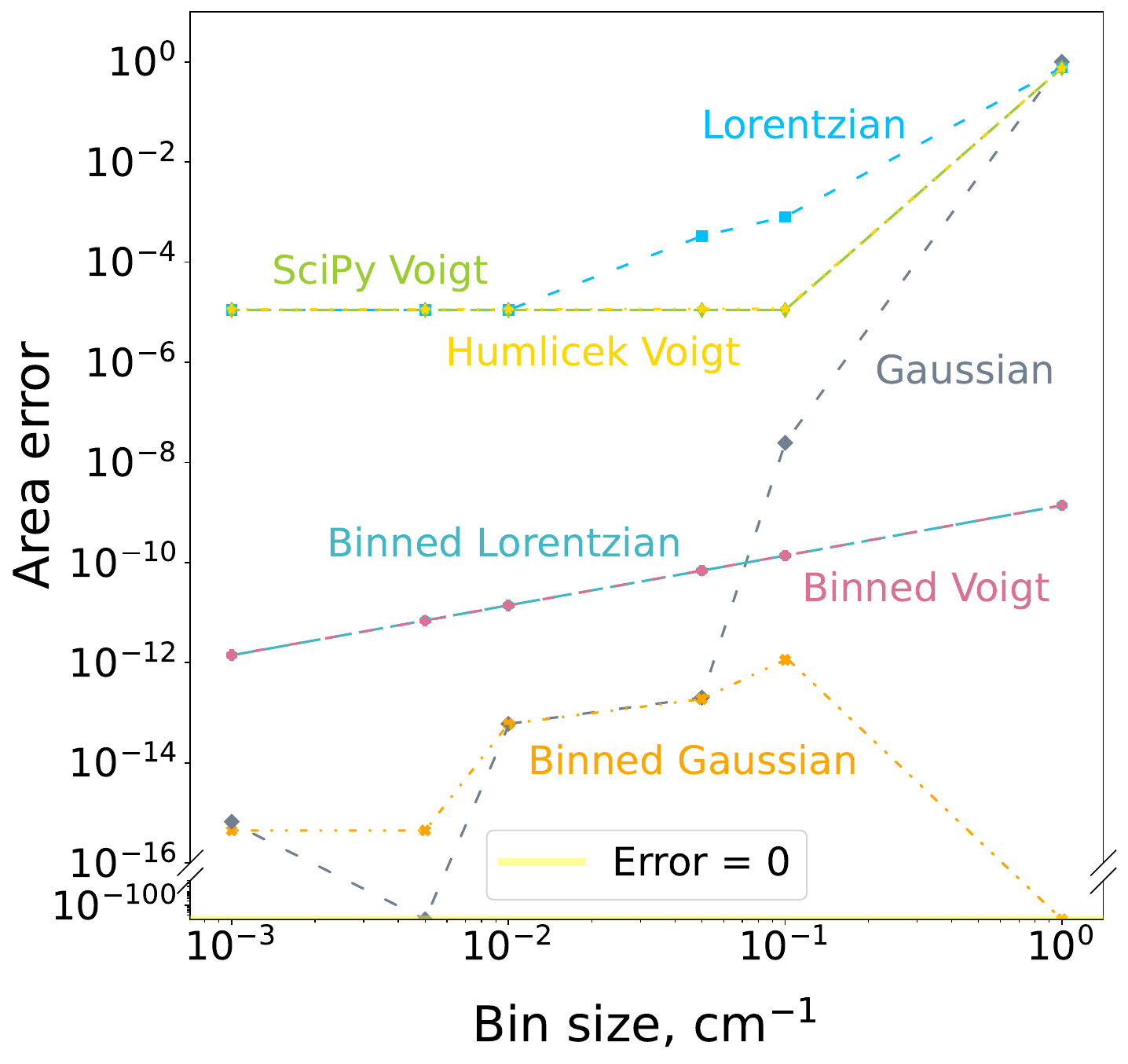}}
{\includegraphics[scale=0.34]{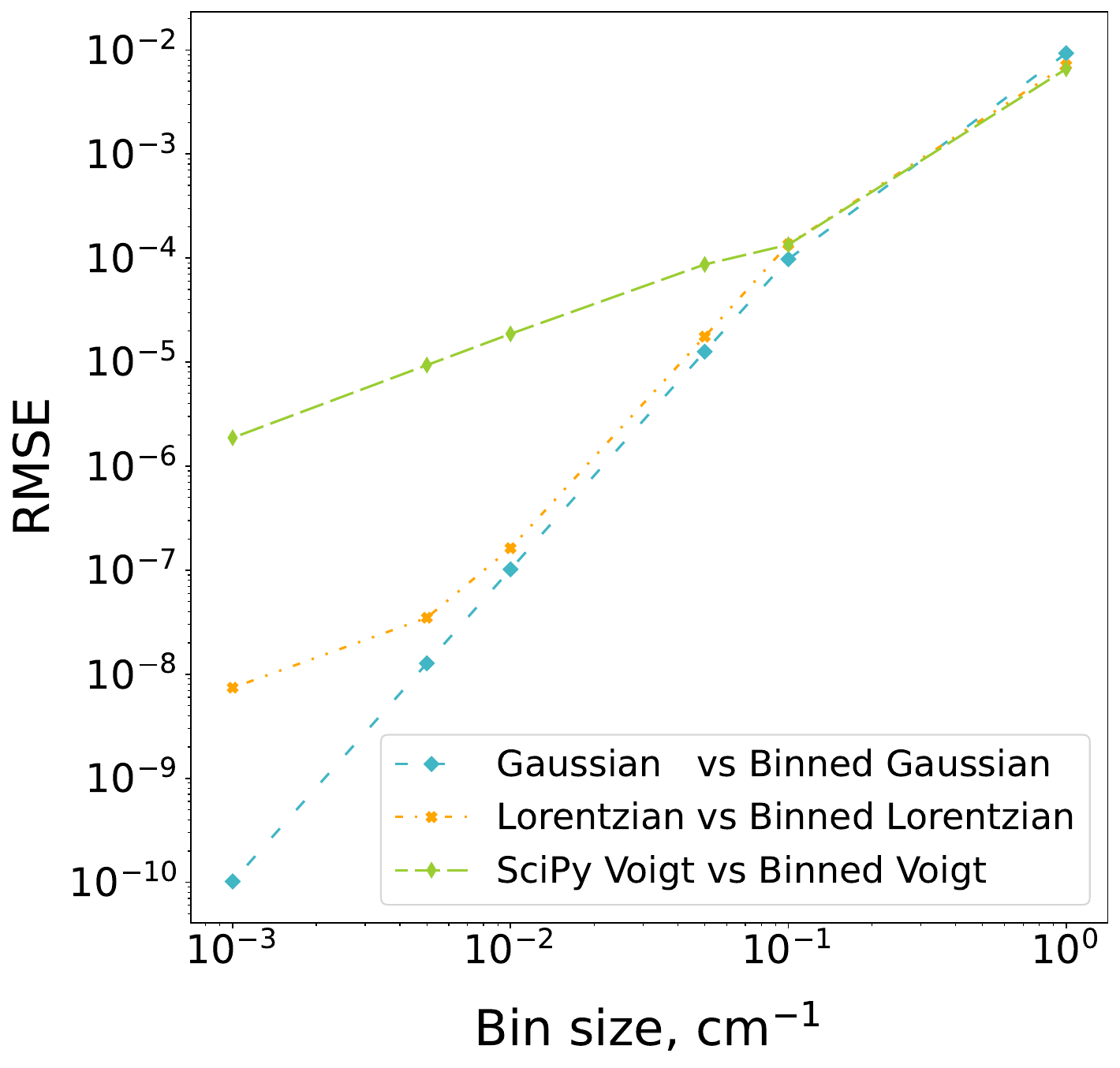}}
\caption{Line profile  errors $\epsilon_{\rm area}$ (Eq.~\eqref{e:eps:area}) and $\epsilon_{\rm sampling}$ (Eq.~\eqref{e:eps:sampl})   for different line profiles representing the same single normalized transition ($\tilde{\nu}_{if}$ = 4064.574702~cm$^{-1}$), including Gaussian ($\alpha_{\rm G}=0.1$), Lorentzian ($\gamma_{\rm L}$ is estimated by $\gamma_{\rm ref}=0.07$ and $n_{L}=0.5$), and Voigt, for the large  wavenumber range of [0, 8000] cm$^{-1}$ at different temperatures $T$, pressures $P$, and bin sizes $\Delta \tilde{\nu}$. 
Top-left panel: $\epsilon_{\rm area}$ for different temperatures $T$ at  $P=1$ bar and number of grid points $N_\textrm{points}=80001$ (bin size $\Delta \tilde{\nu}=0.1 \textrm{cm}^{-1}$) without a cutoff. Top-right panel: $\epsilon_{\rm area}$ for different pressures $P$ at $T=300$~K and $\Delta \tilde{\nu}=0.1 \textrm{cm}^{-1}$ without a cutoff. Bottom-left panel: $\epsilon_{\rm area}$ for different number of grid points $N_\textrm{points}$ (bin sizes $\Delta \tilde{\nu}$) at $T=300$~K and $P=1$ bar without a cutoff. Bottom-right panel: $\epsilon_{\rm sampling}$  for different number of grid points $N_\textrm{points}$ (bin sizes $\Delta \tilde\nu$) at $T=300$ K, $P=1$ bar, and cutoff $=100$~cm$^{-1}$.}
\label{fig:error plot}    
\end{figure}

Figure~\ref{fig:error plot} gives a detailed analysis of the line profile errors  $\epsilon_{\rm area}$ and $\epsilon_{\rm sampling}$  for a single normalized transition at $\tilde{\nu}_{if} = 4064.574702$~cm$^{-1}$ for  different temperatures $T$, pressures $P$, number of points $N_\textrm{points}$, and bin sizes $\Delta \tilde{\nu}$; and a large  wavenumber range of [0, 8000] cm$^{-1}$ is used.
All sampling methods have negligible $\epsilon_{\rm sampling}$, while all binned methods should lead to perfect $\epsilon_{\rm sampling}$, at least within the numerical error. 

For the cross-comparisons, the Gaussian line profile is always the most accurate sampling line profile to preserve the profile area even with smaller number of points due to its compact shape. 
The \textsc{SciPy} Voigt and Huml\'i\v{c}ek Voigt profiles provide similar accuracy. For constant pressure and grid number, the line profiles uncertainty decreases with increasing temperature, see Figure~\ref{fig:error plot} (first panel). In contrast, line profiles show a loss of accuracy as the pressure or grid number increase, see the second and third panels. 
The binned Lorentzian and Voigt profiles appear to degrade at higher pressures due to their non-negligible wings cutoff beyond the interval [0, 8000] cm$^{-1}$.

The area errors $\epsilon_{\rm area}$ and the sampling errors $\epsilon_{\rm sampling}$ of different line profiles are shown in the fourth panel of Figure~\ref{fig:error plot}. While the sampling profiles are almost perfect and perform better with more grid points, the binned profile 
can introduce a substantial error especially when the resolution is low. 
Generally,  in radiative transfer applications direct sampling line profiles are preferred \citep{jt708} especially for coarser grids.

We ignore the pseudo-Voigt profiles in this analysis as having too low accuracy.

\subsubsection{Comparison of  program run times with different line profiles}

To compare the program running times for computing cross-sections with different line profiles, in Table~\ref{tab:time} we give examples of cross-sections calculatons for different numbers of transitions,  bin sizes (numbers of grid points), and filters used. 

\begin{table}
\centering
\setlength{\tabcolsep}{2.35mm}{
\begin{threeparttable}
\caption{ Example of the program run times on 32 cores, in s, for computing cross-sections using different line profiles at temperatures $T=300$ K, pressures $P=1$ bar, and cutoff $=25$ cm$^{-1}$ for the ExoMol H$_2$O POKAZATEL line list \citep{jt734}, for the wavenumber range $[20000,22000]$ cm$^{-1}$. The left-hand panel compares the run times for different numbers of transitions $10^4, 10^6$, and $10^8$ with bin size of $0.1$ cm$^{-1}$ ($20001$ grid points) without any filters. The middle panel compares the run times for different numbers of transitions $10^4, 10^6$, and $10^8$ with bin size of $0.1$ cm$^{-1}$ ($20001$ grid points) with uncertainty and threshold filters. The right four columns compare the run times for different bin sizes (0.1, 0.01, 0.005, and 0.001 cm$^{-1}$) and the numbers of grid points (20 001, 200 001, 400 001, and 2000 001) using $10^8$ transitions with uncertainty and threshold filters.}
\label{tab:time}
\begin{tabular}{l|ccc|ccc|ccc}
\hline
\multirow{6}{*}{\thead{Line profiles}} & \multicolumn{2}{c|}{Uncertainty filter} & \multicolumn{1}{c|}{None} & \multicolumn{1}{c|}{Uncertainty filter} & \multicolumn{2}{c|}{$0.01$ cm$^{-1}$} & \multicolumn{1}{c|}{Uncertainty filter} & \multicolumn{2}{c}{$0.01$ cm$^{-1}$} \\
& \multicolumn{2}{c|}{Threshold filter} & \multicolumn{1}{c|}{None} & \multicolumn{1}{c|}{Threshold filter} & \multicolumn{2}{c|}{$10^{-30}$ cm molecule$^{-1}$} & \multicolumn{1}{c|}{Threshold filter} & \multicolumn{2}{c}{$10^{-30}$ cm molecule$^{-1}$} \\
& \multicolumn{2}{c|}{Bin size $\Delta \tilde\nu =$} & \multicolumn{1}{c|}{$0.1$ \textrm{cm}$^{-1}$} & \multicolumn{1}{c|}{Bin size $\Delta \tilde\nu =$} & \multicolumn{2}{c|}{$0.1$ \textrm{cm}$^{-1}$} & \multicolumn{1}{c|}{$N_{\rm transitions}=$} & \multicolumn{2}{c}{$10^8$} \\
& \multicolumn{2}{c|}{$N_{\rm points}=$} & \multicolumn{1}{c|}{20 001} & \multicolumn{1}{c|}{$N_{\rm points}=$} & \multicolumn{2}{c|}{20 001} & \multicolumn{1}{c|}{$N_{\rm points}=$} & \multicolumn{2}{c}{$2000/\Delta \tilde\nu+1$} \\ \cline{2-10} 
& \multicolumn{3}{c|}{$N_{\rm transitions}=$} & \multicolumn{3}{c|}{$N_{\rm transitions}=$} & \multicolumn{3}{c}{Bin size (\textrm{cm}$^{-1}$) $\Delta \tilde\nu=$} \\
& $10^4$ & \ \ $10^6$ & \ \ $10^8$ & $10^4$ & $10^6$ & \ \ $10^8$ & \ 0.01 & 0.005 & \ 0.001 \\
\hline
Doppler	              & 2.70 & \ \ 6.79 & \ \ 54.63 & 3.01 & 2.64 & \ \ 8.00 & \ 7.66 & \ 9.42 & \ 13.43 \\ 
Binned Doppler	      & 4.37 &    10.52 &    114.23 & 3.98 & 4.00 &    10.12 & \ 9.45 &  12.65 & \ 19.66 \\ 
\hline
Gaussian	          & 2.84 & \ \ 7.43 & \ \ 52.53 & 3.74 & 3.85 & \ \ 9.37 & \ 8.72 &  11.09 & \ 15.00 \\ 
Binned Gaussian	      & 4.42 &    10.85 &    119.77 & 3.85 & 3.96 &    10.08 & \ 9.39 &  12.28 & \ 20.03 \\ 
\hline
Lorentzian	          & 2.83 & \ \ 7.39 & \ \ 51.69 & 3.87 & 3.97 & \ \ 9.47 & \ 8.77 &  10.46 & \ 14.15 \\ 
Binned Lorentzian	  & 4.14 & \ \ 9.16 & \ \ 61.24 & 3.95 & 3.95 & \ \ 9.67 & \ 9.24 &  11.31 & \ 14.73 \\ 
\hline
SciPy Voigt	          & 2.99 & \ \ 9.77 & \ \ 83.19 & 3.87 & 3.91 & \ \ 9.34 & \ 8.85 &  10.83 & \ 13.21 \\
SciPy Wofz Voigt	  & 3.10 &    10.26 &    121.01 & 3.80 & 3.97 & \ \ 9.71 & \ 9.09 &  11.73 & \ 17.27 \\ 
Huml\'i\v{c}ek Voigt  & 3.24 &    10.69 &    161.25 & 3.90 & 4.05 &    10.15 &  12.62 &  20.03 & \ 46.62 \\
Thompson pseudo-Voigt & 2.99 & \ \ 9.89 & \ \ 85.86 & 3.80 & 3.90 &    10.14 & \ 9.27 &  11.63 & \ 19.85 \\ 
Kielkopf pseudo-Voigt & 4.01 &    10.15 & \ \ 85.98 & 3.82 & 3.94 &    10.08 & \ 9.57 &  13.09 & \ 19.72 \\ 
Olivero pseudo-Voigt  & 3.93 & \  9.93 & \ \ 88.55 & 3.81 & 3.89 &    10.01 & \ 9.58 &  12.48 & \ 20.44 \\ 
Liu-Lin pseudo-Voigt  & 3.98 &    10.03 & \ \ 85.20 & 3.88 & 3.99 &    10.18 & \ 9.40 &  14.02 & \ 19.63 \\ 
Rocco pseudo-Voigt	  & 3.97 &    10.09 & \ \ 85.47 & 3.97 & 3.98 & \ \ 9.83 & \ 9.56 &  13.08 & \ 20.16 \\ 
Binned Voigt	      & 6.19 &    13.80 &    657.68 & 3.95 & 4.35 &    10.32 &  22.38 &  38.78 &  107.06 \\
\hline
\end{tabular}
\end{threeparttable}
}
\end{table}

From Table~\ref{tab:time}, we can find that the Lorentzian profile is the fastest followed by the Doppler, Gaussian and \textsc{SciPy} Voigt profiles. Pseudo-Voigt profiles process data with \textsc{NumExpr} \citep{cooke2009numexpr} which is a fast numerical expression evaluator for \textsc{NumPy}. As an approximation method commonly in use, the Huml\'i\v{c}ek Voigt profile runs rather slowly. Because of the increased number of steps in binned profile calculations, the program costs more time than the sampling methods. 
The \textsc{SciPy} wofz Voigt, Huml\'i\v{c}ek Voigt, and Rocco pseudo-Voigt profiles are all have similar accuracy but significantly slower than the \textsc{SciPy} Voigt profile. 
Binned Voigt profiles are more accurate than standard sampling Voigt profiles, but take much more computer time.

To summarize, considering its higher accuracy and relatively short run times, we recommend the \textsc{SciPy} Voigt profile for calculating cross-sections.

%\red{It is a nice comparison for single lines. Can we also provide examples of the total times to produce typical cross-sections, for small, medium and large line lists, different line profiles and different Voigt implementations. How long does it take to generate a full cross section for 10to10 for a given T, P on a 0.01 cm-1 grid using Doppler, Lorentz and Voigt?}

\subsubsection{Comparison of program run times with \textsc{PyExoCross}, \textsc{ExoCross} and \textsc{RADIS}}

We tested run times of \textsc{PyExoCross}, ExoCross, and RADIS programs on 32 cores.
\textsc{PyExoCross} makes good use of the cores and costs 371 s for reading files, processing data, calculating absorption cross-sections, and saving results for H$_2$O at temperature $T=3000$ K, pressure $P=1$ bar, cutoff $=25$ cm$^{-1}$, uncertainty $\le 0.01$ cm$^{-1}$, and threshold $\ge 10^{-40}$ cm molecule$^{-1}$; wavenumber range from $0$ to $41200$ cm$^{-1}$ with \textsc{SciPy} Voigt line profile using the ExoMol $^{1}$H$_{2}^{16}$O line list of dataset POKAZATEL \citep{jt734}. Fortran program \textsc{ExoCross} costs 3600 s for whole processes with the same conditions; however, only about 300 s of this is actually processing data with the wall clock time dominated by reading and unpacking the files.

\textsc{RADIS} downloads ExoMol compressed bz2 files then converts them into HDF5 files and saves both original files and converted files to local for next time calculating. To compare with \textsc{RADIS}, \textsc{PyExoCross} costs 27 s (including reading compressed bz2 files, processing data, calculating cross-sections, and saving results) and \textsc{RADIS} costs 25 s (not including times for downloading, decompressing, and converting files; includes reading HDF5 files, processing data, and calculating cross-sections) on calculating absorption cross-sections of ExoMol $^{1}$H$_{2}^{16}$O line list of dataset POKAZATEL in the wavenumber range {[$20000$,$20500$]} cm $^{-1}$ at temperature $T=300$ K, pressure $P=1$ bar, cutoff $=10^{-27}$ cm$^{-1}$, and threshold $\ge 10^{-40}$ cm molecule$^{-1}$ with Voigt line profile.

\subsection{Additional functionality}
\label{sec:additionlfun}

\textsc{PyExoCross} provides the following  additional functionality for users:  (i) locating line lists in the ExoMol database with  uncertainties  provided as part of the States file; (ii) checking for updated line lists  compared with the last or new version of the master file, or the version the users saved last time; and (iii) downloading the States and Transitions files via the ExoMol API instead of downloading files one by one.

\section{Calculation protocol}
\label{sec:protocol}

\begin{figure}
\centering
\includegraphics[width=0.6\textwidth]{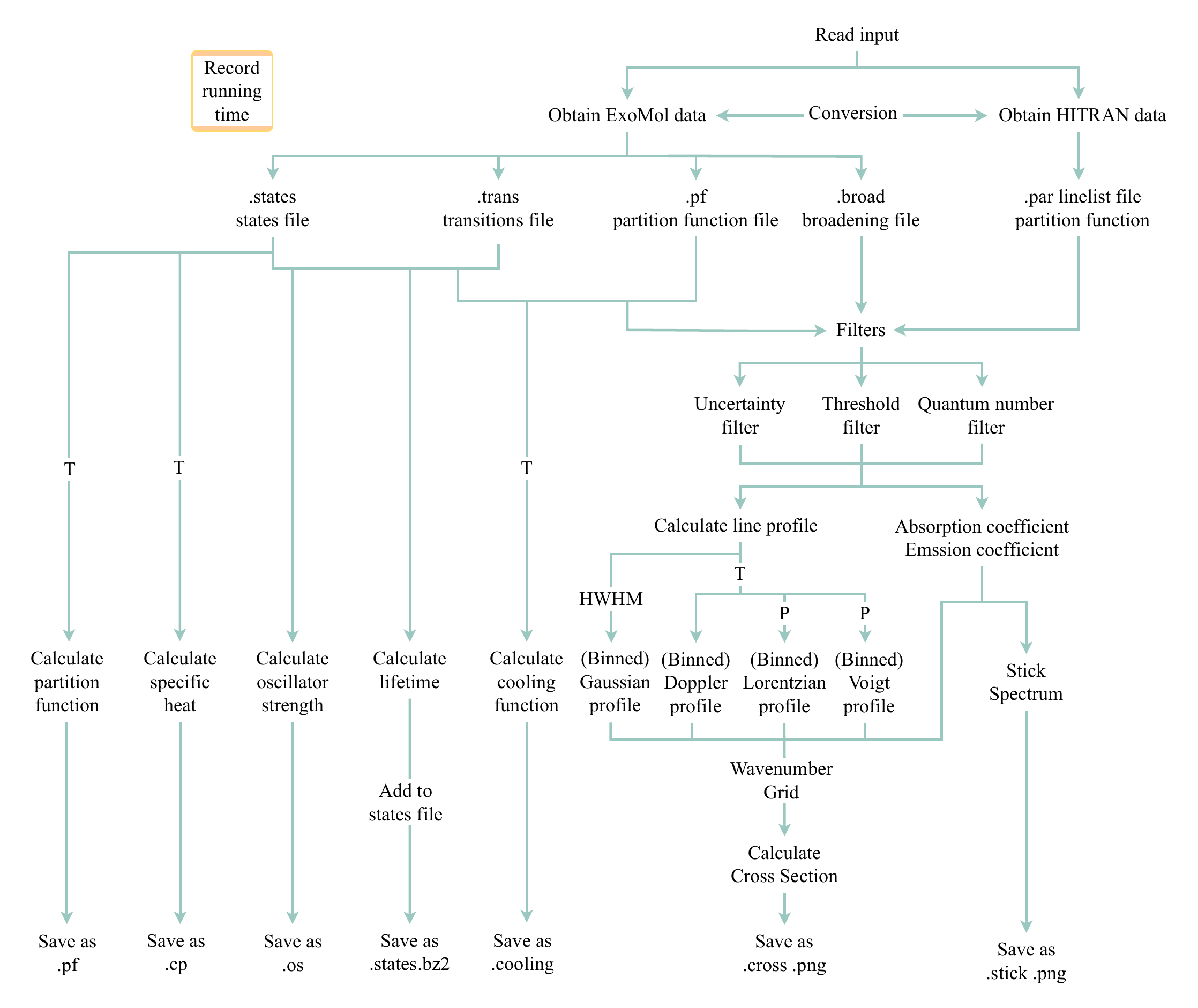}
\caption{Calculation protocol of the \textsc{PyExoCross} program. $T$ and $P$ refer to temperature and pressure, separately.}
\label{fig:pyexocross process}
\end{figure}

The \textsc{PyExoCross} program processes line lists with the following steps (see Figure~\ref{fig:pyexocross process}):
\begin{itemize}[itemindent=-0.5em,leftmargin=1.2em] 
\item[$-$] Populate the input file with the necessary data and settings to be read by \textsc{PyExoCross};
\item[$-$] Obtain (download) data from the ExoMol and/or HITRAN websites (if required) ;
\item[$-$] Read the ExoMol line lists files: (\texttt{.states}) States file, (\texttt{.trans}) Transitions file(s), (\texttt{.pf}) partition function file and (\texttt{.broad}) broadening file or the HITRAN line lists (\texttt{.par}) files and partition function;
\item[$-$] Compute partition function (if required);
\item[$-$] Compute specific heat (if required);
\item[$-$] Compute cooling function (if required);
\item[$-$] Compute radiative lifetime (if required);
\item[$-$] Compute oscillator strength (if required);
\item[$-$] For intensities, apply  the uncertainty or quantum numbers filters  (if required);
\item[$-$] Compute a line intensity (absorption or emission coefficient, if required);
\item[$-$] Compute line profile (if required);
\item[$-$] Compute cross-sections on a grid of wavenumbers (if required);
\item[$-$] Save the results of partition functions, specific heats, cooling functions, radiative lifetimes, and cross-sections into the separate files;
\item[$-$] Plot cross-sections (if required);
\item[$-$] Report program running time.
\end{itemize}

\section{Data format}

\textsc{PyExoCross} can use the input data in either ExoMol or HITRAN format. It can also convert between different formats including   HITRAN or ExoMol. 
 Table~\ref{tab:exomol file types} gives all file types in the ExoMol database. \textsc{PyExoCross} available file types are described in the above sections.

\begin{table}
\centering
\setlength{\tabcolsep}{2.2mm}{
\begin{threeparttable}
\caption{Specification of the ExoMol file types (contents in brackets are optional) \citep{jt898}.}
\label{tab:exomol file types}
\begin{tabular}{lllll}
\toprule
File extension & \makecell[c]{\textsc{PyExoCross} \\ available} & $N_{\rm files}$&File DSname &  Contents \\
\midrule
\texttt{.all} & Input &1& Master& Single file defining contents of the ExoMol database.\\
\texttt{.def} & Input &$N_{\rm tot}$& Definition& Defines contents of other files for each isotopologue.\\
\texttt{.states} & Input/Output &$N_{\rm tot}$& States & Energy levels, quantum numbers, lifetimes, (Land\'e $g$-factors, uncertainties).\\
\texttt{.trans} & Input/Output &$^a$& Transitions & Einstein $A$ coefficients, (wavenumber).\\
\texttt{.broad} & Input/Output &$N_{\rm mol}$& Broadening & Parameters for pressure-dependent line profiles.\\
\texttt{.cross} & Output &$^b$& Cross-sections& Temperature- or temperature and pressure-dependent cross-sections.\\
\texttt{.kcoef}& No &$^c$& $k$-coefficients& Temperature and pressure-dependent $k$-coefficients.\\
\texttt{.pf}   & Input/Output & $N_{\rm tot}$&Partition function&   Temperature-dependent partition function, (cooling function).\\
\texttt{.cf} & Output & $N_{\rm tot}$ & Cooling function & Temperature-dependent cooling function.\\
\texttt{.cp} & Output & $N_{\rm tot}$ & Specific heat & Temperature-dependent specific heat.\\
\texttt{.dipoles}& No &$N_{\rm tot}$& Dipoles & Transition dipoles including phases.\\
\texttt{.super} & No &$^d$& Super-lines & Temperature-dependent super-lines (histograms) on a wavenumber grid.\\
\texttt{.nm} & No &$^e$ & VUV cross-sections& Temperature and pressure-dependent VUV cross-sections (wavelength, nm).\\
\texttt{.fits}, \texttt{.h5}, \texttt{.kta}  & No &$^f$ & Opacities & Temperature and pressure-dependent opacities for radiative-transfer applications.\\
\midrule
\texttt{.overview}& No &$N_{\rm mol}$& Overview & Overview of datasets available.\\
\texttt{.readme}& No &$N_{\rm iso}$& Readme & Specifies data formats.\\
\texttt{.model}& No &$N_{\rm iso}$& Model & Model specification.\\
\midrule
\texttt{.cont}& No &$N_{\rm iso}/0$&Continuum& Continuum contribution to the photoabsorption.\\
\texttt{.pdef}& No &$N_{\rm tot}$& Definition& Defines contents of \texttt{.photo} files for
each isotopologue.\\
\texttt{.photo}& No &$N_{\rm iso}$&Photodissociation& Photodissociation cross-sections for each lower state.\\
\bottomrule
\end{tabular}
$N_{\rm files}$: Total number of possible files, \\
$N_{\rm mol}$: Number of molecules in the database, \\
$N_{\rm tot}$: It is the sum of $N_{\rm iso}$ for the $N_{\rm mol}$ molecules in the database, \\
$N_{\rm iso}$: Number of isotopologues considered for the given molecule, \\
$^a$ There are $N_{\rm tot}$ sets of \texttt{.trans} files but for molecules with large numbers of transitions the \texttt{.trans} files are subdivided into wavenumber regions, \\
$^b$ There are $N_{\rm cross}$ sets of \texttt{.cross} files for isotopologue, \\
$^c$ There are $N_{\rm kcoef}$ sets of \texttt{.kcoef} files for each isotopologue, \\
$^d$ There are $N_T$ sets of $T$-dependent super-lines, \\
$^e$ There are $N_{VUV}$ sets of VUV cross-sections, \\
$^f$ Set of opacity files in the format native to specific radiative-transfer programs. 
\end{threeparttable}
}
\end{table}

\subsection{HITRAN format}

%\red{I was wondering if this section could be moved to the section above where the HITRAN format is introduced. It is a bit redundant. }
Compared to the ExoMol format line list files (\texttt{.states} and \texttt{.trans}), the HITRAN format line list file (\texttt{.par}) includes the parameters related to both upper-state and lower-state. Table~\ref{tab:HITRAN format} gives the format of the HITRAN line-by-line records. The description of the HITRAN parameters is specified in Table~\ref{tab:HITRAN data format}.  \textsc{PyExoCross} processes the HITRAN format data to obtain the parameters for calculating cooling functions, oscillator strength, stick spectra and cross-sections. The partition function, molar mass and fractional abundance are available from \href{https://hitran.org/docs/iso-meta/}{https://hitran.org/docs/iso-meta/}. \textsc{PyExoCross} extracts these parameters from the webpage directly and automatically instead of entering these parameters by hand. The partition functions at different temperatures are recorded in the \texttt{.txt} file online which can be mined by \textsc{PyExoCross}. 

\begin{python}[caption={Python code for calculating the physical and astronomical constants.}, label={listing:constant}]
import astropy.constants as ac
Tref = 296.0                        # Reference temperature is 296 K
Pref = 1.0                          # Reference pressure is 1 bar
N_A = ac.N_A.value                  # Avogadro number (1/mol)
h = ac.h.to('erg s').value          # Planck's constant (erg s)
c = ac.c.to('cm/s').value           # Velocity of light (cm/s)
kB = ac.k_B.to('erg/K').value       # Boltzmann's const (erg/K)
R = ac.R.to('J / (K mol)').value    # Molar gas constant (J/(K mol))
c2 = h * c / kB                     # Second radiation constant (cm K)
\end{python}

\subsection{Units}
\label{sec:units}

The default labels and units used in the \textsc{PyExoCross} program are listed in Table~\ref{tab:units}. The physical and astronomical constants in calculations are shown in Listing~\ref{listing:constant}. \texttt{astropy.constants} is a sub-package of \texttt{Astropy} where \texttt{Astropy} is a community Python package for astronomy \citep{robitaille2013astropy}. Some of the parameters used by \textsc{PyExoCross} are not in SI units but in cgs units (e.g. erg).

\begin{table}
\centering
\setlength{\tabcolsep}{5mm}{
\begin{threeparttable}
\caption{Default units used in the \textsc{PyExoCross} program.}
\label{tab:units}
\begin{tabular}{lll}
\toprule
Quantity & Label & Units \\
\midrule
Partition function        & $\textsl{Q}(T)$   & none \\
Specific heat             & $\textsl{C}_p(T)$ & J K$^{-1}$ mol$^{-1}$ \\
Cooling function          & $\textsl{W}(T)$   & erg (s molecule sr)$^{-1}$ \\
Radiative lifetime        & $\tau_i$          & s \\
Oscillator strength       & $f$               & none \\
Molar mass                & $M$               & Dalton \\
Wavenumber                &                   & cm$^{-1}$ \\
Wavelength                &                   & $\mu$m \\      
Temperature               & $T$               & K \\
Pressure                  & $P$               & bar \\        
Absorption coefficient    & $I_{fi}$          & cm molecule$^{-1}$ \\        
Emission coefficient      & $\epsilon_{if}$   & erg (s molecule sr)$^{-1}$ \\   
Absorption cross-sections & $\sigma_{\rm ab}$     & cm$^2$ molecule$^{-1}$ \\         
Emission cross-sections   & $\sigma_{\rm em}$     & erg cm (s molecule sr)$^{-1}$ \\
\bottomrule
\end{tabular}
\end{threeparttable}
}
\end{table}

% \begin{listing}
% \begin{minted}{python}
% import astropy.constants as ac
% Tref = 296.0                        # Reference temperature is 296 K
% Pref = 1.0                          # Reference pressure is 1 bar
% N_A = ac.N_A.value                  # Avogadro number (1/mol)
% h = ac.h.to('erg s').value          # Planck's constant (erg s)
% c = ac.c.to('cm/s').value           # Velocity of light (cm/s)
% kB = ac.k_B.to('erg/K').value       # Boltzmann's const (erg/K)
% R = ac.R.to('J / (K mol)').value    # Molar gas constant (J/(K mol))
% c2 = h * c / kB                     # Second radiation constant (cm K)
% \end{minted}
% \caption{Python code for calculating the physical and astronomical constants}
% \label{listing:constant}
% \end{listing}

\section{Conclusions}
\label{sec:conclusions}

We have developed \textsc{PyExoCross}, a python new program for calculating the spectroscopic properties of molecules with spectral line lists. \textsc{PyExoCross} is the Python version of the Fortran program \textsc{ExoCross} \citep{jt708}. Some CPU parallelization is available, at present there is no GPU support. The spectral line lists of the ExoMol database are used to generate partition functions, specific heats, cooling functions, radiative lifetimes, oscillator strengths, stick spectra, and cross-sections by \textsc{PyExoCross}. At present, \textsc{PyExoCross} can ingest data from the ExoMol, HITRAN, and HITEMP databases to provide parts of the functionalities supported by the \textsc{ExoCross} program. Other features will be developed later, such as other data formats while this project will also implement the processing of other databases in the future. 

Future enhancements to \textsc{PyExoCross} include non-LTE (non-local thermodynamic equilibrium) calculations, generation of k-tables, and opacity production. A library of fractional isotopic abundances will be built for the \textsc{PyExoCross} program to calculate the intensities for different distributions of isotopologues. \textsc{PyExoCross} will also support more molecular spectral databases in future versions, along with support for data type and data format conversions among different databases.

\section*{Acknowledgements}

We thank Ahmed Al-Refaie for helpful discussions and suggestions on the methods of this work. This work was supported by the European Research Council under Advanced Investigator Project 883830.

%%%%%%%%%%%%%%%%%%%%%%%%%%%%%%%%%%%%%%%%%%%%%%%%%%
\section*{Data Availability}

The \textsc{PyExoCross} program is publicly accessible on GitHub (\href{https://github.com/ExoMol/PyExoCross.git}{https://github.com/ExoMol/PyExoCross.git}) and its program manual webpage is \href{https://pyexocross.readthedocs.io/}{https://pyexocross.readthedocs.io/}.\\
The \textsc{ExoCross} Fortran program is freely available from the GitHub (\href{https://github.com/ExoMol/ExoCross.git}{https://github.com/ExoMol/ExoCross.git}). The \textsc{ExoCross} manual webpage is \href{https://exocross.readthedocs.io/}{https://exocross.readthedocs.io/}.\\ 
The molecular line lists discussed in this paper and many more are available from
the ExoMol  website (\href{https://www.exomol.com/}{https://www.exomol.com/}).


\bibliographystyle{rasti}
% \input{main.bbl}

%%%%%%%%%%%%%%%%%%%%%%%%%%%%%%%%%%%%%%%%%%%%%%%%%%

%%%%%%%%%%%%%%%%% APPENDICES %%%%%%%%%%%%%%%%%%%%%

%\appendix


%%%%%%%%%%%%%%%%%%%%%%%%%%%%%%%%%%%%%%%%%%%%%%%%%%

% Don't change these lines
\bsp	% typesetting comment
\label{lastpage}
\end{document}